\documentclass[aps,prd,superscriptaddress,floatfix,nofootinbib]{revtex4}

\usepackage{verbatim}
\usepackage{mathtools}
\usepackage{ragged2e}
\usepackage{amsmath}
\usepackage{amsfonts}
\usepackage{amssymb}
\usepackage{graphicx}
\usepackage{slashed}
\usepackage{bm}
\usepackage{color}
\usepackage{epsf}
\usepackage{orcidlink}
\usepackage[parfill]{parskip}
\usepackage{nicefrac}
\usepackage{mathrsfs}
\usepackage[percent]{overpic}
\usepackage{hyperref}
\hypersetup{
  colorlinks   = true, 
  urlcolor     = blue, 
  linkcolor    = black, 
  citecolor   = red 
}

\newcommand{\re}{\mathfrak{Re}}
\newcommand{\im}{\mathfrak{Im}}

\newcommand{\beq}{\begin{eqnarray}}
\newcommand{\eeq}{\end{eqnarray}}
\newcommand{\Slash}[1]{{\ooalign{\hfil/\hfil\crcr$#1$}}}

\newcommand{\nn}{\nonumber \\}

\newcommand\Ln[1]{{\,\textrm{ln}\left(#1\right)}}  

\DeclareUnicodeCharacter{2212}{-}

\begin{document}
	
	\title{Photon-nucleon entanglement in Compton scattering  \\ at low and high energies}
    \author{Yoshitaka Hatta \orcidlink{0000-0001-6880-3238}}
    \affiliation{Physics Department, Brookhaven National Laboratory, Upton, NY 11973, USA}
    \affiliation{RIKEN BNL Research Center, Brookhaven National Laboratory, Upton, NY 11973, USA}
    \author{V\'ictor Mart\'inez-Fern\'andez \orcidlink{0000-0002-0581-7154}}
    \affiliation{Universit\'e Paris-Saclay - CEA - IRFU, 91191 Gif-sur-Yvette, France}
    \affiliation{Center for Frontiers in Nuclear Science, Stony Brook University, Stony Brook, NY 11794, USA}
    \date{\today}

\begin{abstract}
We study spin-spin entanglement in the final state photon-nucleon system in  Compton scattering, both at low energy below the pion threshold and at high energy in  perturbative QCD to next-to-leading order. We first establish a no-go theorem showing that, for any spin-$\frac{1}{2}$ target,  entanglement cannot be generated in unpolarized Compton scattering if the scattering amplitudes are real.  
We then consider polarized Compton scattering off the electron, the proton and the neutron. At low energy, we uncover a rich variety of maximally entangled  Bell states and their unitary equivalents realized across different regions of the kinematic plane. Interestingly, the proton and neutron targets exhibit  distinct  patterns of entanglement. In the neutron case, the electric and magnetic polarizabilities dramatically influence  the pattern and even the existence of entanglement.  This suggests that entanglement can serve as  a novel tool for investigating the detailed electromagnetic properties of the nucleons.     

\end{abstract}

\maketitle

\section{Introduction}

Compton scattering off the nucleon (proton and neutron) $\gamma + N \to \gamma' + N'$  is a unique process that provides access to  different aspects of the nucleon structure across a wide range of energy scales  \cite{Drechsel:2002ar,Schumacher:2005an,Hagelstein:2015egb}. At low photon energy $\omega$ much less than the pion mass $m_\pi\sim 100$ MeV, a low-energy theorem \cite{Powell:1949bhl,Gell-Mann:1954wra,Low:1954kd} states that the cross section depends only on the charge and the magnetic moment of the nucleon. At somewhat higher energy, still below the pion threshold, the scattering starts to be sensitive to the electric and magnetic polarizabilities $\alpha_E,\beta_M$ which encode  dynamical information about  the nucleon's internal structure  \cite{Baldin1960,Petrunkin1961}. Above the pion threshold, up to several hundred MeV, various intermediate excited states such as the $\Delta$ resonances come into play. In this regime,  chiral perturbation theory \cite{Bernard:1991rq,Hildebrandt:2003fm,Beane:2004ra,Lensky:2015awa,Hagelstein:2020vog} and dispersion relations \cite{Baldin1960,Babusci:1998ww,Drechsel:2002ar} provide  two complementary theory frameworks for the description of scattering.  In the high energy regime, $\omega\gg 1$ GeV, the process becomes sensitive to the partonic structure of the nucleon. In the so-called wide angle Compton scattering kinematics $s\sim -t\sim -u$,  the cross section can  be calculated within perturbative QCD \cite{Radyushkin:1998rt,Diehl:1998kh,Huang:2001ej} in terms of the generalized parton distributions (GPD)   \cite{Diehl:2003ny,Belitsky:2005qn}. 

In this paper, we study quantum informational aspects of Compton scattering off the nucleon. More specifically, we are interested in the  entanglement between the final state nucleon and photon in spin space. A nucleon has two spin states $\pm \frac{1}{2}$ which can be treated as a qubit. Likewise, a photon has two helicity states $\pm 1$ which can be also regarded as a qubit. Before scattering, the two qubits are unentangled, but  entanglement can be generated in the course of scattering, depending on the energy, scattering angle and initial polarizations. Our goal is to determine the existence and strength of entanglement by combining the established tools of hadron physics/QCD with the machinery of quantum information science.

There is already a significant body of work on  entanglement and Bell's inequality in low-energy nucleon-nucleon scattering, both theoretically and experimentally  \cite{Lamehi-Rachti:1976wey,Sakai2006,Beane:2018oxh,Liu:2022grf,Bai:2022hfv,Miller:2023ujx,Bai:2023hrz,Oishi:2024quz,Shen:2025aqf,Oishi:2025wiw,Witala:2025wvi}. Also, spin entanglement in (virtual) photon-nucleon inelastic scattering has been studied at high energy \cite{Qi:2025onf,Fucilla:2025kit,Hatta:2025obw,Fucilla:2026mkg,Agrawal:2026zwa,Hentschinski:2026otq}. 
However, to our knowledge, no work has been done on entanglement in elastic photon-nucleon scattering at any energy, although it should be mentioned that many authors studied entanglement in electron-target Compton scattering~\cite{Cervera-Lierta:2017tdt,Ahrens:2017zsb,NagChowdhury:2021nme,Fedida:2022izl,Blasone:2024jzv,Smeets:2025wro,Blasone:2025ddi}. We will first establish a `no-go theorem' showing that, in unpolarized Compton scattering off any spin-$\frac{1}{2}$ target, entanglement is impossible in kinematical regimes where the scattering amplitudes are real. To realize entanglement, we  consider polarized Compton scattering in which the incoming photon and nucleon are both polarized. Such  `polarization-assisted' entanglement has been studied  in various contexts~\cite{Cervera-Lierta:2017tdt,Blasone:2024jzv,Cheng:2025zaw,Altakach:2026fpl,Zhang:2026nwm,Fang:2026ddi,Bloss:2026yrf}. Fortunately, there are already facilities capable of performing polarized Compton scattering such as Mainz Microtron (MAMI) \cite{A2:2014iky}
and   Jefferson Laboratory \cite{JeffersonLabHallA:2004bve}.
Moreover, (polarized) Deeply Virtual Compton Scattering is one of the flagship processes of the future Electron-Ion Collider \cite{AbdulKhalek:2021gbh}. At low energy, polarized experiments are especially suited for the extraction of  the nucleon polarizabilities \cite{Babusci:1998ww,Hildebrandt:2003md}. At high energy, polarization provides an additional  lever arm for disentangling different  GPDs \cite{Belitsky:2005qn}.     While it is challenging  to measure final state spin correlations, we propose entanglement as a novel research direction in these experiments, one that naturally connects to the broader quantum information science (QIS) community.     

In Section II, we explain the basic kinematics of Compton scattering and introduce  the helicity amplitudes. In  Section III, we construct the spin density matrix with arbitrary polarizations in the initial state and apply it to electron-target Compton scattering. In Sections IV, we consider the low energy regime $\omega<m_\pi$ and uncover a rich pattern of entanglement in the final state including the emergence of maximally entangled Bell states. We also compare the proton and neutron targets and study the role of the polarizabilities. In Section V, we do the same in the high energy regime $\omega \gg 1$ GeV where the helicity amplitudes are calculable in terms of GPDs.

\begin{figure}[t]
\begin{overpic}[width=0.8\textwidth]{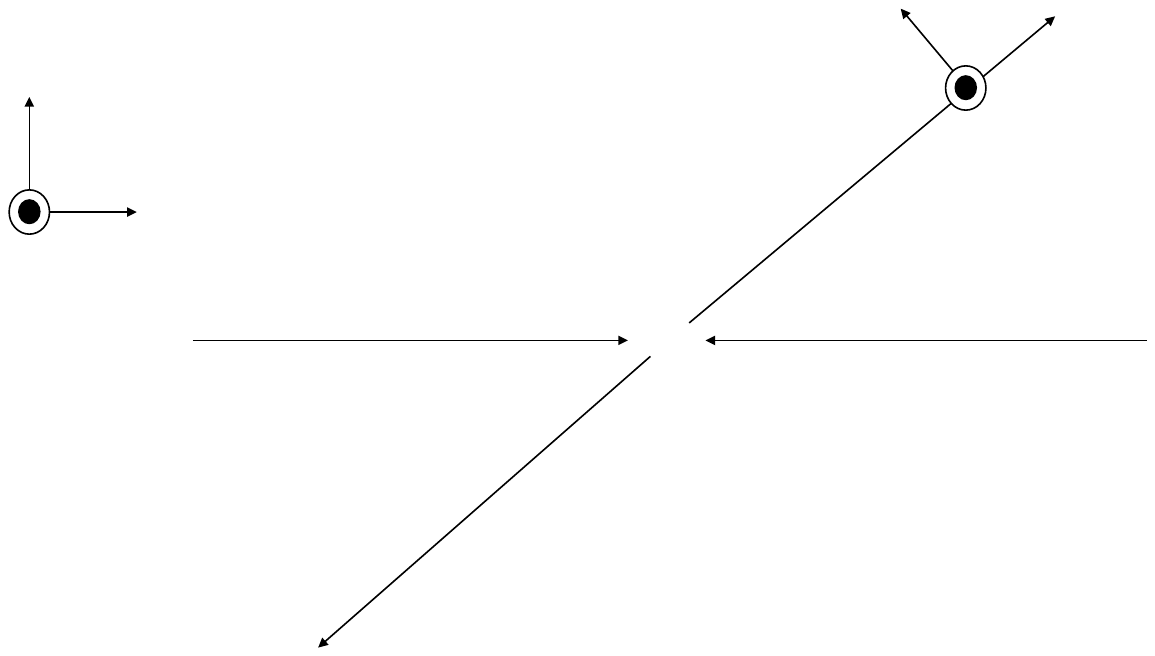}
\put(6,54){\large $x$}
\put(8,46){\large $y$}
\put(17,50){\large $z$}
\put(60,63){\large $x'$}
\put(65,53){\large $y'$}
\put(72,62){\large $z'$}
\put(70,43){\Large{$\gamma$}} 
\put(55,43){\Large{$\theta$}} 
\put(24,43){\Large{$N$}}
\put(53,50){\Large{$N'$}}
\put(35,34){\Large{$\gamma'$}}
        \end{overpic}
        \vspace{-34mm}
 \caption{ Compton scattering in the CM frame. Scattering takes place in the $xz$ plane with scattering angle $\theta$.   
   }. \label{frame}
    \end{figure}

\section{Compton scattering off the nucleon}\label{sec::RCS}

The process of interest is  Compton scattering $N(k)+\gamma(q) \to N(k')+\gamma(q')$ off the nucleon $N$ which can be either the proton $N=p$ or the neutron $N=n$. Many studies in the low-energy literature use  the laboratory frame where the nucleon is initially at rest. For the  purpose of computing the spin density matrix, it is advantageous to work in the center-of-mass (CM) frame where the nucleon   and the photon $\gamma$ move in the $+z$ and $-z$ directions, respectively. The nucleon is deflected by an angle $\theta$, and we define   the $xz$ plane as the scattering plane on an event-by-event basis, see Fig.~\ref{frame}. The initial and final momenta are  parametrized  as 
\beq
k^\mu &=& (E,0, 0,\omega), \nn 
k'^\mu &=& (E,\omega \sin\theta,0,\omega \cos\theta),\nn
q^\mu &=& \omega(1,0, 0,-1)\equiv \omega(1,\hat{q}), \nn 
q'^\mu &=& \omega(1,-\sin\theta,0,-\cos\theta)\equiv \omega(1,\hat{q}') ,\label{cm}
\eeq
where $E=\sqrt{m^2+\omega^2}$ and $m$ is the nucleon mass.  Due to  parity and time-reversal symmetry, there are six independent scattering amplitudes \cite{Hearn:1962zz,Hemmert:1997tj,Babusci:1998ww,Hagelstein:2015egb}.  A common parametrization in the low-energy  ($\omega \ll 1$ GeV) literature is  
\beq
T &=& \xi_{out}^\dagger \Bigg\{A_1\epsilon'^*\cdot \epsilon + A_2\hat{q}'\cdot \epsilon \hat{q}\cdot \epsilon'^* + iA_3 \sigma\cdot (\epsilon'^*\times \epsilon) +iA_4  \sigma\cdot (\hat{q}'\times \hat{q})\epsilon'^*\cdot \epsilon \nn 
&&+iA_5 \sigma\cdot [(\epsilon'^*\times \hat{q}) \epsilon\cdot \hat{q}' -(\epsilon\times \hat{q}')\epsilon'^*\cdot \hat{q}]  +iA_6 \sigma\cdot [(\epsilon'^*\times \hat{q}') \epsilon\cdot \hat{q}' -(\epsilon\times \hat{q})\epsilon'^*\cdot \hat{q}] \Biggr\}\xi_{\rm in}\nn
&\equiv& \epsilon'^*_\lambda\xi_{out}^\dagger T_{\lambda i} \xi_{\rm in} \epsilon_i .\label{as}
\eeq
$\xi_{in}$ and $\xi_{out}$ are  two-component spinors describing the spin state of the initial and final nucleons. $\epsilon$, $\epsilon'$ are the three-dimensional polarization vectors for the initial and final photons satisfying $\hat{q}\cdot \epsilon=\hat{q}'\cdot \epsilon'=0$. The six amplitudes $A_{1,2,3,4,5,6}$ are nonperturbative functions of $\omega$ and $\theta$, or equivalently, the Mandelstam variables 
\beq
s=(E+\omega)^2, \qquad t=-2\omega^2(1-\cos\theta), \qquad u=-2\omega(E+\omega \cos\theta)+m^2.
\eeq 
They are real-valued functions below the pion threshold $\omega \lesssim m_\pi$, and become complex above the threshold.  
For our purpose, it is more convenient to work with the helicity amplitudes $\phi_{1,2,3,4,5,6}$ based on the Jacob-Wick formalism \cite{Jacob:1959at,Hearn:1962zz,Leader:2001nas}
\beq
&& \phi_1=\langle +_{N'}+_{\gamma'}|T|+_N+_\gamma\rangle=\langle --|T|--\rangle, \nn 
&& \phi_2=\langle --|T|++\rangle = -\langle ++|T|--\rangle ,\nn 
&&  \phi_3=\langle +-|T|++\rangle=\langle -+|T|--\rangle=\langle ++|T|+-\rangle=\langle --|T|-+\rangle, \nn
&& \phi_4=\langle -+|T|++\rangle =-\langle +-|T|--\rangle =-\langle ++|T|-+\rangle
=\langle --|T|+-\rangle
\nn &&  \phi_5=\langle -+|T|-+\rangle = \langle +-|T|+-\rangle,
\nn 
&& \phi_6= \langle +-|T|-+\rangle =-\langle -+|T|+-\rangle . \label{heli}
\eeq
$\pm_N$ and $\pm_\gamma$ denote the helicities $\pm \frac{1}{2}$ and $\pm 1$ of the incoming nucleon and  photon, respectively, and similarly for the outgoing particles $\pm_{N'},\pm_{\gamma'}$. The sign differences  when flipping helicities or interchanging the initial and final states are dictated by parity and time-reversal symmetry. 
Our normalization is such that the unpolarized differential cross section in the CM frame reads 
\beq
\frac{d\sigma}{d\Omega}=\frac{1}{128\pi^2 s}\left(|\phi_1|^2+|\phi_2|^2+2|\phi_3|^2+2|\phi_4|^2+|\phi_5|^2+|\phi_6|^2\right) .
\eeq
The explicit relation to the amplitudes in (\ref{as}) is, for example, 
\beq
\phi_1=2m\epsilon'^*_{+\lambda} (\xi^+_{out})^\dagger T_{\lambda i} \xi_{in}^+ \epsilon_{+i},
\eeq
where the helicity eigenstates  are 
\beq
&& \epsilon_\pm=\frac{1}{\sqrt{2}}(1,\mp i,0)=\frac{1}{\sqrt{2}}(\epsilon_x\mp i\epsilon_y), \qquad \epsilon'^{*}_\pm= \frac{1}{\sqrt{2}} (\cos\theta,\pm i,-\sin\theta),  \label{epsilonconvention}
\eeq
\beq
\xi^+_{in}=\begin{pmatrix}1 \\ 0\end{pmatrix}, \qquad \xi^-_{in}=\begin{pmatrix}0 \\ 1\end{pmatrix}, \qquad (\xi^+_{out})^\dagger=\left(\cos\frac{\theta}{2},\sin\frac{\theta}{2}\right),\qquad (\xi^-_{out})^\dagger=\left(-\sin\frac{\theta}{2},\cos\frac{\theta}{2}\right).  \label{eigen}
\eeq 
(Remember that the incoming photon is left-moving.) 
A straightforward calculation gives  (cf. \cite{Hearn:1962zz})
\beq
&&\phi_1=\frac{m}{2} \cos\frac{\theta}{2}\Bigl[2A_1-A_2+6A_3+A_4-6A_5+8A_6+2(A_1-A_3+4A_5-4A_6)\cos\theta +(A_2-A_4-2A_5)\cos 2\theta \Bigr],\nn
&&\phi_2= \frac{m}{2} \sin\frac{\theta}{2}\Bigl[2A_1+A_2+6A_3-A_4+6A_5+8A_6-2(A_1-A_3-4A_5-4A_6)\cos\theta -(A_2-A_4-2A_5)\cos 2\theta \Bigr] ,\nn 
&&\phi_3= -2m\cos\frac{\theta}{2}\sin^2\frac{\theta}{2} \Bigl[A_1+A_2-A_3+A_4-2A_6+(A_2-A_4-2A_5)\cos\theta\Bigr], \nn 
&&\phi_4=  2m \sin\frac{\theta}{2}\cos^2\frac{\theta}{2} \Bigl[-A_1+A_2+A_3+A_4+2A_6+(-A_2+A_4+2A_5)\cos\theta\Bigr], \nn 
&&\phi_5= 2m\cos^3\frac{\theta}{2}\Bigl[A_1-A_3-(A_2-A_4-2A_5)(1-\cos\theta)\Bigr],\nn 
&& \phi_6=2m\sin^3\frac{\theta}{2}\Bigl[-A_1+A_3+(-A_2+A_4+2A_5)(1+\cos\theta)\Bigr]. \label{transf}
\eeq
The inverse transformation is 
\beq
&&A_1=\frac{1}{4m}\left((\phi_1-2\phi_3+\phi_5)\cos\frac{\theta}{2}+(\phi_2-2\phi_4-\phi_6)\sin\frac{\theta}{2}\right), \nn
&&A_2= \frac{1}{8m\sin^2\frac{\theta}{2}}\left((\phi_1-2\phi_3+\phi_5)\cos\frac{\theta}{2}+(\phi_2-2\phi_4-\phi_6)\sin\frac{\theta}{2}-\frac{\phi_1+\phi_5}{\cos\frac{\theta}{2}} +\frac{2\phi_4\sin\frac{\theta}{2}}{\cos^2\frac{\theta}{2}}\right),
\nn
&&A_3=\frac{1}{4m}\left((\phi_1-2\phi_3+\phi_5)\cos\frac{\theta}{2}+\frac{2\phi_6}{\sin\frac{\theta}{2}}-\frac{2\phi_5}{\cos\frac{\theta}{2}} +(\phi_2-2\phi_4-\phi_6)\sin\frac{\theta}{2}\right), \nn 
&&A_4=\frac{1}{8m}\left(\frac{-\phi_2+2\phi_4+\phi_6}{\sin\frac{\theta}{2}} +\frac{\phi_1-2\phi_3+\phi_5}{\cos\frac{\theta}{2}}\right),\nn
&&A_5=\frac{1}{64m\sin^2\frac{\theta}{2}}\Biggl(\frac{8\phi_6\cos^2\frac{\theta}{2}}{\sin\frac{\theta}{2}}-\frac{1}{\cos^3\frac{\theta}{2}}\Bigl\{\phi_1+2\phi_3-3\phi_5+4(\phi_3+\phi_5)\cos\theta -2\phi_2\sin\theta \nn && \qquad  \qquad  -(\phi_1-2\phi_3+\phi_5)\cos 2\theta -(\phi_2-2\phi_4)\sin 2\theta\Bigr\}\Biggr),
\nn 
&&A_6=\frac{1}{8m} \left(\frac{-\phi_6}{\sin^3\frac{\theta}{2}}+\frac{\phi_3}{\sin^2\frac{\theta}{2}\cos\frac{\theta}{2}}+\frac{\phi_5}{\cos^3\frac{\theta}{2}} + \frac{\phi_4}{\sin\frac{\theta}{2}\cos^2\frac{\theta}{2}}\right). \label{inverse}
\eeq

As an example, consider the Born diagrams for Compton scattering off an  electron with mass $m_e$ 
\beq
{\cal M}= -ie^2 \epsilon'^*_\mu \epsilon_\nu\bar{u}(k')\left[\frac{\gamma^\mu (\Slash q+\Slash k+m_e) \gamma^\nu }{2k\cdot q}+\frac{\gamma^\nu (\Slash k-\Slash q'+m_e)\gamma^\mu}{-2k \cdot q'}\right]u(k). \label{born}
\eeq
where $e^2=4\pi \alpha_{em}=\frac{4\pi}{137}$. The helicity amplitudes  can be readily computed \cite{Tsai:1972sg,Hagelstein:2015egb,Blasone:2024jzv} 
\beq
 \phi_1 &=& -2e^2\cos\frac{\theta}{2} \left( 1- \frac{E-3\omega}{E+\omega\cos\theta
}\sin^2\frac{\theta}{2} \right)\to -2e^2 \cos^3\frac{\theta}{2},  \nn 
 \phi_2&=&
 -2e^2\frac{(E-\omega)^2}{m_e(E+\omega \cos\theta)}\sin^3\frac{\theta}{2} \to -2e^2 \sin^3\frac{\theta}{2}  ,   \nn 
 \phi_3&=&2e^2\frac{E-\omega}{E+\omega\cos\theta}\sin^2\frac{\theta}{2}\cos\frac{\theta}{2}\to 2e^2 \sin^2\frac{\theta}{2}\cos\frac{\theta}{2} , 
\nn
 \phi_4&=&2e^2\frac{m_e}{E+\omega\cos\theta}\sin\frac{\theta}{2} \cos^2\frac{\theta}{2}\to  2e^2 \sin\frac{\theta}{2}\cos^2\frac{\theta}{2}   ,
\nn
\phi_5&=&-2e^2 \frac{E+\omega}{E+\omega\cos\theta}\cos^3\frac{\theta}{2} \to -2e^2\cos^3\frac{\theta}{2} ,
\nn
 \phi_6&=&2e^2 \frac{m_e }{E+\omega\cos\theta}\sin^3\frac{\theta}{2} \to  2e^2 \sin^3\frac{\theta}{2},
\label{phiphi}
\eeq
where the arrow means the Thomson scattering limit $\omega\to 0$. In this limit,  the lab frame  and the CM frame are essentially the same, and we recover the familiar formula 
\beq
\frac{d\sigma}{d\Omega}=\frac{\alpha^2}{2m_e^2}(1+\cos^2\theta) .
\eeq

\section{Spin density matrix}

Being a spin-$\frac{1}{2}$ fermion, a nucleon can be considered a  qubit as far as the spin degree of freedom is concerned.  Likewise, since a real photon has two physical polarizations, it can be also treated as a qubit.  Depending on the collision energy $\sqrt{s}=E+\omega$, the  scattering angle $\theta$ and initial polarizations, the two-qubit system in the final state are entangled. The  degree of entanglement can be quantified by computing the spin density matrix 
\beq
\rho= \frac{1}{4}\left(\mathbb{I}\otimes \mathbb{I} +B^a_N \sigma^a\otimes \mathbb{I} + B^b_\gamma \mathbb{I}\otimes \tau^b + C^{ab}\sigma^a\otimes \tau^b\right), \label{density}
\eeq 
where $\mathbb{I}$ is the unit 2x2 matrix and $\sigma^{a}$, $\tau^b$ are the Pauli matrices associated with the outgoing  nucleon and photon qubits, respectively. The vectors $\vec{B}_N$, $\vec{B}_\gamma$ denote the polarizations of the final state nucleon and photon, respectively, and the 3x3 matrix $C$ represents their correlation.  The density matrix is frame-dependent and basis-dependent, although the presence of entanglement is independent of these choices. Here we work in   the $x'y'z'$ frame depicted in Fig.~\ref{frame} which  can be reached from the original $xyz$ frame by rotating around the $y$-axis by angle $\theta$. Correspondingly, $a,b=x',y',z'$ in (\ref{density}) and in what follows.

We have computed the density matrix  (\ref{density}) in two ways,  directly in terms of the $A$'s in (\ref{as}) and in terms of the helicity amplitudes. The second  method is simpler and will be described below. The first method is relegated to the Appendix.  
Let us reorganize the helicity amplitudes in the matrix form 
\beq
T=\begin{pmatrix} \phi_1 & \phi_3 & -\phi_4 & -\phi_2 \\ 
\phi_3 & \phi_5 & \phi_6 & -\phi_4 \\ \phi_4 & -\phi_6 & \phi_5 & \phi_3 \\ \phi_2 & \phi_4 & \phi_3 & \phi_1 \end{pmatrix}, \label{tmatrix}
\eeq
where the rows and columns have labels  $|+_N+_\gamma\rangle, |+_N-_\gamma\rangle, |-_N+_\gamma\rangle,|-_N-_\gamma\rangle$ in this order. 
The unpolarized cross section is proportional to 
\beq
\frac{1}{4}{\rm Tr}[T^\dagger T]=\frac{1}{2}\Bigl(|\phi_1|^2+|\phi_2|^2+2|\phi_3|^2+2|\phi_4|^2+|\phi_5|^2+|\phi_6|^2\Bigr). \label{unpolc}
\eeq
The spin density matrix is  obtained by inserting appropriate Pauli matrices representing the final state nucleon and photon spins.  A care is needed for the photon since it moves in the $-z'$ direction. In (\ref{density}), we have introduced 
the vectors $\vec{B}_{N,\gamma}$ and the matrix $C$ in reference to the $x'y'z'$ coordinate system, with the positive $z'$ direction being chosen as the  spin quantization axis.  The photon density matrix in this frame reads 
\beq
\rho_\gamma={\rm Tr}_N[\rho] =\frac{1}{2}\begin{pmatrix} 1+B_\gamma^{z'} & B^{x'}_\gamma -iB^{y'}_\gamma \\ B_\gamma^{x'}+iB_\gamma^{y'} & 1-B^{z'}_\gamma \end{pmatrix}.   \label{photonrho}
\eeq 
Here, the rows and columns have labels $|\uparrow\rangle, |\downarrow\rangle$ where 
\beq
|\uparrow_{\gamma'}\rangle = \frac{1}{\sqrt{2}}(|x'\rangle +i|y'\rangle)=|-_{\gamma'}\rangle, \qquad |\downarrow_{\gamma'}\rangle = \frac{1}{\sqrt{2}}(|x'\rangle -i|y'\rangle) =|+_{\gamma'}\rangle, \label{down}
\eeq
are the eigenstates with spin projections $\pm 1$ along the $z'$ axis. 
In this basis, $B^{z'}_\gamma=\pm 1$ means  that the spin $z'$-component is $\pm 1$. $B^{x'}_\gamma=1$ means 
100\% linear polarization along the $x'$ axis (horizontal polarization), and $B^{x'}_\gamma=-1$ means 100\% linear polarization along the $y'$ axis (vertical polarization). $B^{y'}_\gamma=\pm 1$ means 100\% linear polarization along the axis that makes an angle $\pm \frac{\pi}{4}$ from the $x'$ axis. The vector $\vec{B}_\gamma$ can be recovered from  the polarization vector $\vec{B}_{hel}$ in the helicity basis as 
\beq
\left(B^{x'}_{hel},B^{y'}_{hel},B^{z'}_{hel} \right)  =\left(B^{x'}_\gamma,-B^{y'}_\gamma,-B^{z'}_\gamma\right),
\label{trick}
\eeq
considering the fact that the right-handed coordinate system for the left-moving photon is $(x',-y',-z')$.

With this caveat, we write down the density matrix  as
\beq
&& C^{ab}= \frac{{\rm Tr}\left[T^\dagger \begin{pmatrix} \sigma^{x'} \\ \sigma^{y'} \\ \sigma^{z'}\end{pmatrix}^a \otimes (\tau^{x'},-\tau^{y'},-\tau^{z'})^b T\right] }{{\rm Tr}[T^\dagger T]}=\frac{1}{|\phi_1|^2+|\phi_2|^2+2|\phi_3|^2+2|\phi_4|^2+|\phi_5|^2+|\phi_6|^2} \label{cmatrix}  \\ &&\qquad  \times    \begin{pmatrix} 0 & 2\im[\phi_1\phi_2^*+\phi_5\phi_6^*]  & -2\re[(\phi_1-\phi_5)\phi_4^*-(\phi_2+\phi_6)\phi_3^*] &  \\ 2\im[\phi_5\phi_6^*-\phi_1\phi_2^*-2\phi_3\phi_4^*] & 0 & 0 \\0&  2\im[(\phi_1-\phi_5)\phi_3^*  +(\phi_2+\phi_6)\phi_4^* ] & |\phi_5|^2+|\phi_6|^2-|\phi_1|^2-|\phi_2|^2 
\end{pmatrix}_{ab}, \notag 
\eeq
where the sign change  (\ref{trick}) is effectively implemented in the array of  $\tau$'s. Below we use a simpler notation $(\tau^{x'},-\tau^{y'},-\tau^{z'})^b=\tau^{x'}\tau^b \tau^{x'}$. 
The nonzero components of the  polarization vectors are 
\beq
B_N^{y'} = \frac{{\rm Tr}[T^\dagger (\sigma^{y'} \otimes 1)T]}{{\rm Tr}[T^\dagger T]} = \frac{2\im[(\phi_2-\phi_6)\phi_3^*-(\phi_1+\phi_5)\phi_4^*]}{ |\phi_1|^2+|\phi_2|^2+2|\phi_3|^2+2|\phi_4|^2+|\phi_5|^2+|\phi_6|^2}, \label{bn}
\eeq
\beq
B_\gamma^{x'} = \frac{{\rm Tr}[T^\dagger (1\otimes \tau^{x'})T]}{{\rm Tr}[T^\dagger T]}  = 
\frac{2\re[(\phi_1+\phi_5)\phi_3^*+(\phi_2-\phi_6)\phi_4^*]}{ |\phi_1|^2+|\phi_2|^2+2|\phi_3|^2+2|\phi_4|^2+|\phi_5|^2+|\phi_6|^2}. \label{bgam}
\eeq
As a consistency check,  consider the Thomson scattering limit $\omega\to 0$ of (\ref{phiphi}) where all the  $\phi$'s are real and $\phi_1\approx \phi_5$, $\phi_2\approx -\phi_6$. The only nonvanishing component is 
\beq
B^{x'}_\gamma=-\frac{\sin^2\theta}{1+\cos^2\theta}. \label{90}
\eeq
This is a well-known result in  electrodynamics  \cite{Berestetskii:1982qgu,RybickiLightman1979}. 
At $\theta=\frac{\pi}{2}$,   $B^{x'}_\gamma=-1$, meaning that the scattered photon is   100\% linearly  polarized in the $y'$ direction.\footnote{The interpretation of $\vec{B}_\gamma$ is convention dependent. Had we chosen  a different convention $\epsilon_+ \to -\epsilon_+$, $\epsilon_- \to \epsilon_-$ in (\ref{epsilonconvention}),  $\phi_{2,3,6}$, hence also $B_\gamma^{x'}$, change signs. In this convention, $B^{x'}_\gamma=+1$ means linear polarization along $y'$, as can be seen by implementing the corresponding change  $|\downarrow_{\gamma'}\rangle \to  -\frac{1}{\sqrt{2}}(|x'\rangle -i|y'\rangle)$ in (\ref{down}).  }

Let us take a first look at entanglement. Assume that the $\phi$'s are real as in electron Compton scattering at tree level. Then the  spin density matrix takes the simple form  
\beq
\rho=\frac{1}{4}\left(\mathbb{I}\otimes \mathbb{I} + B^{x'}_\gamma \mathbb{I}\otimes \tau^{x'}+C^{x'z'}\sigma^{x'}\otimes \tau^{z'}+C^{z'z'}\sigma^{z'}\otimes \tau^{z'}\right). \label{rhogeneral}
\eeq
According to the Peres-Horodecki  criterion \cite{Peres:1996dw,Horodecki:1996nc}, $\rho$  represents an entangled qubit pair iff the partially transposed density matrix $\rho^T$ has a negative eigenvalue. Otherwise, the system is separable (not entangled). Since only the symmetric matrices $\sigma^{x',z'}$ and $\tau^{x',z'}$ are involved, we immediately see that  $\rho^T=\rho$, and the physical density matrix $\rho$ is nonnegative. This can be checked  explicitly. The condition that the matrix (\ref{rhogeneral}) has a negative eigenvalue boils down to 
\beq
(C^{x'z'})^2+(C^{z'z'})^2+(B_\gamma^{x'})^2 - 1>0. \label{ccb}
\eeq
This cannot be satisfied if $\phi$'s are real because  
\beq
(C^{x'z'})^2+(C^{z'z'})^2+(B_\gamma^{x'})^2 - 1 =- \frac{(\phi_3^2-\phi_4^2-\phi_1\phi_5-\phi_2\phi_6)^2+(2\phi_3\phi_4-\phi_2\phi_5+\phi_1\phi_6)^2}{(\phi_1^2+\phi_2^2+2\phi_3^2+2\phi_4^2+\phi_5^2+\phi_6^2)^2}\le 0. \label{neg}
\eeq 
Naively, the relation (\ref{ccb}) seems to be easily satisfied because the maximal value $B_\gamma^{x'}=1$ is possible as we mentioned above. However, the $C$-matrix elements vanish at this point such that no  entanglement is generated.   

We have thus established a {\it no-go theorem} that, in unpolarized Compton scattering and when the amplitudes are real, the final state photon and nucleon (or any other spin-$\frac{1}{2}$ particle)  cannot be entangled. This result was found, although numerically and for the electron target only, in an earlier publication~\cite{Fedida:2022izl}. Here we have given a mathematical proof of this statement regardless of the nature of the spin-$\frac{1}{2}$ target.

\subsection{Polarized Compton scattering}
\label{polcom}

We have seen that, in order to realize an entangled proton-photon pair, we need to work in kinematical regimes where the amplitudes are complex. Alternatively, we can consider polarized Compton scattering    where either the incoming nucleon, or the incoming photon, or both are polarized. In this paper we follow the second approach. 

In the helicity amplitude formalism, it is straightforward to compute the spin density matrix in the presence of initial polarizations. 
We generalize (\ref{cmatrix}) as 
\beq
 C^{ab}(\vec{s}_N,\vec{s}_\gamma)\equiv \frac{1}{{\rm Tr}[T^\dagger T\left(\frac{\mathbb{I}+\vec{s}_N\cdot \vec{\sigma} }{2}\otimes \frac{\mathbb{I}+\vec{s}_\gamma\cdot \tau^x\vec{\tau}\tau^x}{2}\right)]} {\rm Tr}\left[T^\dagger (\sigma^a \otimes \tau^{x'} \tau^b \tau^{x'} )T\left(\frac{\mathbb{I}+\vec{s}_N\cdot \vec{\sigma} }{2}\otimes \frac{\mathbb{I}+\vec{s}_\gamma\cdot \tau^x\vec{\tau}\tau^x}{2}\right)\right] , \label{cpol}
\eeq
where the vectors $\vec{s}_{N,\gamma}$ characterize the direction and strength ($|\vec{s}_{N,\gamma}|\le 1$) of the initial polarizations. 
 The spin density matrix calculated in the previous section corresponds to $C^{ab}(0,0)$. Similarly, the polarization vectors are generalized 
\beq
B^a_N(\vec{s}_N,\vec{s}_\gamma)=\frac{1}{{\rm Tr}\left[T^\dagger T\left(\frac{\mathbb{I}+\vec{s}_N\cdot \vec{\sigma} }{2}\otimes \frac{\mathbb{I}+\vec{s}_\gamma\cdot \tau^x\vec{\tau}\tau^x}{2}\right)\right]}  {\rm Tr}\left[T^\dagger (\sigma^a\otimes \mathbb{I}) T\left(\frac{\mathbb{I}+\vec{s}_N\cdot \vec{\sigma} }{2}\otimes \frac{\mathbb{I}+\vec{s}_\gamma\cdot \tau^x\vec{\tau}\tau^x}{2}\right)\right] 
\eeq
\beq
B^b_\gamma(\vec{s}_N,\vec{s}_\gamma)=\frac{1}{{\rm Tr}\left[T^\dagger T\left(\frac{\mathbb{I}+\vec{s}_N\cdot \vec{\sigma} }{2}\otimes \frac{\mathbb{I}+\vec{s}_\gamma\cdot \tau^x\vec{\tau}\tau^x}{2}\right)\right]}  {\rm Tr}\left[T^\dagger (\mathbb{I}\otimes \tau^{x'}\tau^b \tau^{x'} )T\left(\frac{\mathbb{I}+\vec{s}_N\cdot \vec{\sigma} }{2}\otimes \frac{\mathbb{I}+\vec{s}_\gamma\cdot \tau^x\vec{\tau}\tau^x}{2}\right)\right] . \label{bpol}
\eeq
We restrict ourselves to the cases where the initial polarization vectors are alined with one of the  original orthonormal  axes $x,y,z$, and write $\vec{s}_N=s_N\hat{x}$, $\vec{s}_\gamma=s_\gamma \hat{y}$, etc., with $-1\le s_{N,\gamma}\le 1$. We choose the $+z$ direction to be the  spin quantization axis for both the incoming nucleon and the  photon.  
$\vec{s}_N=\pm\hat{x},\pm\hat{y},\pm\hat{z}$ means that the incoming nucleon is 100\% polarized in the $\pm x,\pm y,\pm z$ directions. Likewise,   $\vec{s}_\gamma=\pm \hat{x}$ means horizontal/vertical linear polarization along the $x/y$ axis, $\vec{s}_\gamma=\pm \hat{y}$ means linear polarization along the axis that makes an angle $\pm \frac{\pi}{4}$ with the $x$-axis, and $\vec{s}_\gamma=\pm \hat{z}$ means circular polarization with spin  $\pm 1$ along the $z$-axis (helicity $\mp 1$). In Appendix B, we present explicit formulas for $\vec{B}_{N,\gamma}$ and $C$ when one of the incoming particles is polarized. Already in this case, more components of $\vec{B}_{N,\gamma}$, $C$ are non-vanishing, increasing  the chance of generating entanglement. 

The density matrix has now become   
\beq
\rho(\vec{s}_N,\vec{s}_\gamma)=\frac{1}{4}\left(\mathbb{I}\otimes \mathbb{I} +B^a_N(\vec{s}_N,\vec{s}_\gamma)\sigma^a\otimes \mathbb{I}+ B^{b}_\gamma(\vec{s}_N,\vec{s}_\gamma)  \mathbb{I}\otimes \tau^{b}+C^{ab}(\vec{s}_N,\vec{s}_\gamma)\sigma^{a}\otimes \tau^{b}\right). \label{newdensity}
\eeq
We will be mostly interested in the case $|\vec{s}_{N,\gamma}|=1$, meaning that the initial state is pure. Due to unitarity, the final density matrix $\rho$ also represents a pure state in this case. Yet, to keep our discussion applicable also to mixed states,   we continue to use the negative eigenvalue of the partially transposed density matrix 
\beq
\rho^T(\vec{s}_N,\vec{s}_\gamma)=\frac{1}{4}\left(\mathbb{I}\otimes \mathbb{I} +B^a_N(\vec{s}_N,\vec{s}_\gamma)\sigma^a\otimes \mathbb{I}+ B^{b}_\gamma(\vec{s}_N,\vec{s}_\gamma)  \mathbb{I}\otimes (\tau^{b})^T+C^{ab}(\vec{s}_N,\vec{s}_\gamma)\sigma^{a}\otimes (\tau^{b})^T\right),  \label{ppt}
\eeq
as a measure of entanglement.   
For a two-qubit system, it is known that there is at most one negative eigenvalue, and its lower bound   is  $-0.5$ \cite{Zyczkowski:1998yd}. This bound is saturated by the  maximally entangled Bell  states \beq
|\Phi^\pm\rangle =\frac{1}{\sqrt{2}}(|\uparrow\uparrow\rangle \pm |\downarrow\downarrow\rangle), \qquad |\Psi^\pm\rangle =\frac{1}{\sqrt{2}}(|\uparrow \downarrow\rangle \pm |\downarrow \uparrow\rangle),
\eeq 
or their unitary equivalents. We have introduced new symbols $\uparrow$ and $\downarrow$ for  spin projections $\pm \frac{1}{2}$, $\pm 1$ along the $z'$ axis, in order to distinguish them from helicities $\pm$. (The difference is important only for the photon.)  
Thus, $|\uparrow\uparrow\rangle$ denotes a final state with $S^{z'}=\frac{3}{2}$,  and $|\uparrow\downarrow\rangle$ is a $S^{z'}=-\frac{1}{2}$ state, etc. 
 
\begin{figure}
\begin{overpic}
[width=0.3\textwidth]{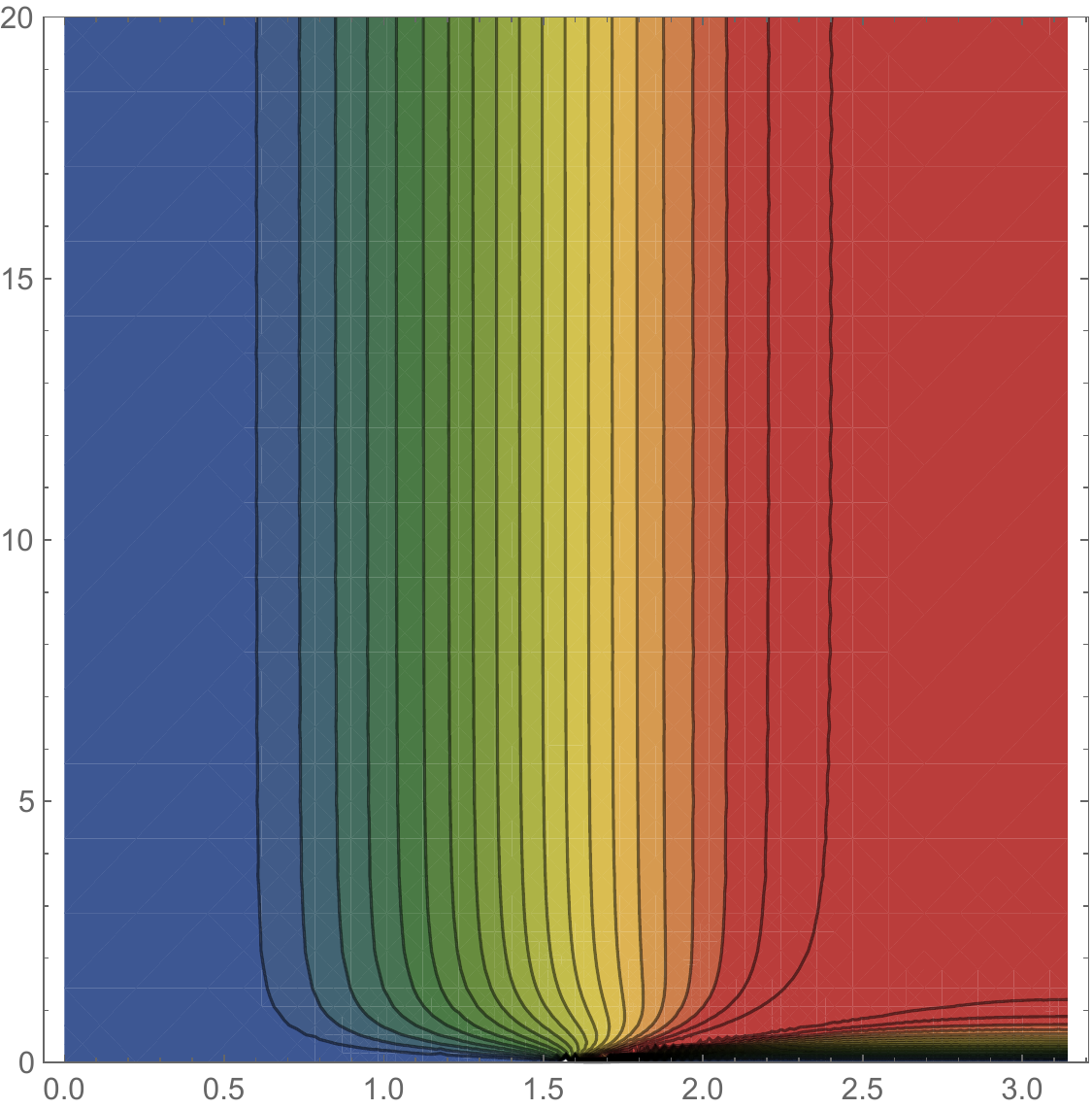}
\put(-10,55){\Large{$\mu$}}
\put(78,89){\color{white} \large{$|\Psi^{+}\rangle$}}
\put(10,8){\color{white} \Large{$\hat{x},\hat{x}$}}
\put(78,36){\color{white} \large{$|\Phi^{-i}_{y'}\rangle$}}
\put(50,-8){\large{$\theta$}}
\put(78,12){\color{white} \large{$|\Phi^-\rangle$}}
\end{overpic}
\begin{overpic}
[width=0.3\textwidth]{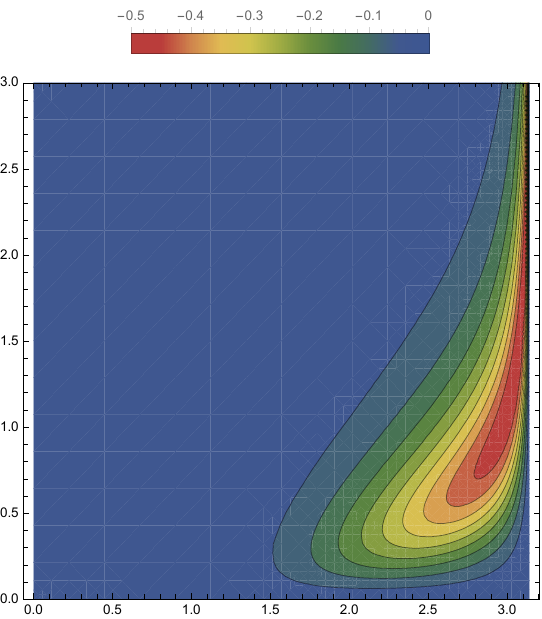}
\put(8,8){\color{white} \Large{$\hat{z},-\hat{z}$}
}
\put(45,-7){\large{$\theta$}}
\put(72,31){\color{white} \large{$|\Psi^+\rangle$}}
\end{overpic}
\begin{overpic}
[width=0.3\textwidth]{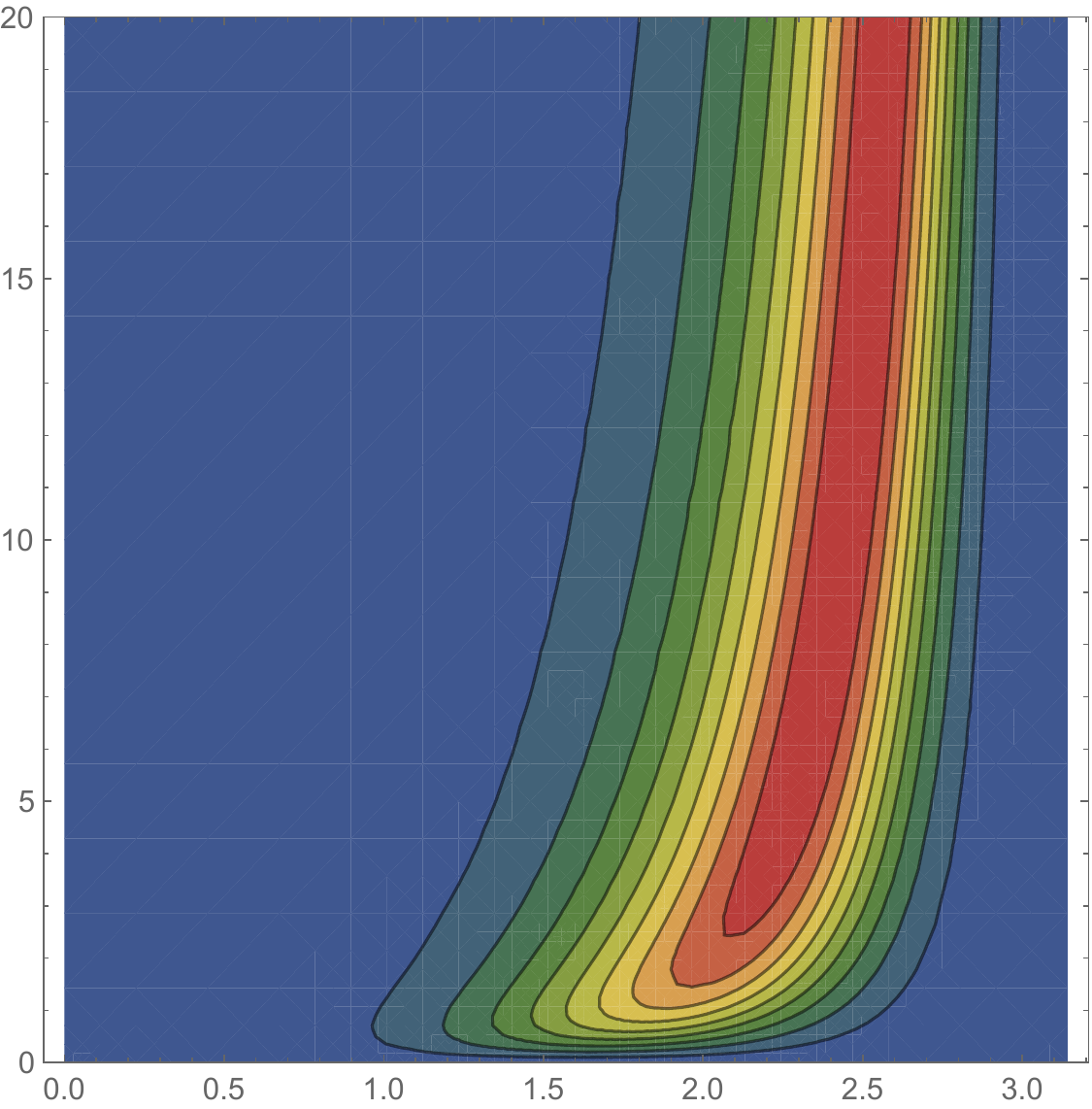}
\put(8,8){\color{white} \Large{$-\hat{z},-\hat{z}$}
}
\put(50,-8){\large{$\theta$}}
\put(70,44){\color{white} \large{$|\Phi^-\rangle$}}
\end{overpic}
\vspace{2mm}
\caption{Contour plot of the minimum eigenvalue  $\lambda_{\rm min}$ in electron-target Compton scattering for three different polarization configurations.  $(\vec{s}_e,\vec{s}_\gamma)=(\hat{x},\hat{x})$ (left), $(\vec{s}_e,\vec{s}_\gamma)=(\hat{z},-\hat{z})$ (middle) and $(\vec{s}_e,\vec{s}_\gamma)=(-\hat{z},-\hat{z})$ (right). $\mu=\frac{\omega}{m_e}$ is the rescaled photon energy. For better legibility, the middle plot shows a different range $\mu<3$.  }
\label{elect}
\end{figure}
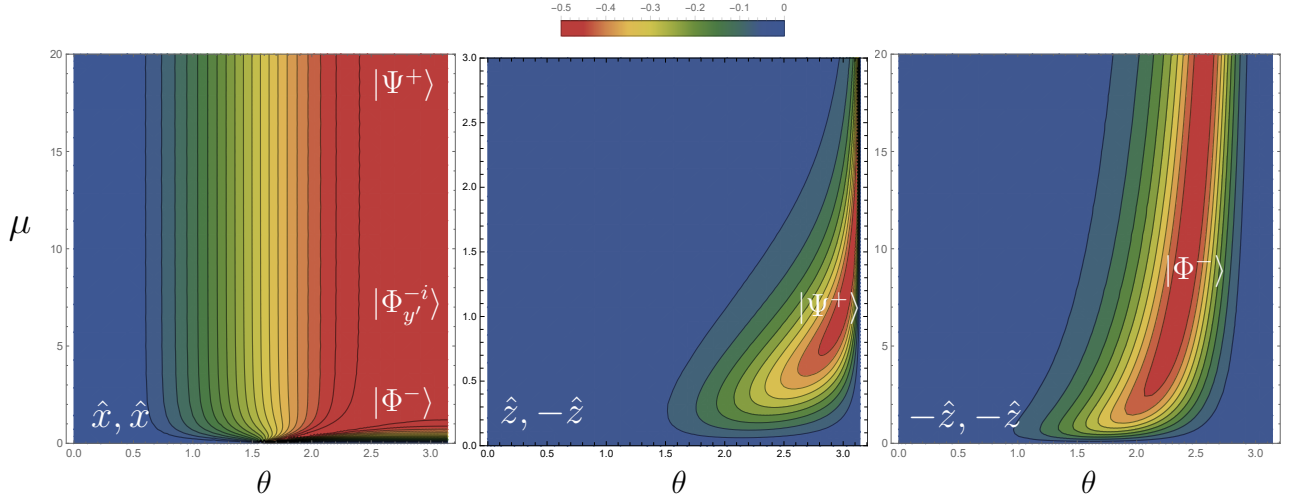

As a warm-up, consider again electron-target Compton scattering. (We write $\vec{s}_N\to \vec{s}_e$, etc. in the rest of this section.) Using  (\ref{phiphi}), it is straightforward to construct the partially transposed density matrix $\rho^T$ (\ref{ppt}) and numerically check if it has a negative eigenvalue. We have studied 18 configurations 
 of  100\% ($|s_{e,\gamma}|=1$) polarized initial states  
 \beq
(\vec{s}_e,\vec{s}_\gamma)=\begin{pmatrix}\hat{x} \\ \hat{y} \\ \hat{z}\end{pmatrix}(\pm \hat{x},\pm \hat{y},\pm \hat{z}),
\eeq  
and detected entanglement $\lambda_{\rm min}<0$ in all but two configurations $(\vec{s}_e,\vec{s}_\gamma)=(\vec{x},-\vec{z}), (\vec{z},-\vec{x})$.  
In Fig.~\ref{elect}, we plot the minimum eigenvalue  $\lambda_{\rm min}$ of  $\rho^T$  as a function of $0\le \theta \le \pi$ and the photon energy $\mu\equiv \frac{\omega}{m_e}$ normalized by the electron mass. We only  show three configurations of $(\vec{s}_e,\vec{s}_\gamma)$ which we find particularly interesting. The same color coding in the range $-0.5\le \lambda_{\rm min}\le 0$ is used here and in all the plots in the following.  Regions of (nearly) maximal entanglement are shown in red,   corresponding to  $-0.5\le \lambda_{\rm min}< -0.45$. Different Bell states are realized in different red regions. This can be judged by computing the $C$-matrix in respective regions and comparing with  
\beq
|\Psi^+\rangle : C={\rm diag}(1,1,-1), \quad |\Psi^-\rangle: (-1,-1,-1), \quad |\Phi^+\rangle: (1,-1,1), \quad |\Phi^-\rangle: (-1,1,1). \label{4bell}
\eeq
Recall that, in our convention,  the  $(\hat{z},-\hat{z})$ and $(-\hat{z},-\hat{z})$ configurations have  initial spins  $S^z=-\frac{1}{2}$ and $S^z=-\frac{3}{2}$, respectively. This is consistent with the generation of  spin-anti-aligned ($\Psi$-type) and spin-aligned ($\Phi$-type) Bell states in the two cases.

However, not all maximally entangled states are of these types. In the $(\vec{s}_e,\vec{s}_\gamma)=(\hat{x},\hat{x})$ plot, we find 
\beq
C\approx \begin{pmatrix} 0 & 0 & -1 \\ 0 & 1 & 0 \\ -1 & 0 & 0\end{pmatrix},
\eeq
in a  small region around $\mu\approx 7$ and $\theta\approx 0.95\pi$,   corresponding to the state 
\beq
\frac{1}{2}\Bigl(|\uparrow\uparrow\rangle -|\uparrow\downarrow\rangle - |\downarrow\uparrow\rangle -|\downarrow\downarrow\rangle
\Bigr) .
\eeq
This is a linear combination of two Bell states. By changing the spin quantization axis,  it can be rewritten in the form 
\beq
|\Phi^{-i}_{y'}\rangle\equiv  \frac{1}{\sqrt{2}}\left(|\uparrow\rangle_{y'}|\frac{\pi}{4}\rangle - i|\downarrow\rangle_{y'}|-\frac{\pi}{4}\rangle\right), \label{bell2}
\eeq
where the subscript $y'$ denotes spin   eigenstates with $y'$ as the quantization axis $|\uparrow(\downarrow) \rangle_{y'} = \frac{1}{\sqrt{2}}(|\uparrow_e\rangle\pm i|\downarrow_e\rangle)$. Likewise, $|\pm \frac{\pi}{4}\rangle=\frac{1}{\sqrt{2}}(|\uparrow_\gamma\rangle\pm i|\downarrow_\gamma\rangle)$ means linearly polarized states along the direction that makes an angle $\pm \frac{\pi}{4}$ from the $x'$ axis. (\ref{bell2}) is equivalent to the $\Phi$-type Bell states up to a local unitary transformation.   Different Bell states are continuously connected. For example, at $\mu=10$ and $\theta=0.9\pi$, 
\beq
C\approx \begin{pmatrix} 0.834 & 0 & -0.552 \\ 0 & 1.000 & 0 \\ -0.551 & 0 & -0.833 \end{pmatrix} \label{five}
\eeq
This is a maximally entangled state (since $(0.552)^2+(0.834
)^2\approx 1$), and is intermediate between $|\Psi^+\rangle$ and $|\Phi^{-i}_{y'}\rangle$.  

It is interesting to note  that, whenever  maximal entanglement occurs, the final state  particles are unpolarized $\vec{B}_{e,\gamma} \approx 0$.  The initial polarizations are `used up' to realize maximal entanglement. Conversely, in the regions where $\vec{B}_{e,\gamma}$ are  nonvanishing, the strength of entanglement is limited \cite{Liu:2026dzv}. The same comments apply to the other cases studied in the next sections.

Let us comment on the previous works on  entanglement in  Compton scattering off an electron \cite{Cervera-Lierta:2017tdt,Ahrens:2017zsb,NagChowdhury:2021nme,Fedida:2022izl,Blasone:2024jzv,Asenov:2025ueu}. Our result is consistent with and extends the work~\cite{Fedida:2022izl}.  The right plot in Fig.~\ref{elect} is essentially a reproduction of Fig.~21 of \cite{Fedida:2022izl}.\footnote{Ref.~\cite{Fedida:2022izl} employs  a convention in optics where a helicity $+1$ photon is called `left-handed,' but uses the standard convention in particle physics  for the electron, i.e., a left-handed electron has helicity $-\frac{1}{2}$. Therefore,  the `$LL$' initial condition studied in \cite{Fedida:2022izl} corresponds to the `$LR$' or   $(-\hat{z},-\hat{z})$ configuration in our convention. Note also that what is actually plotted in Fig.~21 of \cite{Fedida:2022izl} is slightly different from ours. } In addition, we have considered all possible initial polarizations including transverse and linear polarizations, and   identified the nature of maximally entangled states by monitoring the $C$-matrix.

An earlier work \cite{Cervera-Lierta:2017tdt} studied longitudinally polarized Compton scattering which corresponds to the last two plots in Fig.~\ref{elect}.     However, the authors  examined only the  limiting cases $\mu\to \infty$ and $\mu\to 0$ and did not find maximal entanglement. We note that the red band in Fig.~\ref{elect} (right) persists up to arbitrary large values of $\mu$ (even if $\mu> 10^{6}$). However, if we set $m_e=0$ ($\mu=\infty$) from the beginning, we do not see  maximal entanglement, not even entanglement, for any value of $\theta$.   Indeed, in this limit, 
\beq
\phi_1= -2e^2\sec\frac{\theta}{2} ,  \qquad \phi_5=-2e^2\cos\frac{\theta}{2}, \qquad \phi_{2,3,4,6}=0,
\label{massless} 
\eeq
 and we find pure states $\rho(\hat{z},-\hat{z})={\rm diag}(0,1,0,0)=|\uparrow\downarrow\rangle\langle \uparrow\downarrow|$ and $\rho(-\hat{z},-\hat{z})={\rm diag}(0,0,0,1)=|\downarrow\downarrow\rangle\langle \downarrow\downarrow|$   independently of $\theta$. In contrast, the red region in Fig.~\ref{elect} (left) survives in the $\mu\to \infty$ limit. This can be seen analytically. We find  
\beq
\lambda_{\rm min}(\hat{x},\pm \hat{x}) = \frac{-5+4\cos\theta + \cos 2\theta}{16\left(1+\cos^4\frac{\theta}{2} \right)}. \label{always}
\eeq
This approaches $-\frac{1}{2}$ as $\theta\to \pi$. It is essentially the function  plotted in Fig.~\ref{elect} (left) in the region $\mu\gg 1$. The same result is found in the cases $(\hat{x},\pm \hat{y})$, $(\hat{y},\pm \hat{x})$ and $(\hat{y},\pm \hat{y})$. This means  that maximal entanglement can be  always achieved  in the scattering of a 100\% transversely polarized electron and a 100\% linearly polarized photon in the backward region $\theta \sim \pi$, irrespective of the direction of polarizations.  

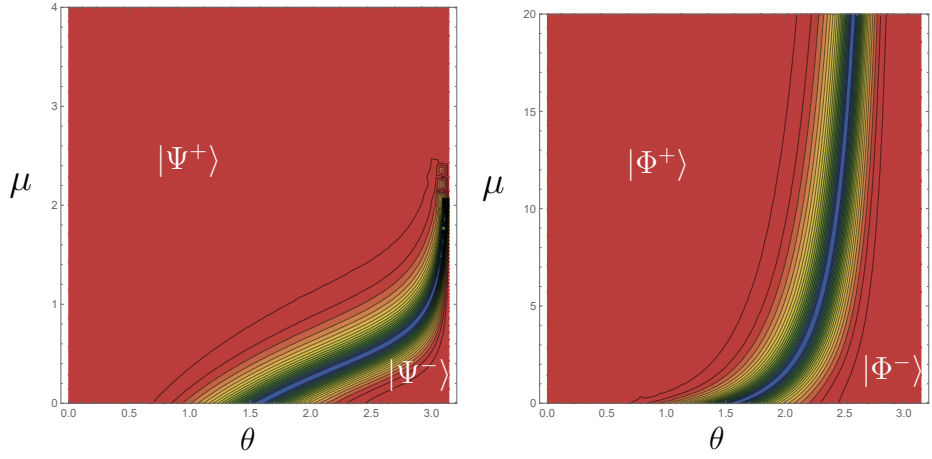
\begin{figure}
\begin{overpic}
[width=0.3\textwidth]{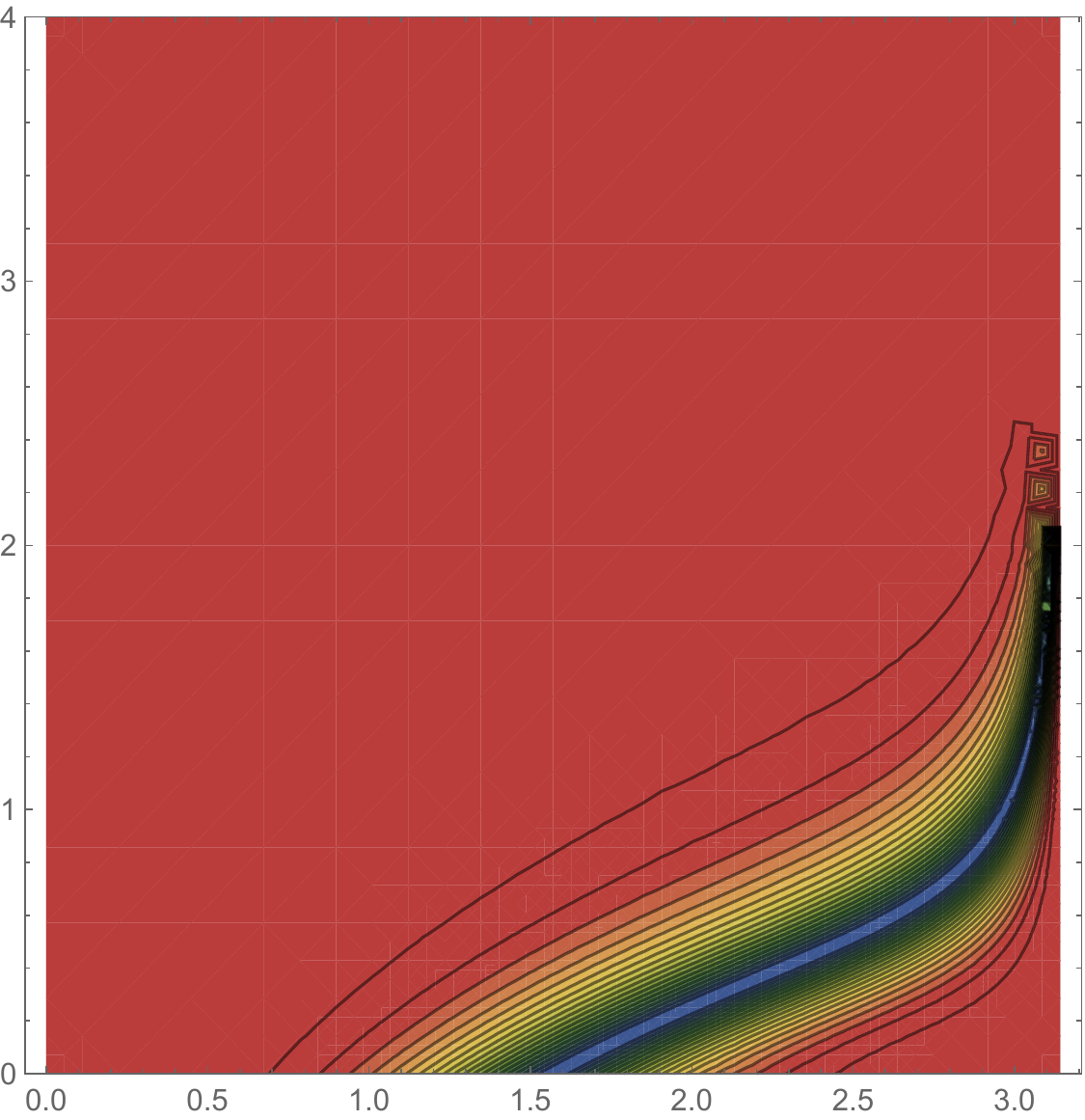}
\put(-10,55){\Large{$\mu$}}
\put(25,60){\color{white} \large{$|\Psi^+\rangle$}}
\put(80,9){\color{white} \large{$|\Psi^-\rangle$}}
\put(45,-7){\large{$\theta$}}
\end{overpic}
\hspace{7mm}
\begin{overpic}
[width=0.3\textwidth]{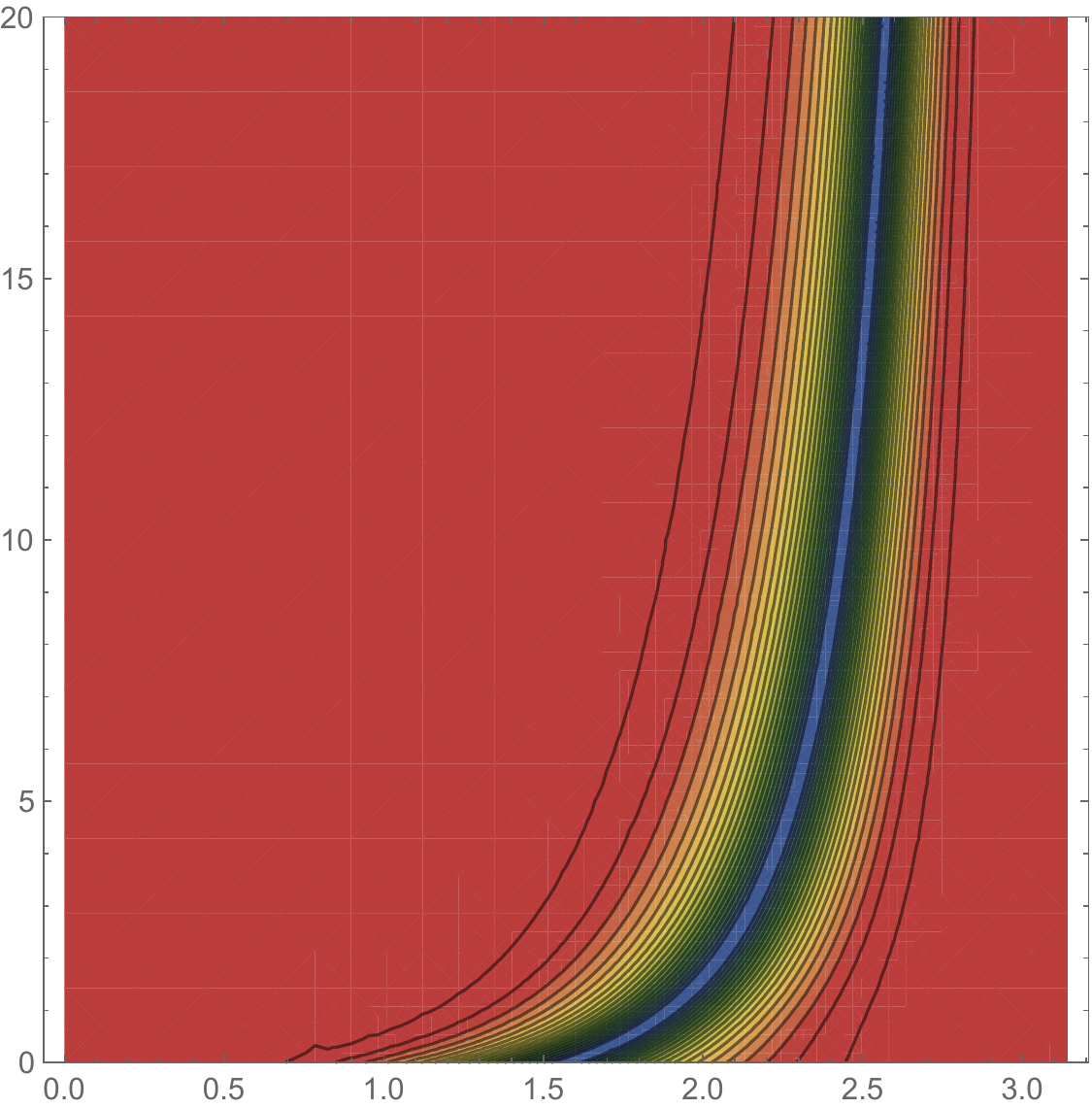}
\put(-10,55){\Large{$\mu$}}
\put(80,109){\color{white} \large{$|\Phi^-\rangle$}}
\put(25,60){\color{white} \large{$|\Phi^+\rangle$}}
\put(82,10){\color{white} \large{$|\Phi^-\rangle$}}
\put(45,-7){\large{$\theta$}}
\end{overpic}
\vspace{2mm}
\caption{Minimal eigenvalue $\lambda_{\rm min}$ of the states resulting from the Bell states $|\Psi^+\rangle$ (left) and $|\Psi^-\rangle$ (right) as the initial condition. Here and in the following, the color coding is the same as in Fig.~\ref{elect}.}
\label{min}
\end{figure}

The authors of \cite{Blasone:2024jzv} assumed that the incoming particles are in one of the Bell states, and concluded  that maximal entanglement cannot survive in the final state.  We have repeated this computation in the present framework.  As an example, in Fig.~\ref{min}, we plot the minimal eigenvalue of the partially transposed density matrix  resulting from   the initial density matrices 
\beq
\rho_{in} = \frac{1}{4}\left(\mathbb{I}\otimes \mathbb{I} + \sigma^x\otimes \tau^x \pm \sigma^y \otimes (-\tau^y)\mp   \sigma^z \otimes (-\tau^z)\right), \label{initialbell}
\eeq
where the upper (lower) sign corresponds to $|\Psi^+\rangle$ ($|\Phi^+\rangle$). 
We see that entanglement is almost maximal, if not perfectly maximal, everywhere except in a narrow strip. In order to better understand this, following \cite{Blasone:2024jzv}, we directly look at the final states resulting from the Bell states 
\beq
&&T *\frac{|++\rangle\pm |--\rangle}{\sqrt{2}} = \frac{\phi_1\mp \phi_2}{\sqrt{2}}|++\rangle + \frac{\phi_3\mp \phi_4}{\sqrt{2}}|+-\rangle +\frac{\phi_4\pm \phi_3}{\sqrt{2}}|-+\rangle + \frac{\phi_2\pm \phi_1}{\sqrt{2}}|--\rangle\,,  \nn
&& T *\frac{|+-\rangle\pm |-+\rangle}{\sqrt{2}} = \frac{\phi_3\mp \phi_4}{\sqrt{2}}|++\rangle + \frac{\phi_5\pm \phi_6}{\sqrt{2}}|+-\rangle +\frac{-\phi_6 \pm \phi_5}{\sqrt{2}}|-+\rangle + \frac{\phi_4\pm \phi_3}{\sqrt{2}}|--\rangle  \,.\label{mas} 
\eeq
In  Fig.~\ref{min} (left),  $|\phi_{1}|\gg |\phi_{2,3,4}|$ on the left hand side of the narrow strip and $|\phi_2|\gg |\phi_{1,3,4}|$ on the right hand side.   Similarly, in the right plot, $|\phi_5|\gg |\phi_{3,4,6}|$ on the left hand side of the strip and $|\phi_6|\gg |\phi_{3,4,5}|$ on the right hand side. Therefore,  (\ref{mas}) are  very close to the original Bell state everywhere on  the left hand side of the strip, and to the other Bell state on the right hand side. However, perfect maximal entanglement is recovered only in the limits $\theta\to 0,\pi$.   

Ref.~\cite{Ahrens:2017zsb} identified a maximally entangled state in the Breit frame with a linearly polarized initial photon. We presume this is consistent with our observation below (\ref{always}), although a direct comparison is difficult because of the different frame and basis choices.  Finally, Ref.~\cite{NagChowdhury:2021nme} claimed that maximal entanglement is realized in unpolarized Compton scattering. In our view, this cannot be correct since it contradicts  our no-go theorem. 

\section{Low energy regime}

In this and the next sections, we study Compton scattering off the nucleon in the low energy and high energy regimes, respectively. As mentioned in the Introduction, this is the first study of its kind.  By low energy, we mean below the pion threshold $\omega\lesssim m_\pi = 140$ MeV where all the six amplitudes are real. In this regime, the scattering amplitudes can be expanded in powers of $\frac{\omega}{m}$. According to the low energy theorem, the coefficients of the first two terms  $\omega^0,\omega^1$ in this expansion are universally given by the charge $Q$  and the  magnetic moment of the target   \cite{Powell:1949bhl,Gell-Mann:1954wra,Low:1954kd}. At higher orders,  new constants due to the nontrivial nucleon  structure come into play.  At ${\cal O}(\omega^2)$,  the electric $\alpha_E$ and magnetic $\beta_M$  polarizabilities need to be included. At ${\cal O}(\omega^3)$, the so-called spin polarizabilities  appear \cite{Ragusa:1993rm}. In this paper, we neglect the latter and use the following formulas
\beq
 \phi_1 &=& e^2 \left( -2Q^2-\frac{\kappa^2 \omega(E+\omega)}{m^2}+ \frac{2Q^2(E-3\omega)-\kappa(4Q+\kappa)\omega}{E+\omega\cos\theta
}\sin^2\frac{\theta}{2} \right)\cos\frac{\theta}{2} +8\pi m \omega^2\left(\alpha_E+\beta_M\right)\cos^3\frac{\theta}{2}, \nn 
 \phi_2&=&e^2  \left( \frac{-2Q^2(E-\omega)+\kappa(4Q + \kappa)\omega}{m} + \frac{2m^2Q^2+\kappa^2 \omega (E+\omega)}{m(E+\omega\cos\theta)}\cos^2\frac{\theta}{2}\right)\sin\frac{\theta}{2} +8\pi m \omega^2(\alpha_E- \beta_M)\sin^3\frac{\theta}{2} , \nn 
 \phi_3&=&\left(e^2\frac{2Q^2(E-\omega) +\kappa(2Q + \kappa)\omega}{E+\omega\cos\theta} -8\pi m \omega^2(\alpha_E-\beta_M)\right)\sin^2\frac{\theta}{2}\cos\frac{\theta}{2} ,
\nn
 \phi_4&=&\left(e^2\frac{2Q^2m^2 -\kappa(\kappa+2Q)(E\omega+\omega^2)}{m(E+\omega\cos\theta)}-8\pi m \omega^2 (\alpha_E+\beta_M)\right)\sin\frac{\theta}{2} \cos^2\frac{\theta}{2}, 
\nn
\phi_5&=&\left(e^2 \frac{-2Q^2m^2(E+\omega)+\kappa^2(2\omega^2(E+\omega)+\omega m^2)}{m^2(E+\omega\cos\theta)}+8\pi m \omega^2 (\alpha_E+\beta_M) \right)\cos^3\frac{\theta}{2}, 
\nn
 \phi_6&=& \left(e^2 \frac{2Q^2m^2 +\kappa(\kappa+4Q)(E\omega -\omega^2)}{m(E+\omega\cos\theta)}-8\pi m \omega^2(\alpha_E-\beta_M)\right)\sin^3\frac{\theta}{2}.  \label{fullphi}
\eeq
To obtain this, we recalculated the Born diagrams (\ref{born}) using the following Feynman rules  for the  photon-nucleon  vertices  
\beq
\gamma^\mu \to Q\gamma^\mu+i\kappa\frac{\sigma^{\mu\lambda}q_\lambda}{2m}, \qquad \gamma^\nu\to  Q\gamma^\nu-i\kappa\frac{\sigma^{\nu\lambda}q'_\lambda}{2m},
\eeq
where  $Q$ is the fractional electric charge ($+1$ for the proton and 0 for the neutron) and  $\kappa$ is  the  anomalous magnetic moment ($1.79$ for the proton and $-1.91$ for the neutron). Since we neglect the spin polarizabilities appearing at ${\cal O}(\omega^3)$, for consistency  the result needs to be expanded to ${\cal O}(\omega^2)$. But the difference is numerically small for $\omega<m_\pi$, and we prefer to use the full expressions which are actually more concise. As for the polarizabilities, in the literature they are  often presented as  a part of $A_i=A_i^{\rm Born}+A_i^{pol}$ 
\beq
A^{pol}_1=4\pi (\alpha_E+\beta_M\cos\theta)\omega^2 +{\cal O}(\omega^3), \qquad 
A^{pol}_2=-4\pi \beta_M \omega^2 +{\cal O}(\omega^3) ,\qquad A_{3,4,5,6}^{pol}=0, \label{a12}
\eeq
The ${\cal O}(\alpha_E\omega^2), {\cal O}(\beta_M\omega^2)$ terms in (\ref{fullphi}) have been derived by substituting (\ref{a12}) 
into  (\ref{inverse}).\footnote{For completeness, we show the scattering amplitude $A$'s in the CM frame obtained from (\ref{inverse}) and  expanded to $\omega^2$    
\beq
A_1&=&\frac{4\pi\alpha_{em}}{m} \left(-Q^2+ \bigl( (Q+\kappa)^2(1+\cos\theta)-Q^2\bigr) (1-\cos\theta) \frac{\omega^2}{4m^2}\right) +4\pi (\alpha_E+\beta_M\cos\theta)\omega^2 +{\cal O}(\omega^3)\nn
A_2&=&\frac{4\pi\alpha_{em}}{m} \left(\frac{Q^2\omega}{m} -\bigl(Q^2+(Q-\kappa)(3Q+ \kappa)\cos\theta\bigr)\frac{\omega^2}{4m^2}\right) -4\pi \beta_M \omega^2 +{\cal O}(\omega^3) \nn 
A_3&=& \frac{4\pi\alpha_{em}}{m} \left(((Q+\kappa)^2(1-\cos\theta)-\kappa^2 )\frac{\omega}{2m} -\bigl((Q+\kappa)^2(1-\cos\theta)+\kappa^2\bigr)(1+\cos\theta)\frac{\omega^2}{4m^2} \right)+{\cal O}(\omega^3) \nn 
A_4&=& \frac{4\pi\alpha_{em}}{m}  \left(-\frac{(Q+\kappa)^2\omega}{2m}+ \bigl((Q+\kappa)^2\cos\theta -\kappa(2Q+\kappa)\bigr)\frac{\omega^2}{4m^2}\right) +{\cal O}(\omega^3)\nn 
A_5&=& \frac{4\pi\alpha_{em}}{m} \left(\frac{(Q+\kappa)^2\omega}{2m} -\bigl( (Q+\kappa)^2\cos\theta +Q^2+Q\kappa-\kappa^2\bigr)\frac{\omega^2}{4m^2}\right) +{\cal O}(\omega^3)\nn 
A_6&=& \frac{4\pi\alpha_{em}}{m} \left( -\frac{Q(Q+\kappa)\omega}{2m} + (Q+\kappa)(2Q\cos\theta+\kappa)\frac{\omega^2}{4m^2}\right)+{\cal O}(\omega^3)
\eeq 
The ${\cal O}(\omega^2)$ terms do not agree with those in \cite{Hemmert:1997tj}. $A_1$ agrees with  \cite{Griesshammer:2012we}. 
}

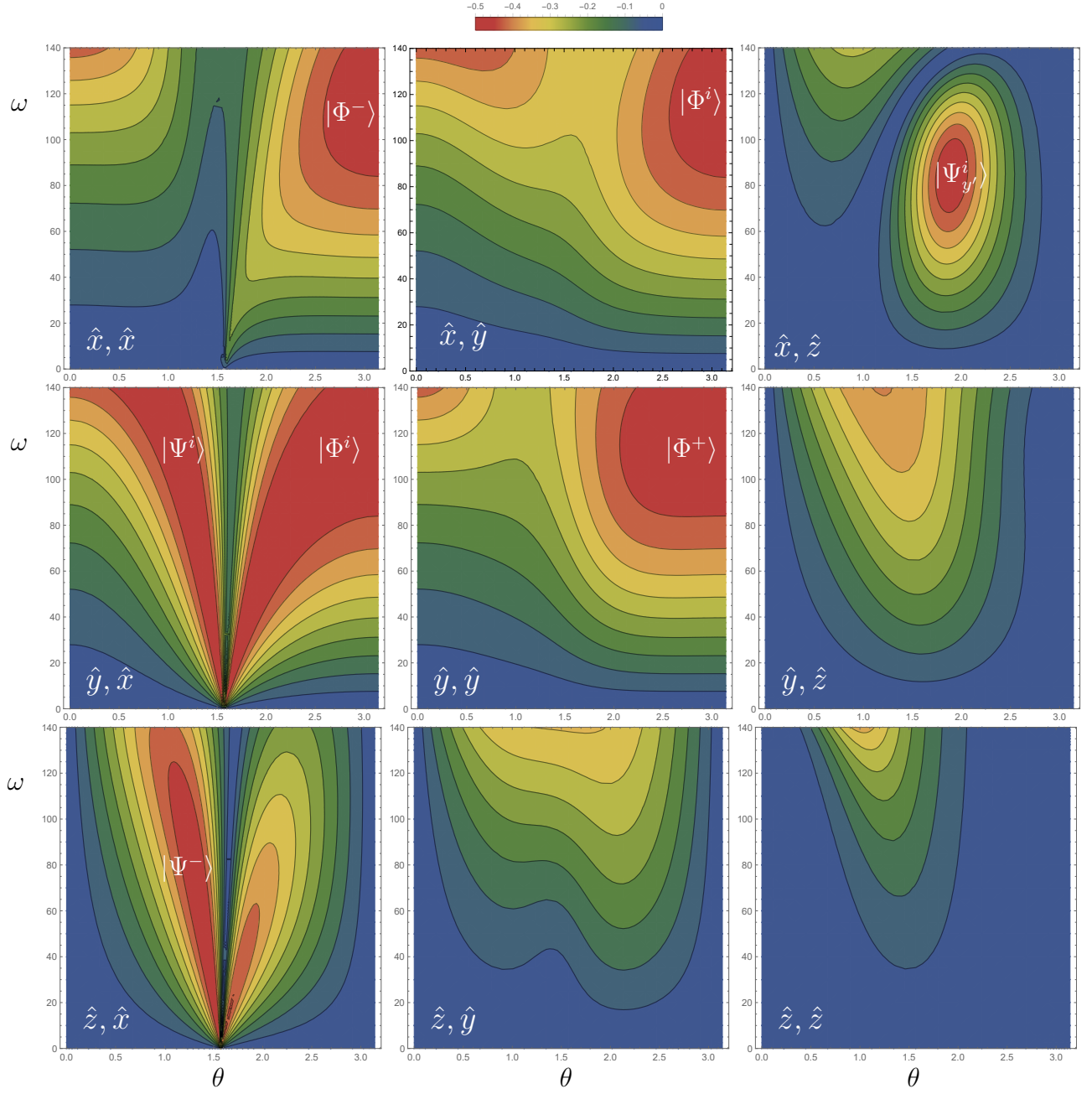
\begin{figure}
\begin{overpic}
[width=0.3\textwidth]{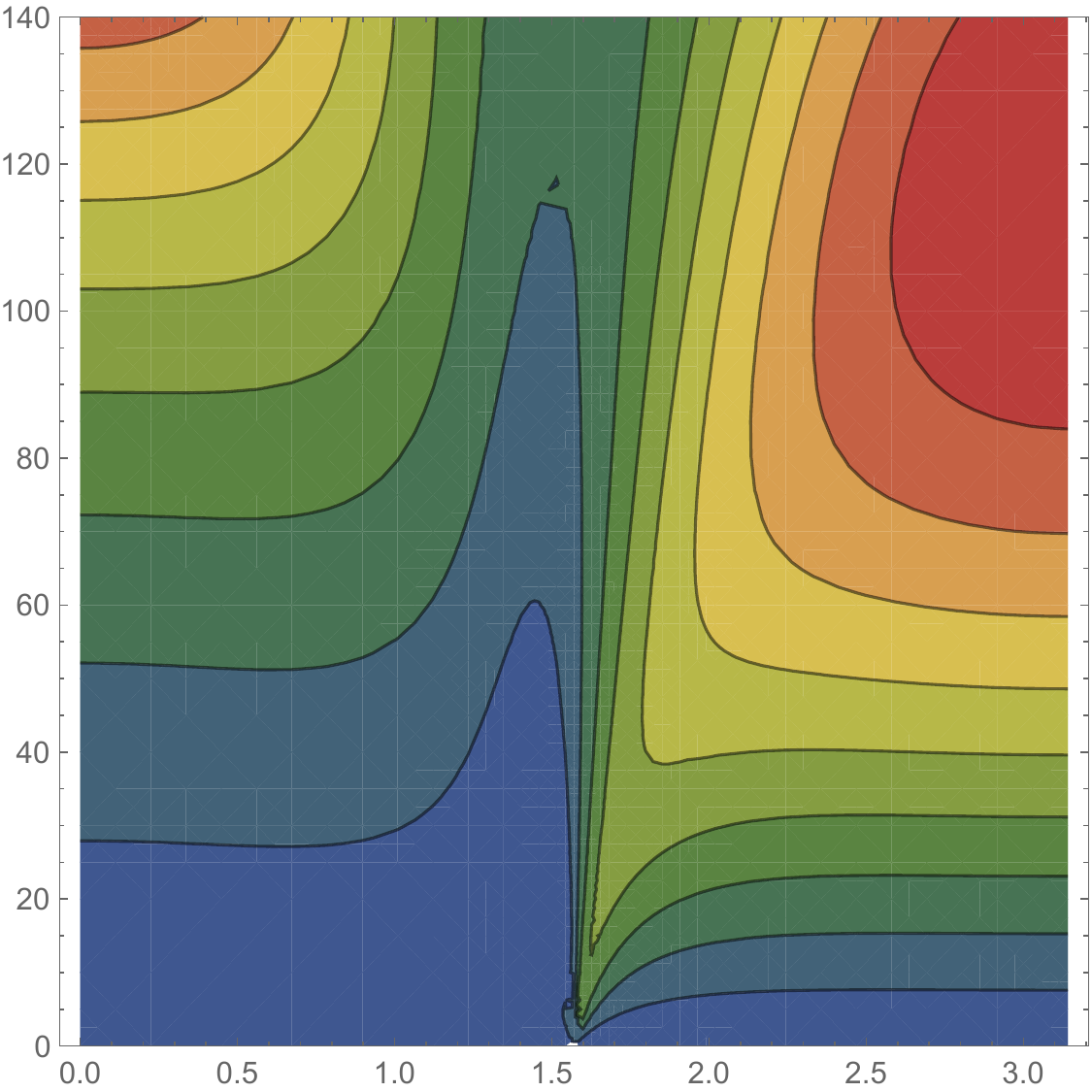}
\put(12,10){\color{white} \Large{$\hat{x},\hat{x}$}
}
\put(-10,80){\large{$\omega$}}
\put(82,77){\color{white} \large{$|\Phi^-\rangle$}}
\end{overpic}
\begin{overpic}
[width=0.3\textwidth]{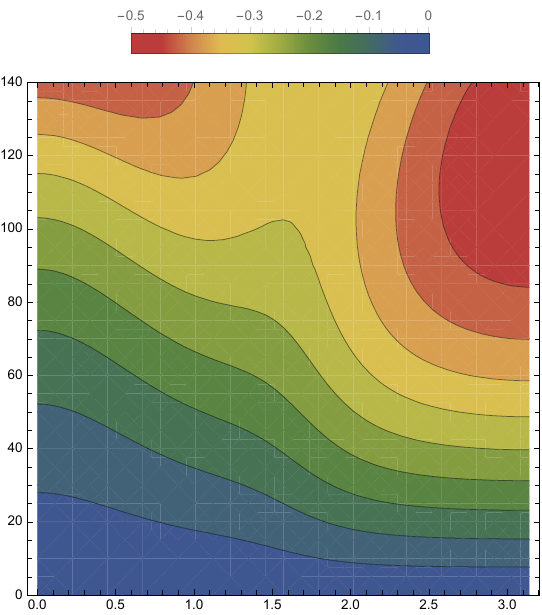}
\put(12,10){\color{white} \Large{$\hat{x},\hat{y}$}}
\put(74,71){\color{white} \large{$|\Phi^{i}\rangle$}}
\end{overpic}
\begin{overpic}
[width=0.3\textwidth]{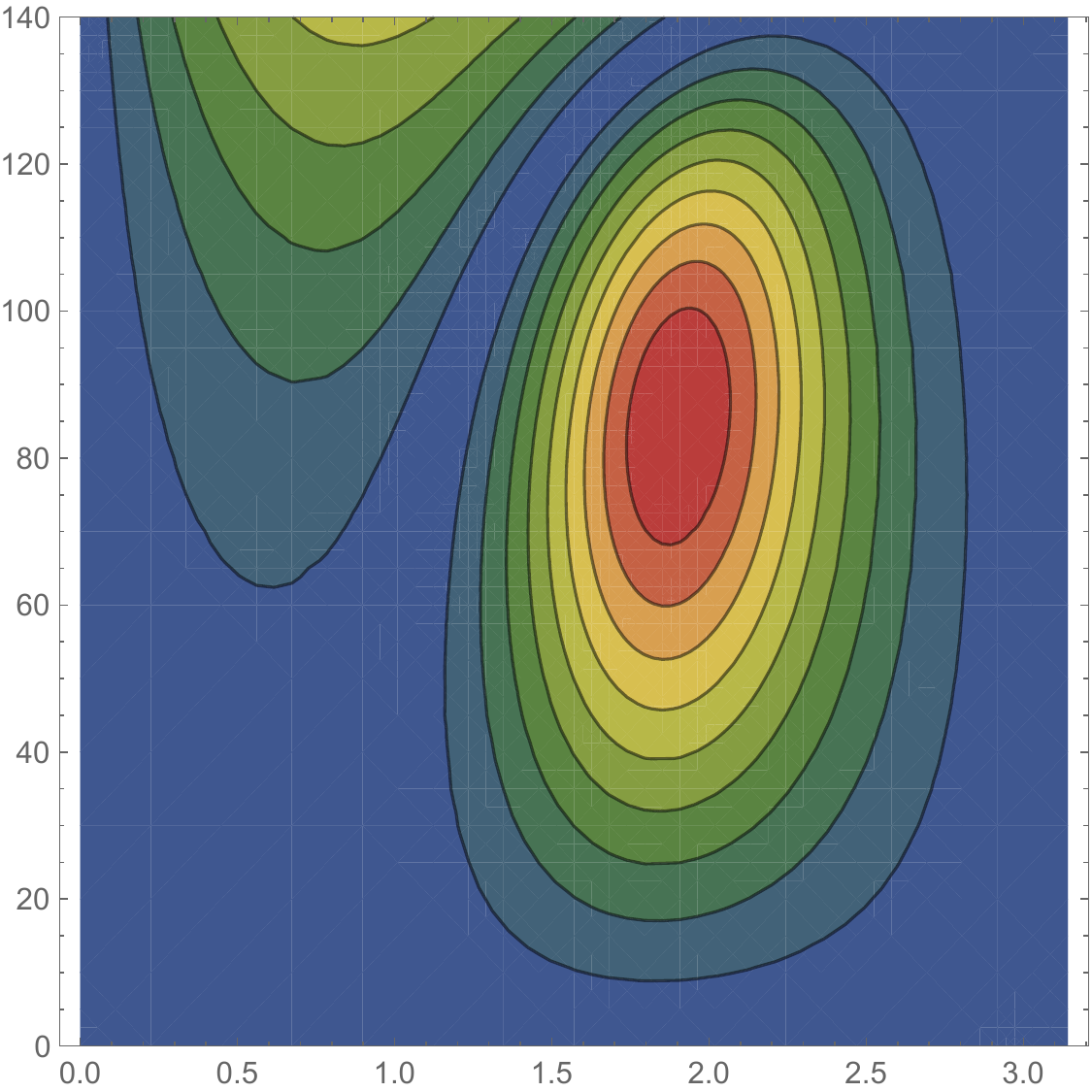}
\put(10,8){\color{white} \Large{$\hat{x},\hat{z}$}
}
\put(57,59){\color{white} \large{$|\Psi^i_{y'}\rangle$}}
\end{overpic}
\begin{overpic}
[width=0.3\textwidth]{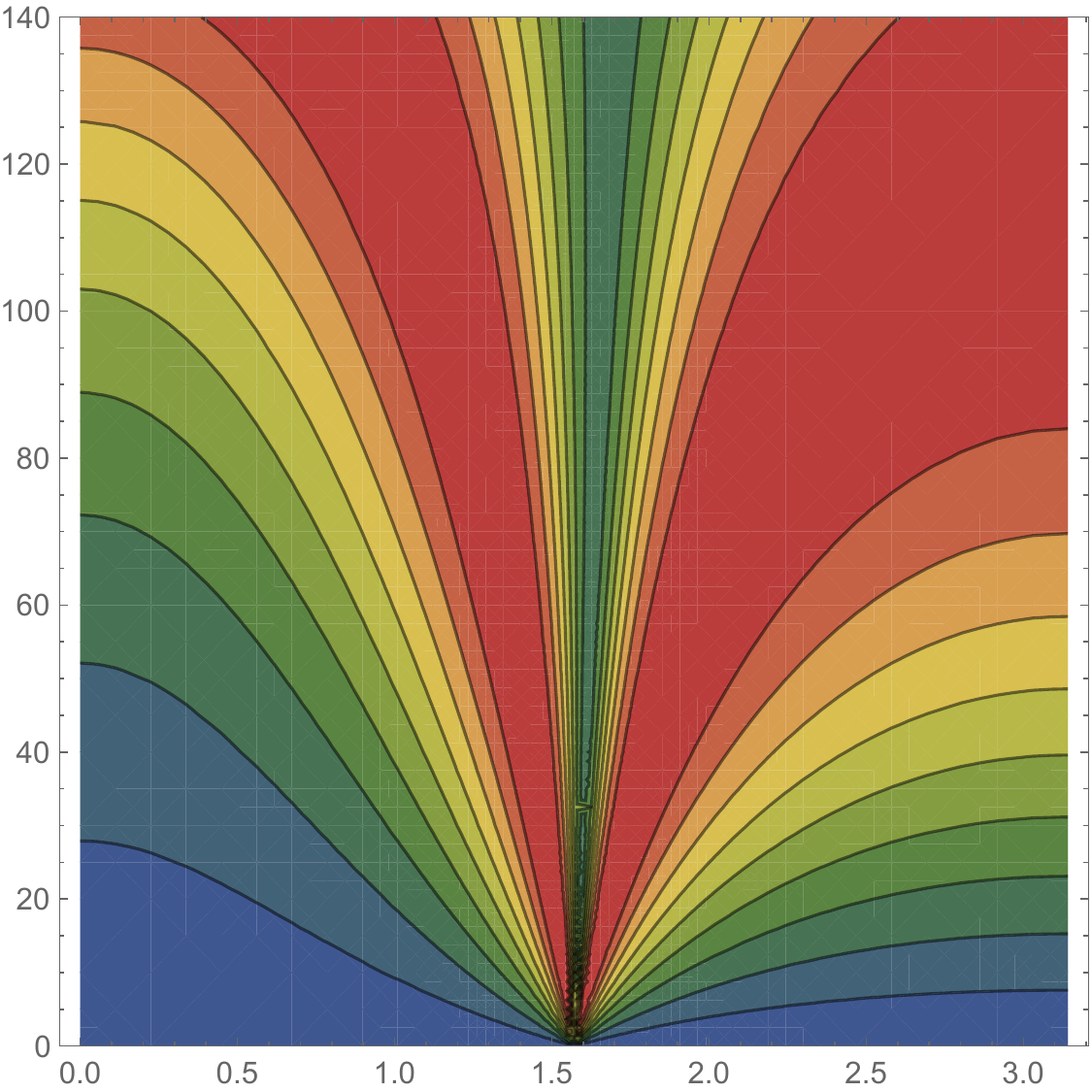}
\put(-10,80){\large{$\omega$}}
\put(12,10){\color{white} \Large{$\hat{y},\hat{x}$}}
\put(80,78){\color{white} \large{$|\Phi^i\rangle$}}
\put(34,78){\color{white} \large{$|\Psi^i\rangle$}}
\end{overpic}
\begin{overpic}
[width=0.3\textwidth]{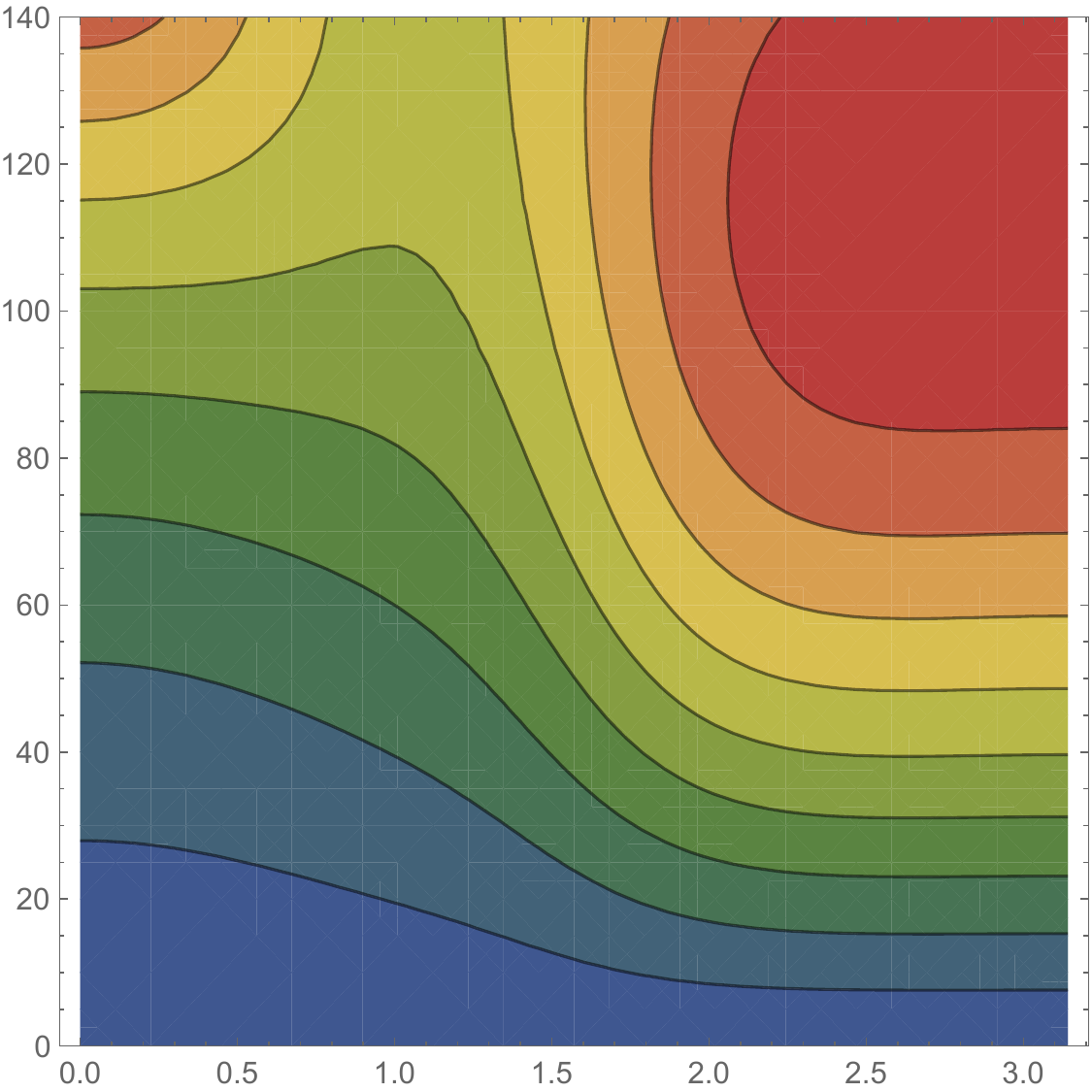}
\put(12,10){\color{white} \Large{$\hat{y},\hat{y}$}}
\put(80,78){\color{white} \large{$|\Phi^+\rangle$}}
\end{overpic}
\begin{overpic}
[width=0.3\textwidth]{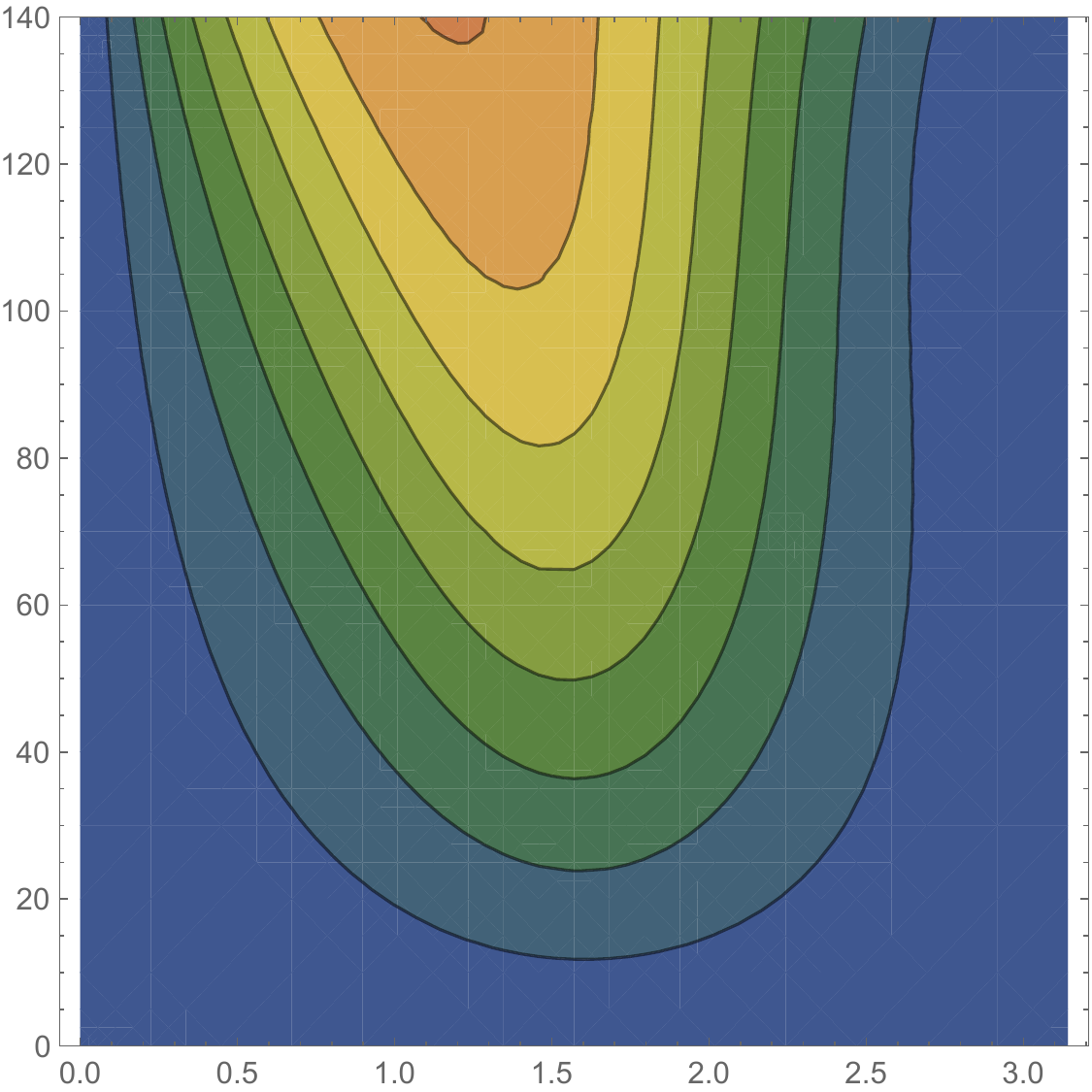}
\put(12,10){\color{white} \Large{$\hat{y},\hat{z}$}}
\end{overpic}
\begin{overpic}
[width=0.3\textwidth]{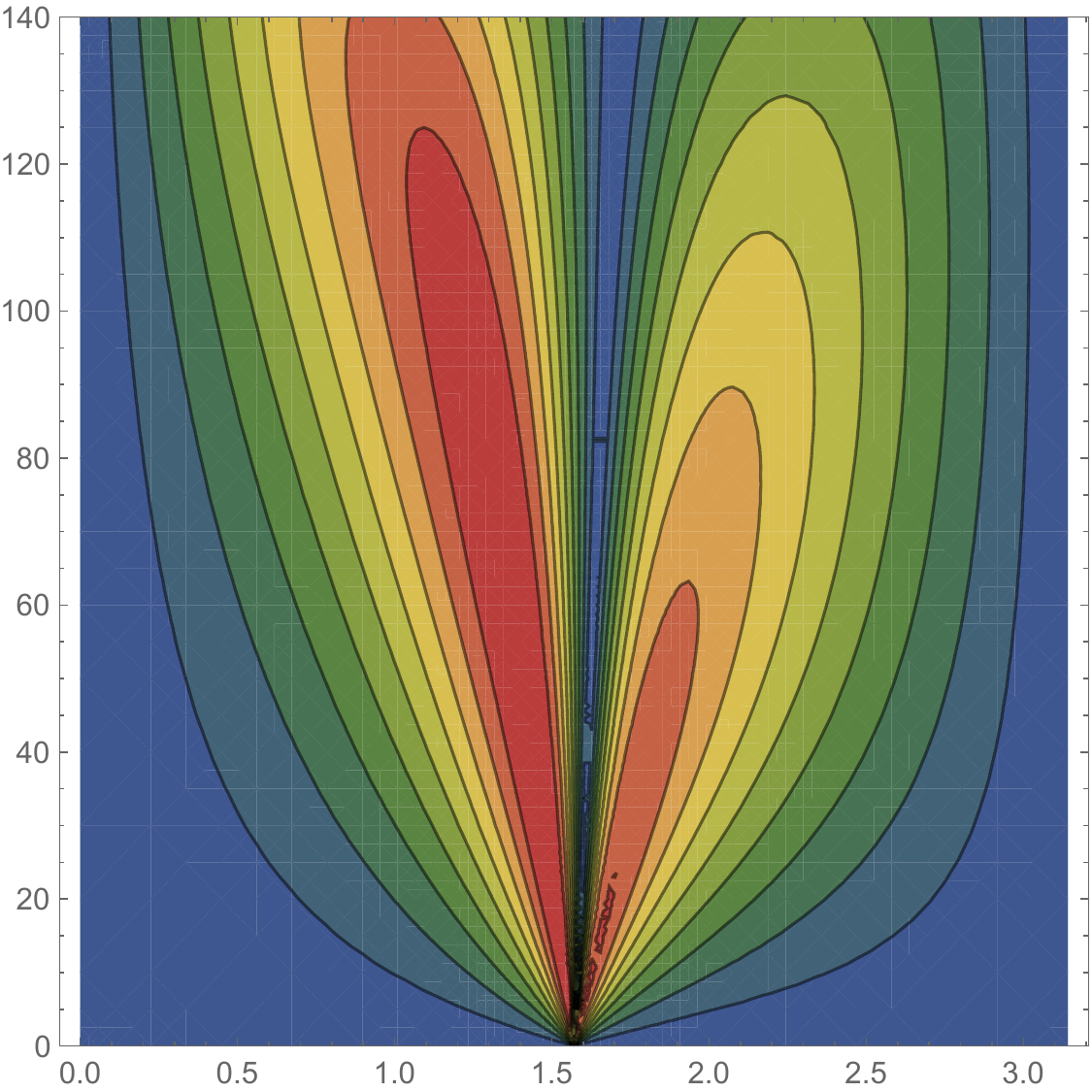}
\put(-10,80){\large{$\omega$}}
\put(12,10){\color{white} \Large{$\hat{z},\hat{x}$}}
\put(35,55){\color{white} \large{$|\Psi^-\rangle$}}
\put(50,-7){\large{$\theta$}}
\end{overpic}
\begin{overpic}
[width=0.3\textwidth]{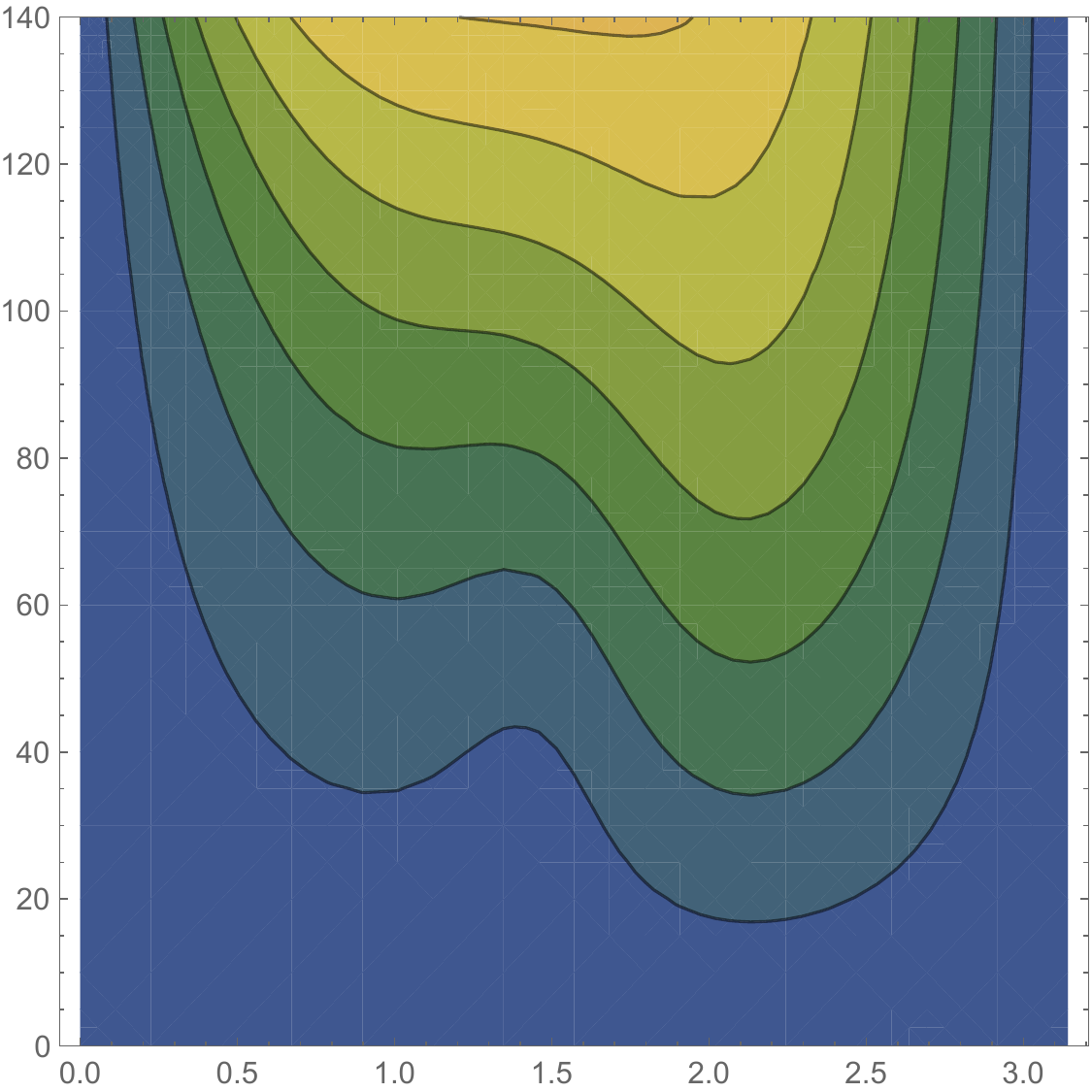}
\put(12,10){\color{white} \Large{$\hat{z},\hat{y}$}}
\put(50,-7){\large{$\theta$}}
\end{overpic}
\begin{overpic}
[width=0.3\textwidth]{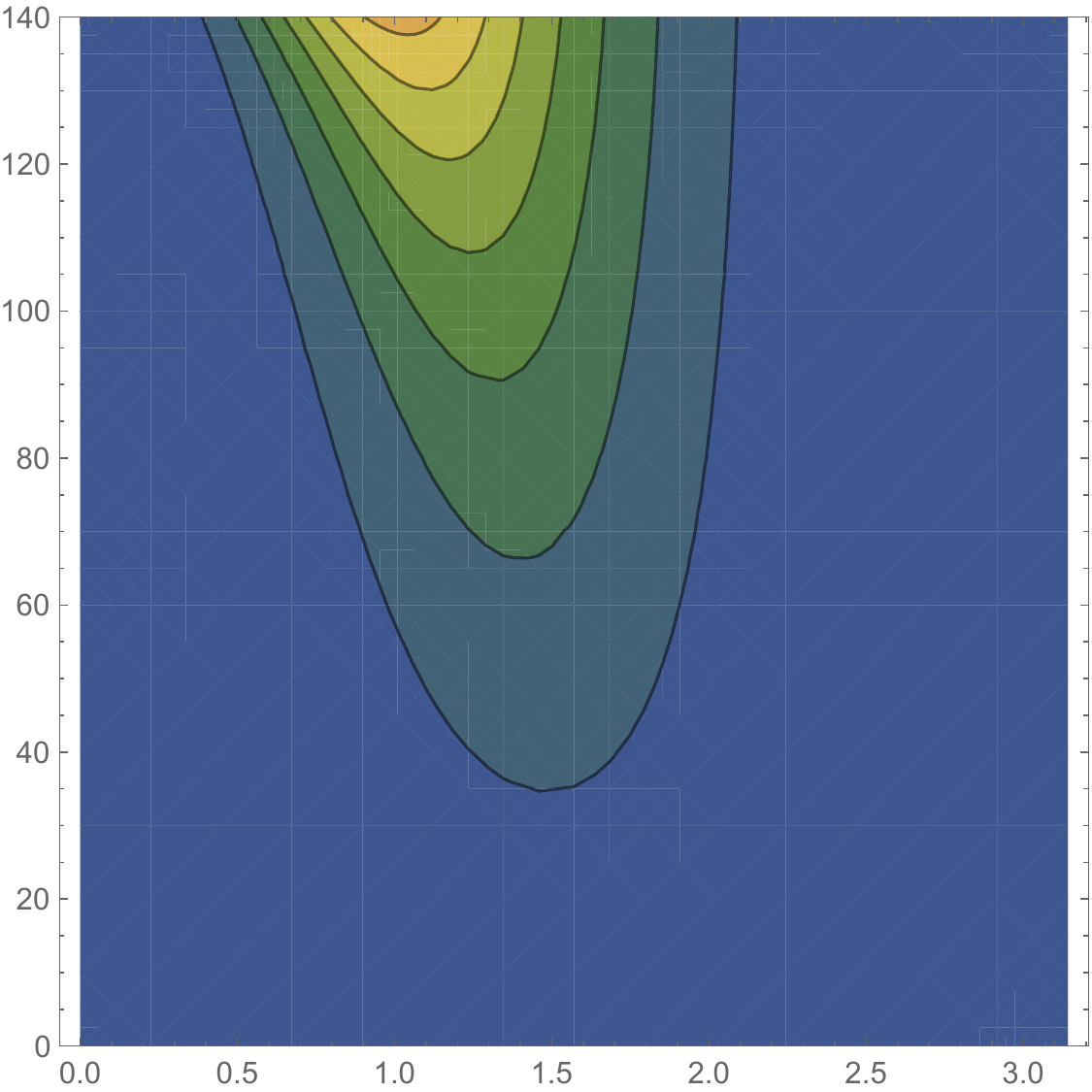}
\put(12,10){\color{white} \Large{$\hat{z},\hat{z}$}}
\put(50,-7){\large{$\theta$}}
\end{overpic}
\vspace{3mm}
\caption{The lowest eigenvalue of the partially transposed density matrix $\rho^T(\vec{s}_N,\vec{s}_\gamma)$ in the ($\theta,\omega$) plane ($\omega$ in units of MeV). The label $\hat{x},\hat{x}$ means $\vec{s}_N=\hat{x}$ and $\vec{s}_\gamma=\hat{x}$, etc.  } 
\label{x}
\end{figure}

In Fig.~\ref{x}, we plot the minimum eigenvalue $\lambda_{\rm min}(\vec{s}_N,\vec{s}_\gamma)$ of the partially transposed density matrix  $\rho^T(\vec{s}_N,\vec{s}_\gamma)$ for the proton ($N=p$) target as a function of $0\le \theta\le \pi$ and $0\le \omega< 140$ MeV. We assume   100\%  polarized initial states ($s_N=s_\gamma=1$) and consider all possible combinations of polarization axes $(\hat{s}_N,\vec{s}_\gamma)=(\hat{x},\hat{x})$, $(\hat{x},\hat{y})$, etc. 
 We have used the central values from a recent extraction \cite{Mornacchi:2022cln} 
\beq
\alpha^p_E&=&12.7\times 10^{-4}{\rm fm}^3 = 1.65 \times 10^{-10}{\rm MeV}^{-3}   , \nn 
\beta^p_M&=& 2.1\times 10^{-4}{\rm fm}^3 = 3.1\times 10^{-11} {\rm MeV}^{-3}, \nn 
\alpha^n_E&=&11.6\times 10^{-4}{\rm fm}^3 = 1.5 \times 10^{-10}{\rm MeV}^{-3}   , \nn 
\beta^n_M&=& 3.7\times 10^{-4}{\rm fm}^3 = 4.8\times 10^{-11} {\rm MeV}^{-3}.
\eeq
We find that $\lambda_{\rm min}$ is everywhere negative, meaning that the outgoing proton and photon are always entangled, in striking contrast to the unpolarized case where there is no entanglement anywhere as we have seen.   In Fig.~\ref{x}, we indicated the region $-0.5\le \lambda_{\rm min}< -0.45$ by red color. Different Bell states are realized in different red regions.  We find the proper Bell states $|\Phi^-\rangle$, $|\Phi^+\rangle$ and $|\Psi^-\rangle$ in the red regions of the  $(\hat{x},\hat{x})$ $(\hat{y},\hat{y})$, $(\hat{z},\hat{x})$ plots, respectively. There are also maximally entangled states equivalent to the Bell states up to a local unitary transformation. 
In the  $(\hat{x},\hat{y})$ plot, we find 
\beq
C\approx\begin{pmatrix} 0 & 1 & 0 \\ 1 & 0 & 0 \\ 0 & 0 & 1\end{pmatrix}, \qquad \rho\approx |\Phi^{i}\rangle\langle \Phi^{i}|, \qquad |\Phi^{i}\rangle \equiv \frac{1}{\sqrt{2}}(|\uparrow\uparrow\rangle +i|\downarrow\downarrow\rangle)
\eeq
In the $(\hat{y},\hat{x})$ plot, there are two branches where entanglement is strong. The right branch is $|\Phi^i\rangle$, while in the left branch,  
\beq
C^{\rm left}\approx \begin{pmatrix} 0 & -1 & 0 \\ 1 & 0 & 0 \\ 0 & 0 & -1\end{pmatrix}  \qquad \rho^{\rm left} \approx |\Psi^{i}\rangle\langle \Psi^i|, \qquad |\Psi^i\rangle \equiv \frac{1}{\sqrt{2}}(|\uparrow\downarrow\rangle+i|\downarrow\uparrow\rangle). 
\eeq
In the red region of the  $(\hat{x},\hat{z})$ plot,   the correlation matrix takes the form 
\beq
C\approx \begin{pmatrix} 0 & 0 & 1 \\ 0 & -1 & 0\\
-1 & 0 & 0 \end{pmatrix}
\eeq
This corresponds to another Bell state 
\beq
\rho\approx |\Psi^i_{y'}\rangle\langle  \Psi_{y'}^i|, \qquad |\Psi^i_{y'}\rangle \equiv 
\frac{1}{\sqrt{2}}\left(|\uparrow\rangle_{y'} \left|-\frac{\pi}{4}\right\rangle+i|\downarrow\rangle_{y'}\left|+\frac{\pi}{4}\right\rangle\right)
\eeq
where the subscript $y'$ denotes spin   eigenstates with $y'$ as the quantization axis. $|\pm \frac{\pi}{4}\rangle$ means linearly polarized states along the direction that makes an angle $\pm \frac{\pi}{4}$ from the $x'$ axis.  
We have also studied the case   $s_N=-s_\gamma=1$ which represents  physically inequivalent configurations. We omit the results since we did not find particularly interesting new structure. 

If only one of the incoming particles is polarized, entanglement is overall weaker. When  the proton is maximally polarized $s_N=1$ but the photon is unpolarized  $s_\gamma=0$, we find entanglement everywhere in all cases $\vec{s}_N=\hat{x},\hat{y},\hat{z}$, but  maximal entanglement is  nowhere found. On the other hand, when the photon is 100\% polarized $s_\gamma=1$ but the proton is unpolarized $s_N=0$, the system is entangled everywhere for  $\vec{s}_\gamma=\hat{z}$, but nowhere entangled for  
$\vec{s}_\gamma=\hat{x},\hat{y}$. Needless to say, if the initial polarizations are weaker $|s_{N,\gamma}|<1$, entanglement is weaker. 

\begin{figure}
\begin{overpic}
[width=0.35\textwidth]{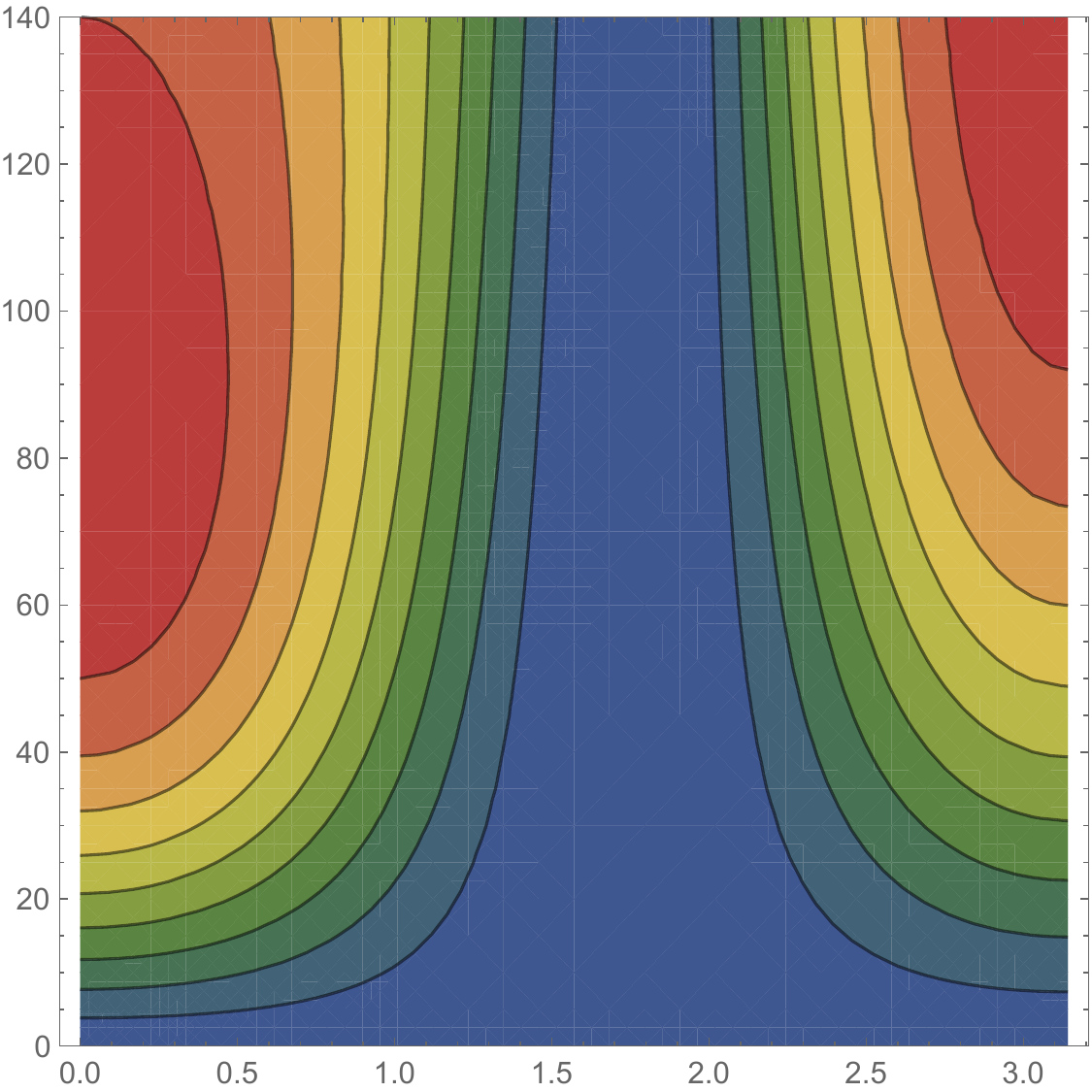}
\put(10,8){\color{white} \Large{$\hat{x},\hat{x}$}
}
\put(-10,80){\large{$\omega$}}
\put(84,85){\color{white} \large{$|\Psi^{-}\rangle$}}
\put(7,60){\color{white} \large{$|\Phi^+\rangle$}}
\end{overpic}
\begin{overpic}
[width=0.35\textwidth]{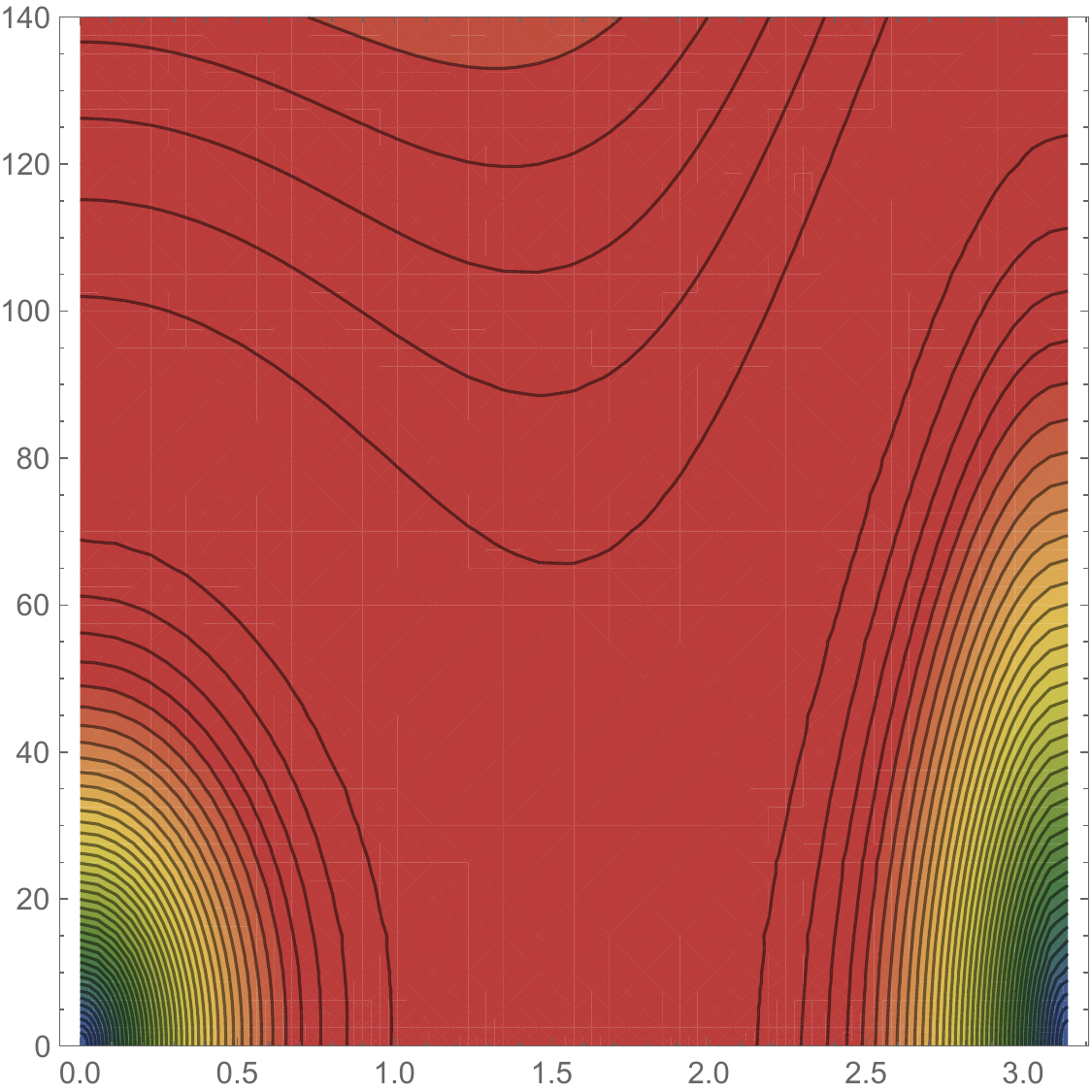}
\put(8,57){\color{white} \large{$|\Phi^{-i}\rangle$}}
\put(35,35){\color{white} \large{$|\Psi^{\circlearrowright}\rangle$}}
\put(80,85){\color{white} \large{$|\Psi^{-i}\rangle$}}
\put(45,11){\color{white} \large{$|\Psi^{-i}\rangle$}}
\put(60,42){\color{white} \large{$|\Psi^{-i}\rangle$}}
\put(10,10){\color{white} \Large{$\hat{y},-\hat{x}$}}
\end{overpic}
\begin{overpic}
[width=0.35\textwidth]{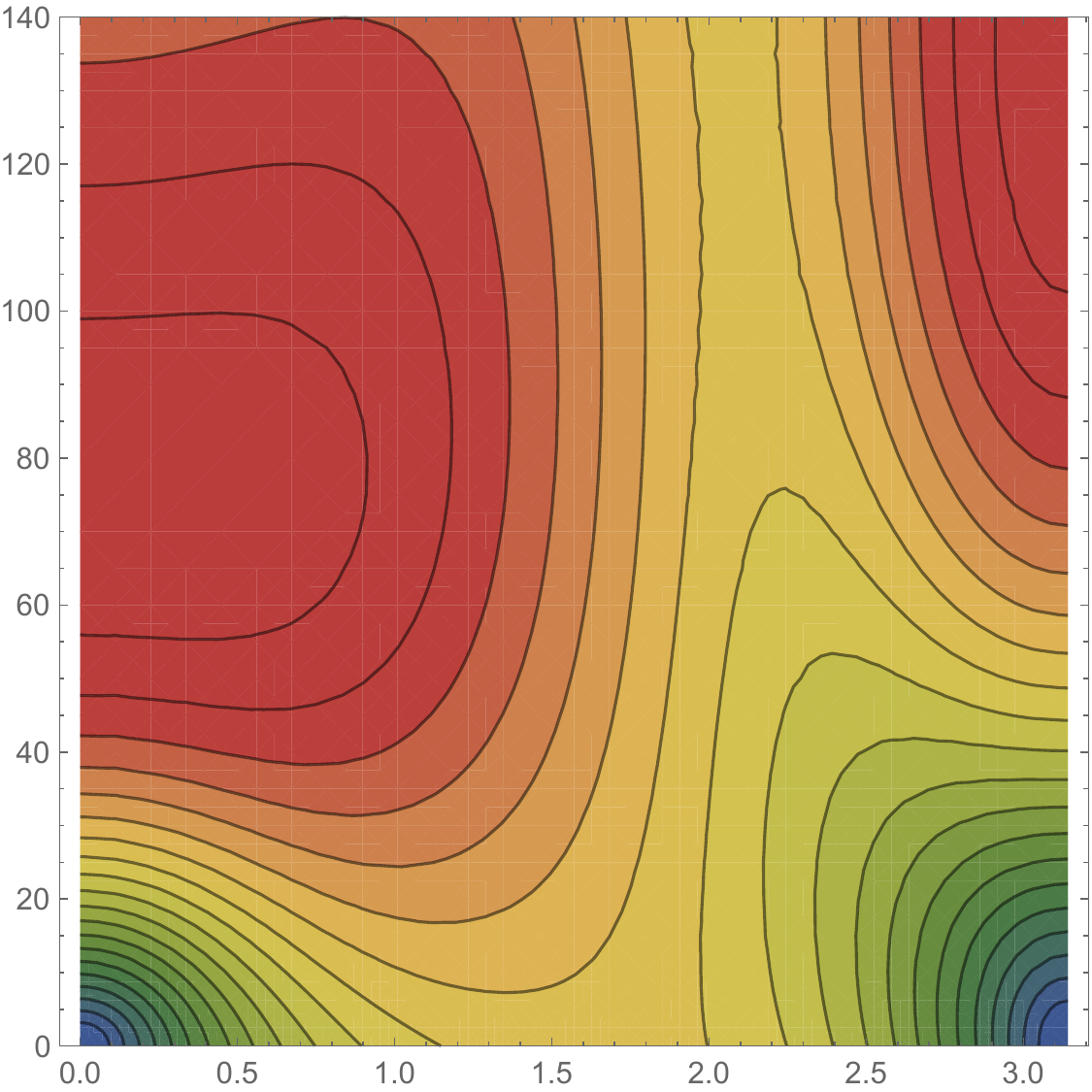}
\put(-10,80){\large{$\omega$}}
\put(10,54){\color{white} \large{$|\Phi^{+}\rangle$}}
\put(84,88){\color{white} \large{$|\Psi^{+}\rangle$}}
\put(10,8){\color{white} \Large{$\hat{y},-\hat{y}$}}
\put(50,-7){\large{$\theta$}}
\end{overpic}
\begin{overpic}
[width=0.35\textwidth]{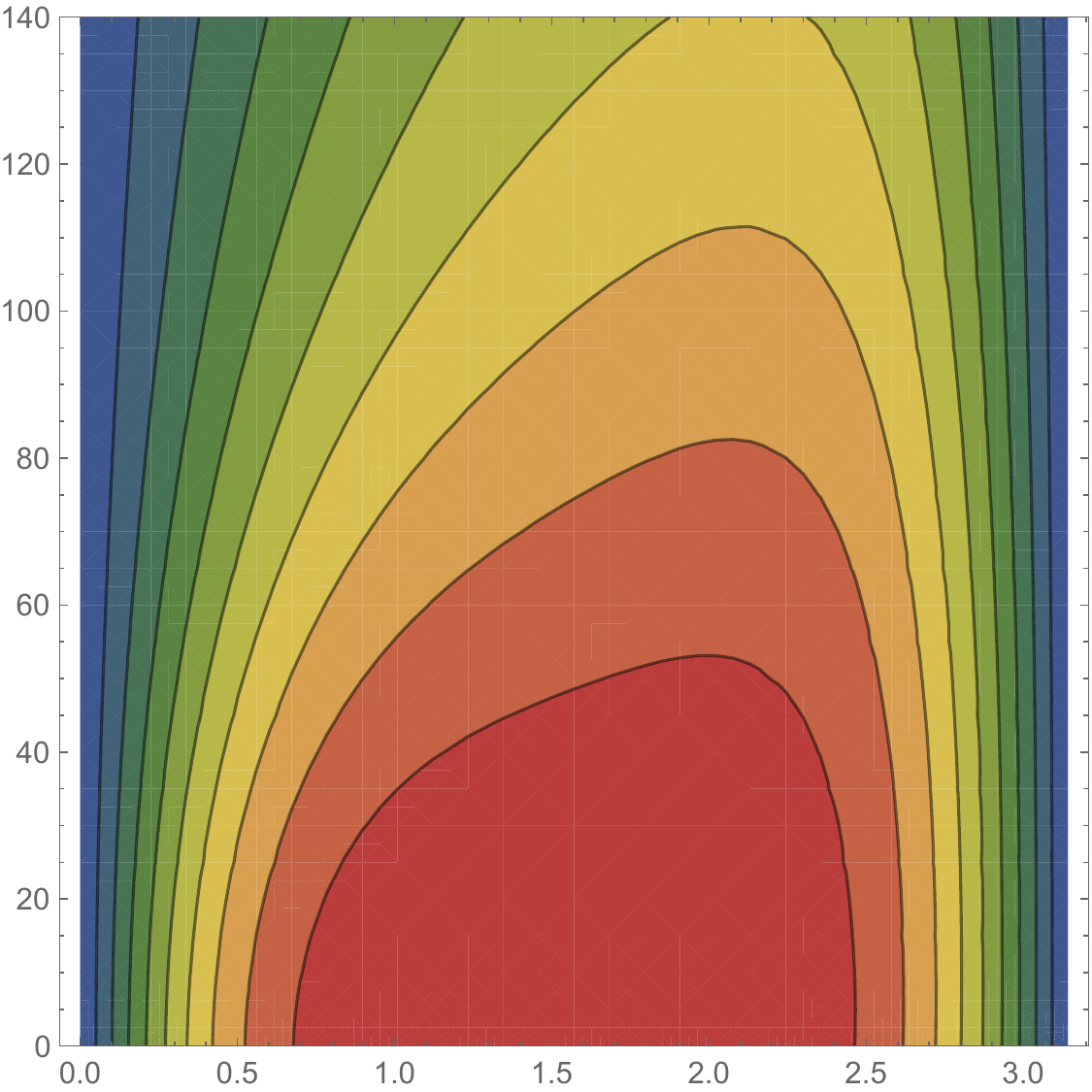}
\put(54,10){\color{white} \large{$|\Psi^{-}\rangle$}}
\put(10,8){\color{white} \Large{$\hat{z},-\hat{x}$}}
\put(50,-7){\large{$\theta$}}
\end{overpic}
\vspace{2mm}
\caption{The same as Fig.~\ref{x}, but for the neutron target. In the $(\hat{y},-\hat{x})$ plot, we have increased the number of contours since the region of strong entanglement $\lambda_{\rm min}<-0.45$ is broad.}
\label{neutron}
\end{figure}
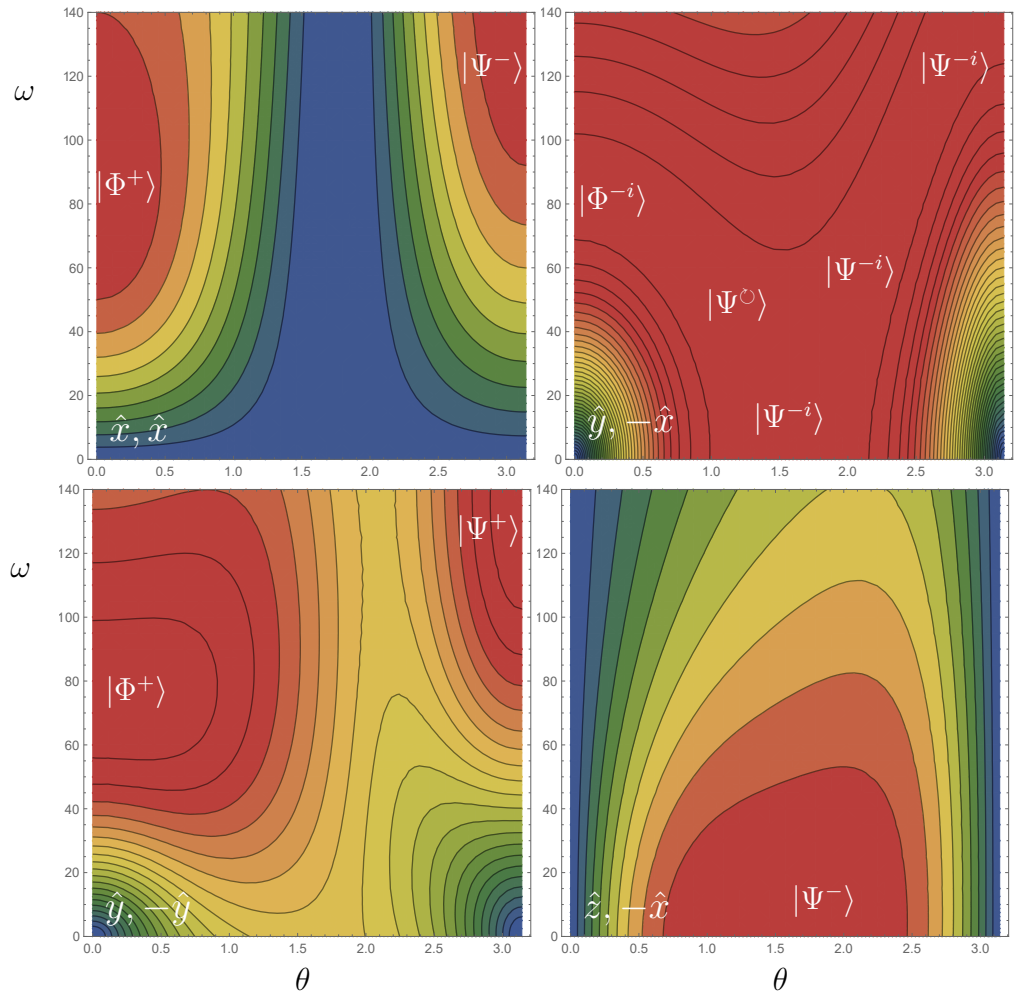

We now turn to the neutron target case $N=n$.   
Since the dominant Thomson scattering term is absent, the scattering is weak, and  one might naively expect weaker entanglement. To our surprise, however, we find a rich pattern of  entanglement, including maximal entanglement in many initial spin configurations.  
In Fig.~\ref{neutron}, we plot four representative results.  In the $(\vec{s}_N,\vec{s}_\gamma)=(\hat{y},-\hat{y})$ plot, we find the $|\Psi^+\rangle$ state which we did not find for  the proton target.   The case $(\vec{s}_N,\vec{s}_\gamma)=(\hat{y},-\hat{x})$ is most exotic. The system is almost everywhere  maximally entangled $\lambda_{\rm min}<-0.45$. Moreover,  different Bell states are realized in different regions. We find three new variants of the Bell states 
\beq
&& C\approx \begin{pmatrix} 0 & -1 & 0 \\ -1 & 0 & 0 \\ 0 & 0 & 1\end{pmatrix}  \qquad \rho \approx |\Phi^{-i}\rangle\langle \Phi^{-i}|, \qquad |\Phi^{-i}\rangle = \frac{1}{\sqrt{2}}(|\uparrow\uparrow\rangle-i|\downarrow\downarrow\rangle) \nn 
&& C\approx \begin{pmatrix} 0 & 1 & 0 \\ -1 & 0 & 0 \\ 0 & 0 & -1\end{pmatrix}  \qquad \rho \approx |\Psi^{-i}\rangle\langle \Psi^{-i}|, \qquad |\Psi^{-i}\rangle = \frac{1}{\sqrt{2}}(|\uparrow\downarrow\rangle-i|\downarrow\uparrow\rangle) \nn 
&& C\approx \begin{pmatrix} 0 & 0 & -1 \\ -1 & 0 & 0 \\ 0 & -1 & 0\end{pmatrix}  \qquad \rho \approx |\Psi^{\circlearrowright}\rangle\langle \Psi^\circlearrowright|.
\eeq
The last state, which we call $\Psi^\circlearrowright$, is quite intriguing. It is unitary-equivalent to a Bell state and describes anticorrelation between spin components along different axes related by  cyclic permutations
\beq
\langle \sigma^{x'}\tau^{z'}\rangle=\langle\sigma^{y'}\tau^{x'}\rangle=\langle \sigma^{z'}\sigma^{y'}\rangle=-1.
\eeq
The different Bell states in the $(\hat{y},-\hat{x})$ plot  are continuously connected.  For example, at $\omega=70$ MeV and $\theta=0.6$, the $C$-matrix takes the form 
\beq
C\approx \begin{pmatrix} 0 & -0.410 & -0.912 \\ -1 & 0 & 0 \\ 0 & -0.912 & 0.410 
\end{pmatrix}. \label{inter}
\eeq
This is a maximally entangled state ($0.410^2+0.912^2\approx 1$) and interpolates between $|\Psi^\circlearrowright\rangle$ and $|\Phi^{-i}\rangle$. The red region at the bottom of the $(\hat{x},\hat{z})$ plot is maximally entangled, but  does not correspond to a  Bell(-like) state. The $C$-matrix have five ${\cal O}(1)$ matrix elements, similar to (\ref{five}).  
Another interesting observation is that we find $\Phi$-type Bell states (spins aligned) in the forward region $\theta \approx 0$ and $\Psi$-type Bell states (spins anti-aligned) in the backward region $\theta \approx \pi$. This is opposite from the proton target case. This may be because the anomalous magnetic moment interaction, dominant in the neutron case, can flip the nucleon spin. 

\begin{figure}
\begin{overpic}
[width=0.35\textwidth]{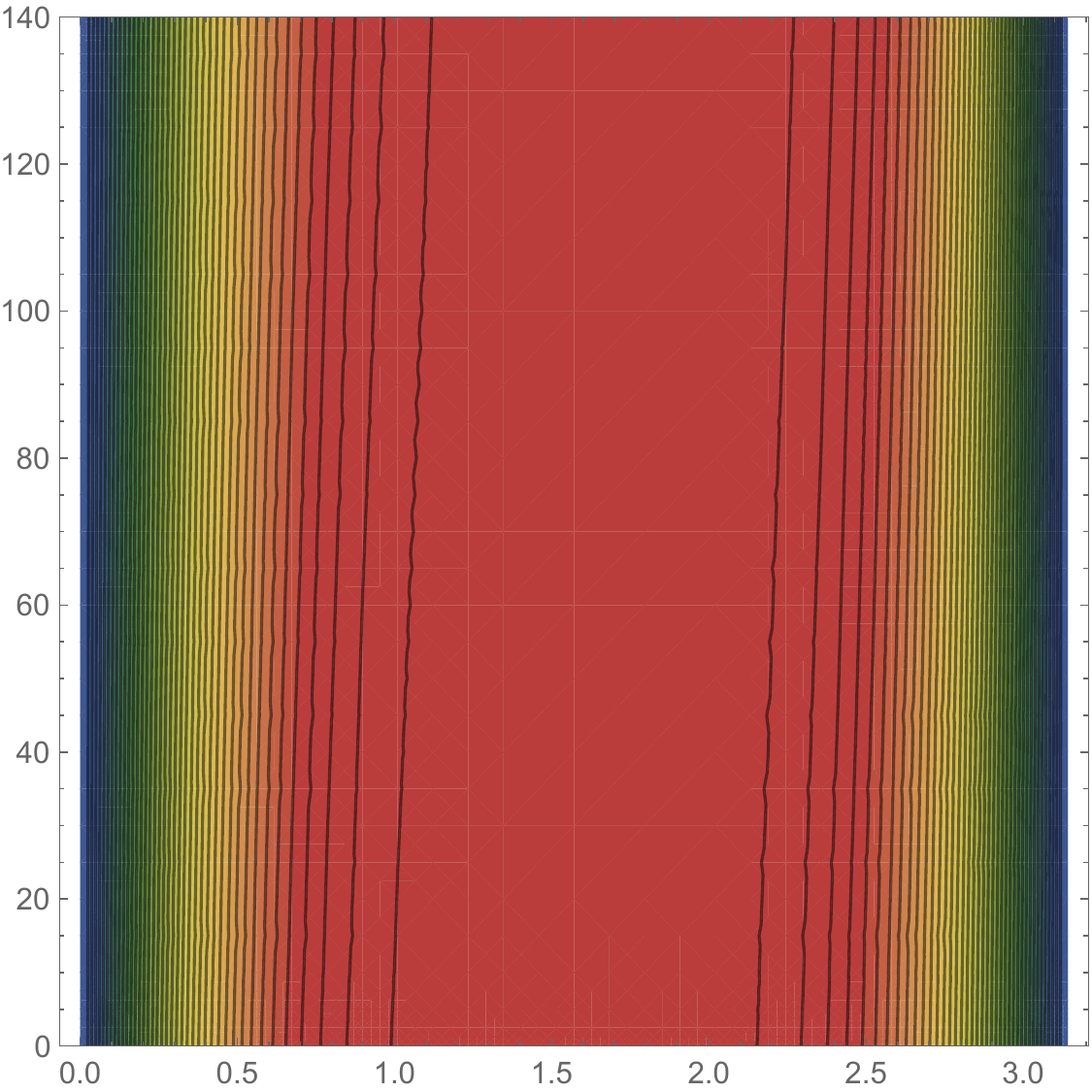}
\put(-10,80){\large{$\omega$}}
\put(46,47){\color{white} \large{$|\Psi^{-i}\rangle$}}
\put(10,8){\color{white} \Large{$\hat{y},-\hat{x}$}}
\put(50,-7){\large{$\theta$}}
\end{overpic}
\begin{overpic}
[width=0.35\textwidth]{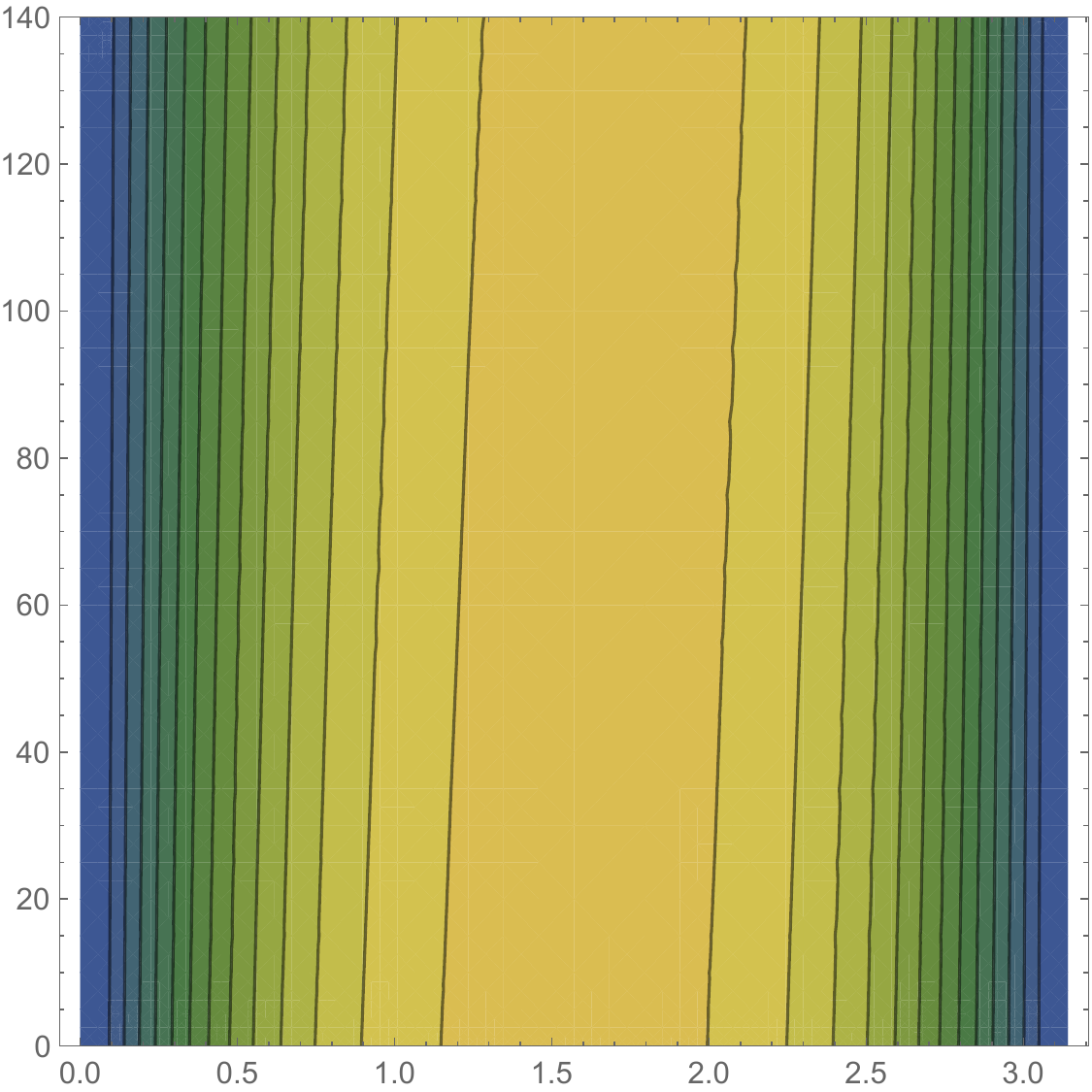}
\put(10,8){\color{white} \Large{$\hat{y},-\hat{y}$}}
\put(50,-7){\large{$\theta$}}
\end{overpic}
\vspace{2mm}
\caption{Same as the $(\hat{y},-\hat{x})$ and $(\hat{y},-\hat{y})$ plots in Fig.~\ref{neutron}, except that the polarizabilities are set to zero $\alpha^n_E=\beta^n_M=0$ by hand. }
\label{neutron2}
\end{figure}
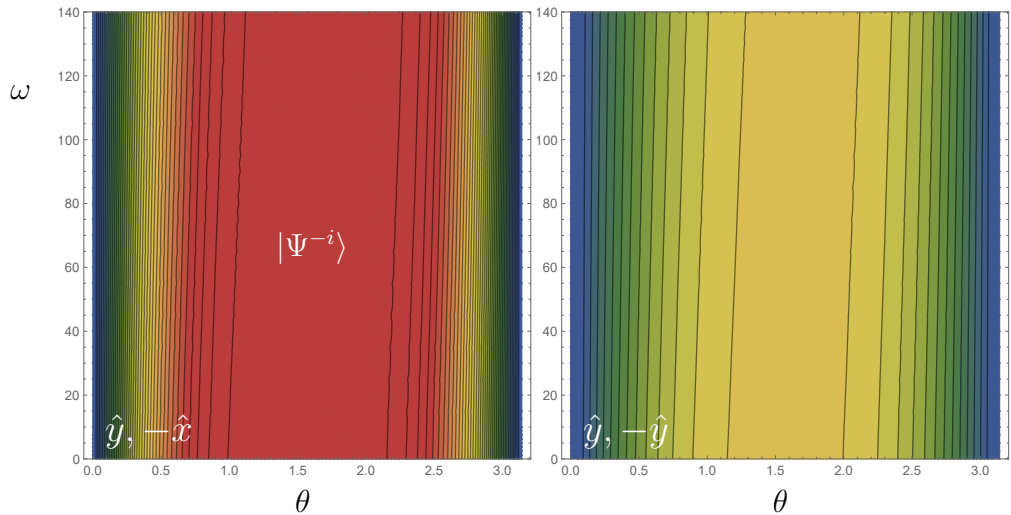

Finally we comment on the role of  the polarizabilities $\alpha_E,\beta_M$. In the proton target case, their numerical impact is minor. The plots in Fig.~\ref{x} do not change appreciably if we set $\alpha^p_E=\beta^p_M=0$. However, in the neutron case, due to the absence of the Thomson scattering term, entanglement is strongly affected by $\alpha^n_E,\beta^n_M$.  For example, if we set $\alpha^n_E=\beta^n_M=0$, the structures in the $(\hat{x},\hat{x})$ plot of Fig.~\ref{neutron} disappear completely. The system becomes separable (not entangled). Thus, the   polarizabilities are essential to generate  entanglement in this case.  In  Fig.~\ref{neutron2}, we show  the $(\hat{y},-\hat{x})$ and $(\hat{y},-\hat{y})$ results again, but this time setting $\alpha^n_E=\beta^n_M=0$.  We see that the anomalous magnetic moment is enough to generate entanglement, but only one Bell state is found in the $(\hat{y},-\hat{x})$  case and  the maximally entangled regions disappear in the $(\hat{y},-\hat{y})$ case. Therefore, the role of the polarizabilities in these cases is to  strengthen,  reshape and diversify the pattern of entanglement.    

Since a free neutron target is not available,  the extraction of the neutron polarizabilities in practice  relies on Compton scattering off the deuteron followed by the  subtraction of the proton contribution   \cite{Kossert:2002jc,COMPTONMAX-lab:2014cve}. 
We leave this more complicated problem for future work. 

\section{Entanglement in Wide angle Compton scattering}

We now turn to the high energy regime defined here as $\omega\gg 1$ GeV.  In the so-called wide angle Compton scattering (WACS) kinematics $\sqrt{-t}, \sqrt{-u}\sim {\cal O}(\sqrt{s})$, QCD factorization is applicable~\cite{Radyushkin:1998rt,Diehl:1998kh,Huang:2001ej}. We will consider that this condition on the Mandelstam variables is met by introducing the following cuts of acceptance to our numerical studies: $-u/s,-t/s \geq 0.2$. The helicity amplitudes for this process have been computed to next-to-leading order (NLO) in \cite{Huang:2001ej,Huang:2003uy}, at which the amplitudes acquire an imaginary part suppressed by the QCD coupling $\alpha_s\ll 1$.
The nucleon mass is treated as a small parameter $m\ll \sqrt{s}$, and we include its effect to  ${\cal O}(m/\sqrt{s})$. 

In perturbative QCD, the photon-nucleon  Compton amplitude is factorized into  the perturbatively calculable photon-parton helicity amplitudes ($\phi_i^{q,g}$ with $q,g$ standing for quarks and gluons, respectively) and the non-perturbative part given by  certain moments of the generalized parton distributions (GPDs)~\cite{Diehl:2003ny}. In the context of WACS, these moments are referred to as soft form factors and will be  represented by $R_i^{q,g}$ with $i\in\{V,A,T\}$ for vector, axial-vector and tensor structures of the nucleon matrix elements, respectively. 

\begin{figure}
    \centering
    \includegraphics[scale=1.2]{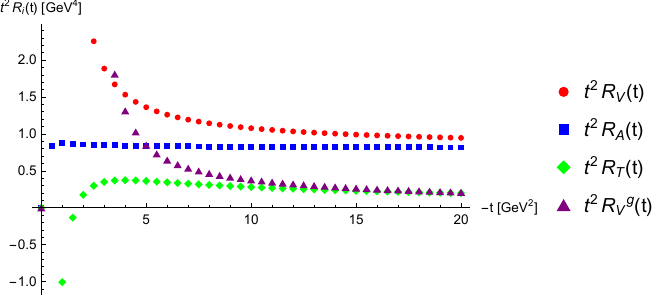}
    \caption{Soft form factors for quarks and gluons weighted by $t^2$ factor, as a function of $-t$ and at a scale $\mu^2=s/2=8~\mathrm{GeV^{2}}$. At large $-t$, quark vector and axial form factors are dominant. We consider summation over $u,d,s$ flavors.}
    \label{fig::t^2*Ri_vs_-t}
\end{figure}

The helicity amplitudes at the quark level are given by (\ref{massless}), whose Lorentz invariant generalization including $\mathcal{O}(\alpha_s)$ effects read\footnote{Refs.~\cite{Huang:2001ej,Huang:2003uy} use a different overall  sign convention for the $\phi$'s from ours. Also, the polarization vector $\epsilon_+=-\frac{1}{\sqrt{2}}(1,-i,0)$ used in~\cite{Huang:2001ej,Huang:2003uy}  has an opposite sign relative to~(\ref{epsilonconvention})  (but $\epsilon_-$ is the same), further flipping the sign of $\phi_{2,3,6}$. As a result, our $\phi$'s are related to those in~\cite{Huang:2001ej} as 
$(\phi_1,\phi_2,\phi_3,\phi_4,\phi_5,\phi_6)_{\textrm{\cite{Huang:2001ej}}}=(-\phi_1,\phi_2,\phi_3,-\phi_4,-\phi_5,\phi_6)_{\rm here}$.}
\footnote{The {\it ad hoc} subtraction of the infrared double pole $\frac{1}{\epsilon^2}$ and the Sudakov double logarithm $\ln^2 (-t)$ done in   \cite{Huang:2001ej} can be fully justified in view of the results in \cite{Bhattacharya:2023wvy,Hatta:2025xuf}.}

\begin{align}\label{massless2}
    \phi^q_1 = & - e^2  2\sqrt{\frac{s}{-u}} \Bigg\{ 1 + \frac{\alpha_s}{4\pi}C_F \Bigg[ \frac{\pi^2}{3} - 7 + \frac{2t-s}{s}\Ln{\frac{t}{u}} + \ln^2\left(\frac{-t}{s}\right) +\frac{t^2}{s^2}\left( \ln^2\left(\frac{t}{u}\right) + \pi^2 \right) - 2i\pi\Ln{\frac{-t}{s}} \Bigg] \Bigg\} \,,\\
    \phi^q_3 = &\, -e^2 \frac{\alpha_s}{2\pi}C_F\left[ \sqrt{\frac{s}{-u}} + \sqrt{\frac{-u}{s}} \right] \,, \\
    \phi^q_5 = & - e^2 2\sqrt{\frac{-u}{s}} \Bigg\{ 1 + \frac{\alpha_s}{4\pi}C_F\Bigg[ \frac{4\pi^2}{3}-7+\frac{2t-u}{u}\Ln{\frac{-t}{s}} + \ln^2\left(\frac{t}{u}\right)+\frac{t^2}{u^2}\ln^2\left(\frac{-t}{s}\right) \nonumber\\
    & \phantom{- e_q^2e^2 2\sqrt{\frac{-u}{s}} \Bigg\{ 1 + \frac{\alpha_s}{4\pi}C_F\Bigg[ } - 2i\pi\left( \frac{2t-u}{2u}+\frac{t^2}{u^2}\Ln{\frac{-t}{s}} \Bigg) \right] \Bigg\} \,,
\end{align}
where $C_F=\nicefrac{4}{3}$. Notice the imaginary part in $\phi_{1,5}^q$. Since quarks are treated as massless, no quark-helicity flip is allowed, leading to $\phi^q_{2,4,6}=0$ to all orders. For gluons, 
\begin{align}
    \phi^g_1 = & -e^2 \frac{\alpha_s}{\pi} \left[ \frac{t^2+u^2}{2s^2}\left( \ln^2\left( \frac{t}{u} \right) + \pi^2 \right) + \frac{t-u}{s}\Ln{\frac{t}{u}} + 1 \right] \,, \label{g1}\\
    \phi^g_3 = &\, -e^2\frac{\alpha_s}{\pi} \,,\\
    \phi^g_5 = & -e^2 \frac{\alpha_s}{\pi} \left[ \frac{s^2+t^2}{2u^2}\ln^2\left(\frac{-t}{s}\right) + \frac{t-s}{u}\Ln{\frac{-t}{s}} +1 -i\pi\left( \frac{t-s}{u} + \frac{s^2+t^2}{u^2}\Ln{\frac{-t}{s}} \right) \right] \label{g3} \,,
\end{align}
where $\phi_5^g$ contains an imaginary part. Gluon helicity-flip amplitudes are actually non-zero, but we neglect them (see below). 
Note that the Mandelstam variables $s,t,u$ in (\ref{massless2}) are those for the photon-nucleon scattering. Although the quark carries a fraction $x$ of the nucleon momentum, $x$ drops out in the ratios such as $s/u$. More importantly, arguments for factorization are justified if the `active' parton carries a dominant fraction of the nucleon momentum $x\approx 1$ \cite{Diehl:1998kh}, and the region $x\ll 1$ must somehow be suppressed (see below). 
Moreover, one has to choose a `symmetric frame' \cite{Diehl:1998kh} in which the `plus' component $k^+=\frac{1}{\sqrt{2}}(k^0+k^z)$ of the nucleon does not change before and after the scattering. In this frame, the photon-nucleon helicity amplitudes read~\cite{Diehl:1998kh,Huang:2001ej}
\beq
\phi_1^{\rm sym} &=& \frac{1}{2}\left(\phi_1^q(R_V+R_A) +\phi_5^q(R_V-R_A)\right) + \frac{1}{2}(\phi_1^g+\phi_5^g)R_V^g\,, \\
\phi_2^{\rm sym} & = & -\frac{\sqrt{-t}}{2m}\phi_3^q R_T\,,\\
\phi_3^{\rm sym} & = &  \phi_3^q R_V + \phi_3^gR_V^g\,,\\
\phi_4^{\rm sym}&=& -\frac{\sqrt{-t}}{4m}\left(\phi_1^q+\phi_5^q\right)R_T\,, \\ 
\phi_5^{\rm sym} &=& \frac{1}{2}\left(  \phi_5^q(R_V+R_A) + \phi_1^q(R_V-R_A) \right) + \frac{1}{2}(\phi_1^g+\phi_5^g)R_V^g\,, \\
\phi_6^{\rm sym} & = & \frac{\sqrt{-t}}{2m}\phi_3^q R_T\,,
\label{kro}
\eeq 
where 
\beq
R_i(t)=\sum_{q\in\{u,d,s\}} e_q^2 \, R_i^q(t) \qquad (i=V,A,T)\,,
\eeq
and $e_q$ is the quark's electromagnetic charge in units of the positron's: $e_u=\nicefrac{2}{3}$ for the up quark, and $e_d,\,e_s=-\nicefrac{1}{3}$ for the down and strange quarks. Note that $\phi_2^{\rm sym}=-\phi_6^{\rm sym}$ to this order~\cite{Huang:2001ej}. 
The soft form factors $R^{q,g}_{V,A,T}$ are given by the `$(-1)$-th' moment of the unpolarized GPD $H$, the polarized GPD $\widetilde{H}$, and the nucleon helicity-flip (`$E$-type') GPD $E$:  
\beq
&&R^q_V(t)= \int_{-1}^1 \frac{dx}{x}\ H^q(x,\eta=0,t)\,, \\ 
&&R^q_A(t) = \int_{-1}^1 \frac{dx}{x}\ {\rm sgn}(x) \widetilde{H}^q(x,0,t)\,, \\
&& R^q_T(t) = \int_{-1}^1 \frac{dx}{x}\ E^q(x,0,t)\,, \\
&& R^g_V(t) = \sum_{q\in\{u,d,s\}} e_q^2\int_0^1 \frac{dx}{x^2}\ H^g(x,0,t)\,, 
\label{gpdint}
\eeq
all evaluated at vanishing skewness $\eta\propto k^+-k'^+=0$. As in \cite{Diehl:1998kh}, we have  neglected the polarized gluon GPD $\widetilde{H}^g$ and the $E$-type gluon GPD $E^g$ since they are poorly known and are anyway expected to be numerically small compared to the quark ones at large $-t$, see Fig.~\ref{fig::t^2*Ri_vs_-t}.\footnote{This is why we neglected gluon helicity-flip amplitudes in (\ref{g1})-(\ref{g3}).} The nucleon spin-flip amplitudes $\phi^N_{2,4,6}$ are non-vanishing because the nucleon is massive. The large prefactor $\sqrt{-t}/m$ is compensated by the relative suppression $R_T\sim \frac{m^2}{-t}R_{V,A}$ at large $-t$~\cite{Huang:2001ej}. The above-mentioned condition $x\approx 1$ is effectively enforced by the integrals~(\ref{gpdint}) because GPDs are increasingly localized in the large-$x$ region as $|t|$ gets larger~\cite{Radyushkin:1998rt,Hatta:2025xuf}. 
Finally, 
the photon-nucleon helicity amplitudes in the original CM frame of Fig.~\ref{frame} are obtained from~(\ref{kro}) after applying  non-trivial rotations to the initial and final nucleon spinors~\cite{Diehl:2001pm,Huang:2001ej,Kroll:2021ecb}   
\beq
&& \phi_1^N= \phi_1^{\rm sym}+ \beta \phi_4^{\rm sym}\,, \nn 
&& \phi_2^N = \phi_2^{\rm sym} - \beta\phi_3^{\rm sym} \,,\nn
&& \phi_3^N = \phi_3^{\rm sym} + \frac{\beta}{2}\left( \phi_2^{\rm sym} - \phi_6^{\rm sym} \right) \,,\nn
&& \phi_4^N = \phi_4^{\rm sym} - \frac{\beta}{2}\left( \phi_1^{\rm sym} + \phi_5^{\rm sym} \right) \,,\nn
&& \phi_5^N= \phi_5^{\rm sym} + \beta \phi_4^{\rm sym} \,, \nn
&& \phi_6^N = \phi_6^{\rm sym} + \beta \phi_3^{\rm sym}\,,
\eeq
where
\begin{equation}
    \beta = \frac{2m\sqrt{-t}}{\sqrt{s}(\sqrt{s}+\sqrt{-u})}\,.
\end{equation} 

From now on, we will focus on the proton target $N=p$. 
To evaluate the moments $R_{V,A,T}$, following \cite{Burkardt:2004bv,Selyugin:2009ic,Hatta:2025xuf}, we adopt a model\footnote{This is different from the model  $H(x,t)=e^{a^2t\frac{1-x}{2x}}f(x)$ adopted  in \cite{Huang:2001ej} and in many older papers. See \cite{Burkardt:2004bv,Hatta:2025xuf} for arguments in favor of  the quadratic form $(1-x)^2$. The factor of $x^{-b}$ is to improve the fit to the electromagnetic form factor in the small $-t$ region~\cite{Selyugin:2009ic}. }
\begin{align}
\label{modeltrue}
    H^q(x,0,t) = \exp\left\{a^2 t (1-|x|)^2|x|^{-b}\right\}f_q(x) \,, & \ \ & \widetilde{H}^q(x,0,t) = \exp\left\{a^2 t (1-|x|)^2|x|^{-b}\right\}\Delta f_q(x) \,,\nonumber\\
    H^g(x,0,t) = \exp\left\{a^2 t (1-|x|)^2|x|^{-b}\right\}x g(x)\,, & \ \ & 
    E^q(x,0,t)=\exp\left\{c^2 t (1-|x|)^2|x|^{-d}\right\} e_q(x)\,.
\end{align}
Here, $f_q(x),\,\Delta f_q(x)$ stand for the unpolarized and polarized quark PDFs; $g(x)$ for the unpolarized gluon PDF of the proton; and $e_q(x) = E^q(x,0,0)$. For these distributions, we use the parameterization of the GK model~\cite{Kroll:2012sm,Goloskokov:2006hr,Goloskokov:2008ib}, defined at a reference scale $\mu_0^2=4~\mathrm{GeV^2}$. Following GK, we approximate the renormalization group evolution $\mu_0^2\to \mu^2$ of GPDs by that of PDFs, and set the scale to $\mu^2=s/2$, corresponding to the typical virtuality in the NLO corrections~\cite{Diehl:1998kh}. The scale-dependent parameters $a^2,b$ (assumed to be common to all the $H$-type GPDs) and $c^2,d$ have been fitted to reproduce the proton electromagnetic form factors $F_1(t)=\sum_q e_q\int dx\ H^q(x,0,t)$, $F_2(t)=\sum_q e_q \int dx\ E^q(x,0,t)$ in the range $2.07 \mbox{\ GeV}^2 < |t| < 11.99 \mbox{\ GeV}^2$ from the data in Ref.~\cite{Arrington:2007ux}. This fitting is performed at selected values of $\mu^2$ and interpolated in-between, hence providing the evolution of the $t$-profile in the above GPD model. For instance, at a scale $\mu^2 = 8\mbox{\ GeV}^2$ we find $a^2=1.072 \mbox{\ GeV}^{-2}$, $b=0.239$, $c^2=0.917 \mbox{\ GeV}^{-2}$, and $d=0.340$. With these values, one reproduces $F_1$ and $F_2$ with a maximum relative error around 5\% in the aforementioned $t$-range. For larger values of  $|t|$, the exponential $t$-profile in (\ref{modeltrue}) must be substituted by a rational dependence~\cite{Hatta:2025xuf,Hoodbhoy:2003uu} $H\sim 1/t^2$. Moreover, the cross section is strongly suppressed by the form factors $F_{1,2}(t)$.  For this reason, we restrict our fitting in the range $|t|<12$ GeV$^2$ and $10<s<25$ GeV$^2$. 
Finally,  the running of $\alpha_s(\mu^2)$ including the quark mass threshold effect is computed from \texttt{PARTONS}~\cite{Berthou:2015oaw} for selected values of $\mu^2 = s/2 \in [ 10/2, 25/2 ]~\mbox{GeV}^2$ and interpolated in-between.

\subsection{Results at NLO}\label{sect::wacs_nlo}

\begin{figure}
\begin{overpic}
[width=0.35\textwidth]{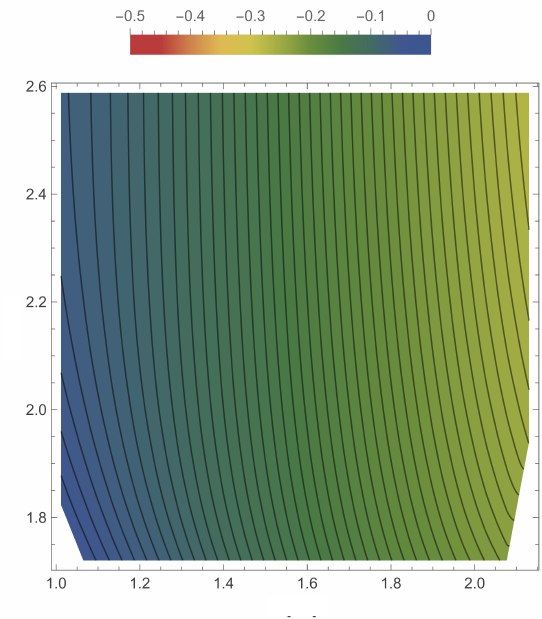}
\put(-5,70){\large{$E$}}
\put(17,13){\color{white} \Large{$\hat{x},\hat{x}$}}
\put(45,0){\large{$\theta$}}
\end{overpic}
   \begin{overpic}
[width=0.35\textwidth]{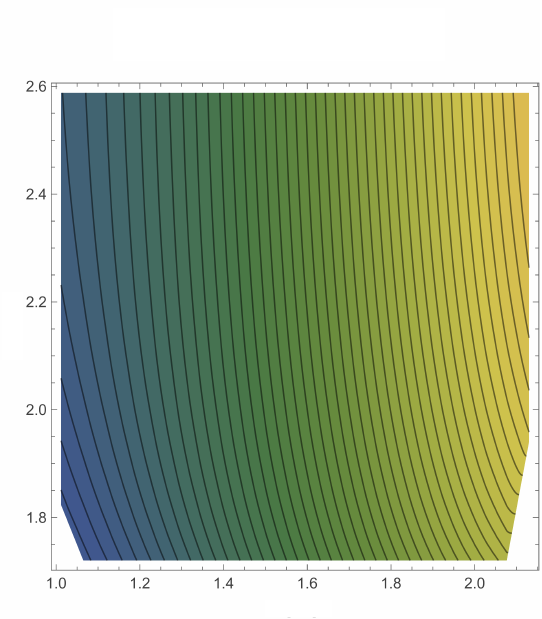}
\put(17,13){\color{white} \Large{$\hat{y},\hat{x}$}}
\put(45,-2){\large{$\theta$}}
\end{overpic}
\vspace{3mm}
   \begin{overpic}
[width=0.35\textwidth]{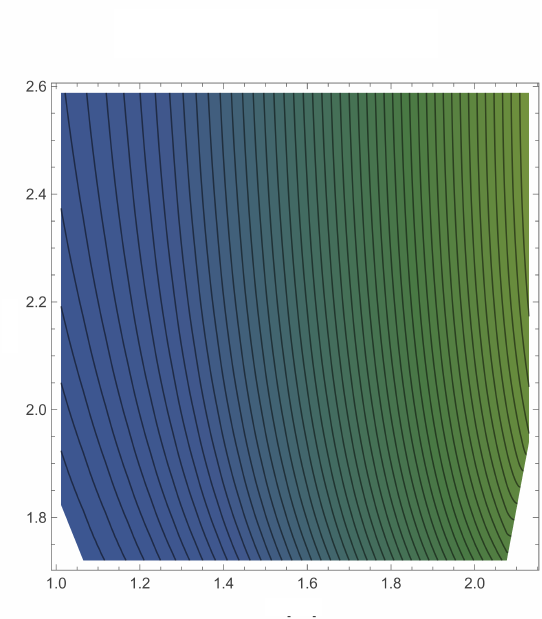}
\put(-5,70){\large{$E$}}
\put(17,13){\color{white} \Large{$\hat{z},\hat{x}$}}
\put(45,-2){\large{$\theta$}}
\end{overpic}
\vspace{2mm}
    \caption{The lowest  eigenvalue of $\rho^T$ in WACS off the proton at NLO with polarization setups $(\vec{s}_N,\vec{s}_\gamma)=(\hat{x},\hat{x})$, $(\hat{y},\hat{x})$, and $(\hat{z},\hat{x})$. $E$ is the proton energy in the CM frame in units of GeV, and $\theta$ the scattering angle in radians.  Other configurations with the same proton polarization but different photon polarization produce similar results, except for $\vec{s}_\gamma \parallel \hat{z}$, see the main text. The white areas in the lower part of the plots relate to kinematic points that 
    do not satisfy the conditions $-u/s,-t/s > 0.2$.}
    \label{fig::plots_minEigenvalue_NLO}
\end{figure}
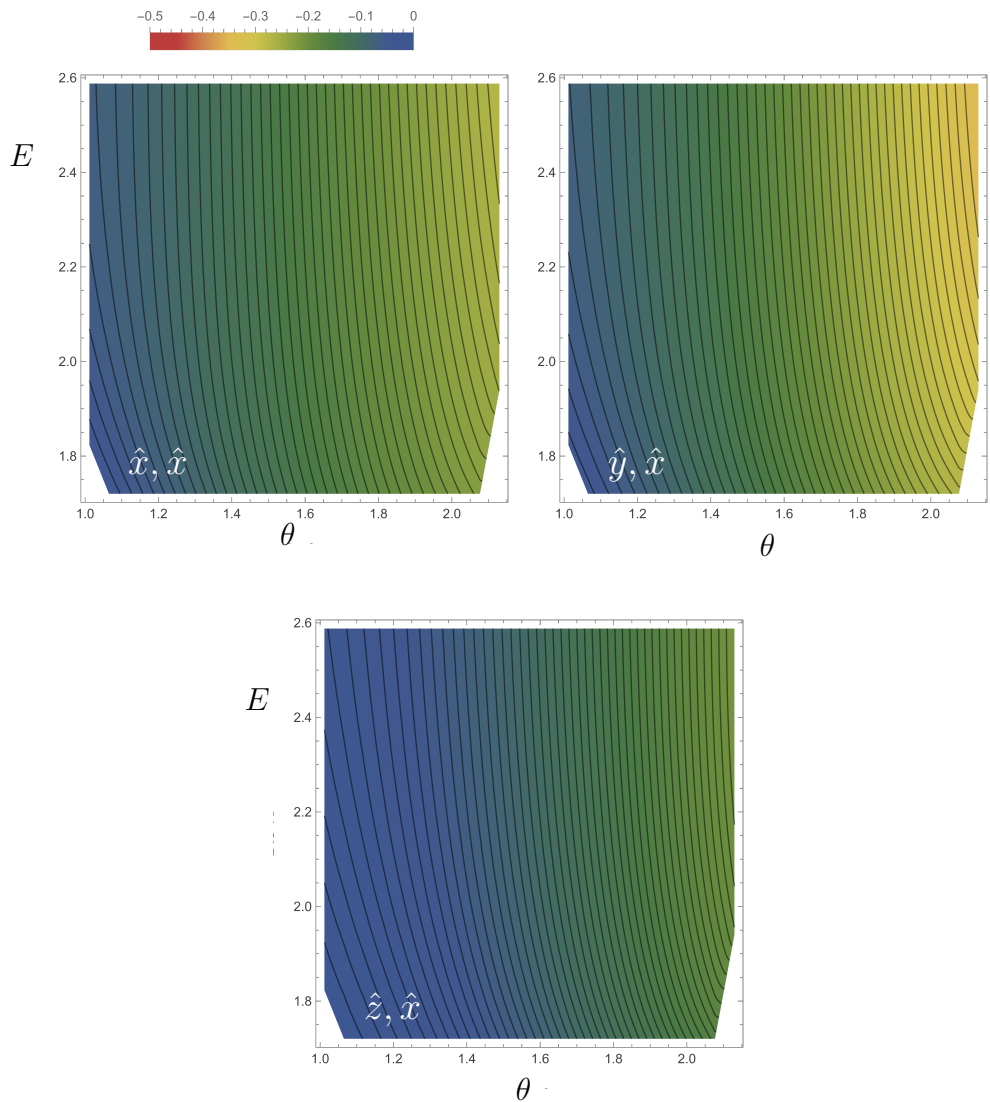

In Fig.~\ref{fig::plots_minEigenvalue_NLO}, we plot the minimum eigenvalue $\lambda_{\rm min}$ of the partially transposed density matrix $\rho^T(\vec{s}_N,\vec{s}_\gamma)$ obtained for different initial polarization configurations. The range of $\theta$ is constrained by the conditions $-t/s,-u/s>0.2$ and $E$ is the CM proton energy. 
The lowest eigenvalues of $\rho^T$ are found only in the most backward ($\theta\sim 120^\circ$) and higher energy  region of the $(\theta,E)$ phase space. The photon-proton pair is entangled everywhere, but in contrast to the low energy regime, no maximal entanglement is found. The final states are typically linear combinations of the four Bell states.

The configurations $(\vec{s}_N,\vec{s}_\gamma) = (\hat{x},\pm\hat{x})$, $(\hat{x},\pm\hat{y})$ have $\lambda_{\rm min}$ slightly above $-0.3$. The cases $(\hat{y},\pm\hat{x}), (\hat{y},\pm\hat{y})$ present the strongest entanglement with  $\lambda_{\rm min}\approx -0.36$. 
For the  longitudinally polarized photon  $(\hat{z},\pm\hat{x}), (\hat{z},\pm\hat{y})$,   $\lambda_{\rm min}\approx -0.22$. 
However, entanglement is particularly weak $\lambda_{\rm min}> -0.032$ if the photon is circularly polarized, i.e.~$\vec{s}_\gamma = \pm\hat{z}$. In fact, in this case entanglement exactly vanishes at leading order (LO). At LO, only $\phi^N_{1,4,5}$ are non-vanishing and real, and the density matrix takes the form 
\begin{flalign}
    & \left.\rho(\vec{s}_N,s_\gamma\hat{z})\right|_{\rm LO} = && \nonumber \\
    & \quad  = \begin{dcases}
        \frac{1}{4} \left[ \mathbb{I}\otimes\mathbb{I} + B_\gamma^{z'}\,\mathbb{I}\otimes\tau^{z'} + B_N^{x'}\,\sigma^{x'}\otimes\mathbb{I} + B_N^{z'}\,\sigma^{z'}\otimes\mathbb{I} + C^{x'z'}\,\sigma^{x'}\otimes\tau^{z'} + C^{z'z'}\,\sigma^{z'}\otimes\tau^{z'} \right] \,, \quad\mbox{for } \vec{s}_N=s_N\hat{x},\,s_N\hat{z} \,, \\
        \frac{1}{4} \left[ \mathbb{I}\otimes\mathbb{I} + s_\gamma\,\mathbb{I}\otimes\tau^{z'} + B_N^{a}\,\sigma^{a}\otimes\mathbb{I} + C^{a z'}\,\sigma^{a}\otimes\tau^{z'} \right] \,, \quad\mbox{for }\vec{s}_N=s_N\hat{y}\,,
    \end{dcases} 
    &&
\end{flalign}
where the different polarization coefficients and correlation matrices are dependent on the particular choice of $\vec{s}_N$ and the $s_\gamma$ value. Since there is no $\tau^{y'}$ involved, it is immediate that $\left.\rho^T\right|_{\rm LO}=\left.\rho\right|_{\rm LO}$. Thus, the partially transposed density matrix is positive-definite at LO and  consequently, all final states are separable for any kinematics. This is a generalization of the no-go theorem in the unpolarized case. When the NLO corrections are turned on, all helicity amplitudes are nonzero  and imaginary parts are developed. However, the purely NLO  helicity amplitudes $\phi^N_{2,3,6}$ are numerically 1--2 orders of magnitude lower than the photon helicity non-flip amplitudes $\phi^N_{1,4,5}$.  Entanglement requires communications between different helicity states, but such effects are suppressed either by $m/\sqrt{-t}$ or $\alpha_s$. 


\section{Discussions and conclusions}

In this paper,  we have for the first time studied entanglement in Compton scattering off the nucleon. Our analysis is restricted to the low and high energy regimes where the scattering amplitudes are (approximately) real. In these regimes, the polarization of the incoming  beams are crucial for generating  entanglement. We have shown that the degree of entanglement is strongly influenced  by  fundamental constants such as the magnetic moment and  polarizabilities at low energy, and by the partonic structure at high energy. We have considered all possible configurations of the initial beam polarizations and found all the four Bell states $|\Phi^\pm\rangle$, $|\Psi^\pm\rangle$ and their unitary rotations.   

At high energy, we find that, conversely to the low-energy scattering, entanglement is never maximally realized and the highest entangled states prefer the most possible backward region ($\theta\sim 120^\circ$) as well as linearly polarized initial photons. The NLO effects are small due to the difference in orders of magnitude between the helicity amplitudes $\phi^N_{2,3,6}\sim\mathcal{O}(\alpha_s)$ and $\phi^N_{1,4,5}$ that are non-vanishing at LO. The impact of NLO QCD corrections on QIS observables is still relatively unexplored \cite{Brandenburg:1998xw,Bernreuther:2001rq,Aguilar-Saavedra:2025byk,DelGratta:2025xjp,Jiang:2026sfw,Goncalves:2026njf}. Our work provides a new contribution to this list. 

In the intermediate region where  $\omega$ is a few hundred MeV, the amplitudes are complex. There  is then  a possibility to realize entanglement even in unpolarized scattering. We leave this to future work.  Another direction is to consider Compton scattering initiated by a virtual, spacelike photon, as in Deeply Virtual Compton scattering.

Finally, it would be very interesting if the predicted entanglement could be confirmed in experiments. This requires the coincident measurement of final state spin-spin correlations.  Up to energies of order 100 MeV, the polarization of the outgoing photon can be  measured via  secondary Compton scattering inside a detector that acts as a polarimeter. At higher energy, the photon conversion process $\gamma \to e^+e^-$ could be utilized, see  \cite{Defurne:2025dkv,Lv:2026qmu,Cheng:2026ktp,deLima:2026jbb} for recent discussions and references therein.  Also,  the polarization of the outgoing nucleon can be measured by a polarimeter exploiting  single spin  asymmetries (analyzing power) in  secondary scattering \cite{Sakai2006,JeffersonLabHallA:2007ati,BessidskaiaBylund:2022qgg,BESIII:2026yyv}. The simplest experimental setup would be to place detectors that serve as polarization analyzers in the far-forward $\theta \approx 0$ and far-backward $\theta\approx \pi$ regions, and measure their  spin correlations. For example, the $|\Phi^+\rangle$ state realized  in the very forward  region $\theta\approx 0$ in the top-left plot in Fig.~\ref{neutron} can be rewritten in terms of the transverse/linear-polarization basis  
\beq
|\Phi^+\rangle = \frac{|\uparrow\rangle_N|\uparrow\rangle_\gamma + |\downarrow\rangle_N|\downarrow\rangle_\gamma}{\sqrt{2}} =\frac{|x'\rangle_N |x'\rangle_\gamma +i|-x'\rangle_N |y'\rangle_\gamma}{\sqrt{2}}.
\eeq
The transversely polarized nucleon  undergoes a secondary scattering inside the detector, generating the $\sin\phi_N$ asymmetry where  $\phi_N$ is the azimuthal angle with respect to the $\hat{x}'\approx \hat{x}$ axis.  In terms of the density matrix, 
\beq
D_N\propto 1+A_N (\sin\phi_N\sigma^x-\cos\phi_N \sigma^y),
\eeq
where $A_N$ is the analyzing power ($N$ here stands for `normal'). 
Likewise, the linearly polarized photon will be distributed inside the detector with  $\cos2\phi_\gamma$ asymmetry described by the density matrix 
\beq
D_\gamma \propto 1+A_\gamma \left( \sigma^x\cos2\phi_\gamma+\sigma^y\sin2\phi_\gamma\right). 
\eeq
The angular distribution of  nucleon-photon pairs arriving in coincidence is therefore  
\beq
{\rm Tr}[(D_N\otimes D_\gamma)\rho]\propto A_NA_\gamma \sin(2\phi_\gamma \pm \phi_N),
\eeq
where the $\pm$ sign is for the $\Phi$-type and $\Psi$-type Bell states, respectively. 
This is analogous to the $\cos 2(\phi_1-\phi_2)$ correlation originally studied for entangled photon pairs in $e^+e^-\to 2\gamma$ \cite{Wu:1950zz}. 
Polarization measurements of  final state photons and protons have been separately done at low energy facilities. However, combining the two  in a single experimental setup is unprecedented, and requires  significant experimental innovations. We hope our work triggers discussions in this direction.

\section*{Acknowledgments}
We thank correspondence with Preslav Asenov, Peter Kroll, and Pawe{\l} Sznajder.  V.M.-F. thanks RIKEN BNL Research Center, where this work was initiated, for hospitality, and the Center for Frontiers in Nuclear Science for funding during his time at the laboratory. For the purpose of open access, the authors have applied a CC-BY copyright licence to any Author Accepted Manuscript (AAM) version arising from this submission. Y.H.~was supported by the U.S. Department of Energy under Contract No. DE-SC0012704, and also by LDRD funds from Brookhaven Science Associates. The research of V.M.-F.~was funded in part by l’Agence Nationale de la Recherche (ANR), project ANR-23-CE31-0019.

\appendix

\section{Alternative derivation of the density matrix for unpolarized initial states}\label{app::density}

In this Appendix, we outline the calculation of the density matrix directly in terms of the amplitudes $A_i$     commonly used in the low-energy literature. It is convenient to work in the frame $x'y'z'$ in which the nucleon and photon momenta read 
\beq
k^\mu &=& (E,-\omega\sin\theta, 0,\omega\cos\theta), \qquad 
k'^\mu = (E,0,0,\omega)\nn
q^\mu &=& \omega(1,\sin\theta, 0,-\cos\theta), \qquad 
q'^\mu = \omega(1,0,0,-1).\label{cm}
\eeq
The nucleon two-component spinor in this frame takes the form
\beq
\xi^+_{\rm in}=\begin{pmatrix} \cos\frac{\theta}{2} \\ e^{i\phi}\sin\frac{\theta}{2}\end{pmatrix} \,, \qquad \xi_{\rm in}^-= \begin{pmatrix}  -e^{-i\phi} \sin\frac{\theta}{2} \\ \cos\frac{\theta}{2} \end{pmatrix}, 
\eeq
with $\phi=\pi$, for the two helicity eigenstates. Consider the spin-dependent amplitude $T_{\lambda i}$ introduced in  (\ref{as}) 
\beq
T_{\lambda i} &=& A_1\delta_{\lambda i} + A_2\hat{q}_\lambda \hat{q}'_i + iA_3\epsilon_{\lambda ik} \sigma_k + iA_4\sigma\cdot (\hat{q}'\times \hat{q})\delta_{\lambda i} \nn && +iA_5\left( \epsilon_{\lambda jk}\hat{q}_j\hat{q}'_i - \epsilon_{ijk}\hat{q}'_j   \hat{q}_\lambda \right)\sigma_k + iA_6 \left(  \epsilon_{\lambda jk}\hat{q}'_j \hat{q}'_i - \epsilon_{ijk}\hat{q}_j \hat{q}_\lambda \right)\sigma_k. 
\eeq
We find it convenient, and also instructive to work in the linear polarization basis for the outgoing photon
\beq
\epsilon'_{x'}=(1,0,0), \qquad \epsilon'_{y'}=(0,1,0),
\eeq
along the $x'$ and $y'$ directions, 
instead of the helicity basis along $z'$ 
\beq
\epsilon'_{\pm} = \frac{1}{\sqrt{2}}(1,\mp i, 0), 
\eeq
adopted in the main text. The density matrix for the outgoing photon takes the same form as (\ref{photonrho})  
\beq
\rho_\gamma = \frac{1}{2}\begin{pmatrix} 1+B_{lin}^{z'} & B_{lin}^{x'}-iB_{lin}^{y'} \\ B_{lin}^{x'}+iB_{lin}^{y'} & 1-B_{lin}^{z'}\end{pmatrix},
\eeq
but now the interpretation is different. $B_{lin}^{z'}=\pm 1$ mean linear polarization in the $x'$ and $y'$ directions, respectively. 
For a left-moving photon, $B^{y'}_{lin}=\pm 1$ means circular polarization $|\mp_\gamma\rangle$ (spin $\pm 1$ in the $z'$ direction).

In this basis, the relevant components are 
\beq
&&T_{x'x'}= A_1-iA_4\sin\theta \sigma_2 -iA_5\sin\theta \sigma_2-iA_6\sin\theta\cos\theta \sigma_2, \nn 
&&T_{x'y'}= iA_3\sigma_3 +iA_5\sin\theta \sigma_1 +iA_6\sin\theta(\sin\theta \sigma_3+\cos\theta \sigma_1), \nn
&& T_{x'z'}= -A_2\sin\theta -iA_3\sigma_2 -iA_5\cos\theta \sigma_2-iA_6(1+\sin^2\theta)\sigma_2,  \nn
&&T_{y'x'}=
-iA_3\sigma_3, \nn 
&&T_{y'y'}=A_1-iA_4\sin\theta \sigma_2, \nn
&& T_{y'z'}=iA_3\sigma_1 +iA_5(\sin\theta\sigma_3+\cos\theta \sigma_1) +iA_6\sigma_1.
\eeq

Squaring the amplitude and averaging  over the spins of the incoming nucleon and the photon, we obtain 
\beq
\frac{1}{4}\sum_{spin,pol}|T|^2&=& \frac{1}{4}\sum_{pol}  \xi^\dagger \epsilon'^*_\lambda T_{\lambda i} \epsilon_i\epsilon^*_j T_{j\rho}^\dagger \epsilon'_\rho\xi  \nn 
&=& \frac{1}{4}\xi^\dagger \epsilon'^*_\lambda T_{\lambda i} g_{ij}T_{j\rho}^\dagger \epsilon'_\rho \xi \nn 
&=&\frac{1}{4} \xi^\dagger_\alpha \epsilon'^*_\lambda \left(A \delta_{\alpha\beta}\delta_{\lambda\rho} +\tilde{B}_N^a\sigma_{\alpha\beta}^a\delta_{\lambda\rho} +\tilde{B}_{lin}^b\delta_{\alpha\beta}\tau^b_{\lambda\rho} + \tilde{C}^{ab}_{lin}\sigma^a_{\alpha\beta}\tau^b_{\lambda\rho}  \right) \xi_\beta \epsilon'_\rho 
\nn
&=& \frac{A}{4} \xi^\dagger \epsilon'^* \left[ \mathbb{I}\otimes \mathbb{I} +B_N^a\sigma^a \otimes \mathbb{I} + B^b_{lin} \mathbb{I}\otimes \tau^b + C^{ab}_{lin}\sigma^a\otimes \tau^b\right]\xi \epsilon',
\eeq
where we defined 
\beq
g_{ij}=\sum_{pol} \epsilon_i \epsilon_j^* = \delta_{ij}-\hat{q}_i\hat{q}_j = \begin{pmatrix} \cos^2\theta & 0 & \sin\theta\cos\theta \\ 0 & 1 & 0 \\ 
\sin\theta\cos\theta & 0 & \sin^2\theta 
\end{pmatrix}_{ij}.
\eeq 
The components of the spin density matrix are computed from 
\beq
A=\frac{1}{4}{\rm Tr}[(Tg T^\dagger)_{\lambda\lambda}].
\eeq
\beq
\tilde{B}_N^a=AB_N^a=\frac{1}{4}{\rm Tr}[(TgT^\dagger)_{\lambda\lambda}\sigma^a], \qquad \tilde{B}_{lin}^b =AB_{lin}^b= \frac{1}{4}\tau^b_{\rho\lambda}{\rm Tr}[(TgT^\dagger)_{\lambda\rho}].
\eeq
\beq
\tilde{C}^{ab}_{lin}=AC^{ab}_{lin}=\frac{1}{4}{\rm Tr}[(TgT^\dagger)_{\lambda\rho}\sigma^a]\tau^b_{\rho\lambda}.
\eeq
After a straightforward but tedious calculation, we find  
\begin{align}
    A = &\ \frac{3}{4}\left|A_1\right|^2
        + \frac{3}{16}\left|A_2\right|^2
        + \frac{5}{4}\left|A_3\right|^2
        + \frac{5}{16}\left|A_4\right|^2
        + \frac{3}{4}\left|A_5\right|^2
        + \frac{3}{2}\left|A_6\right|^2
        + \frac{1}{4}\,\re\left[A_4 A_5^* + 8 A_3 A_6^*\right] \nonumber
        \\[6pt]
        &+ \frac{1}{4}\,\re\left[
        - A_1 A_2^* + A_3 A_4^* + 2 A_3 A_5^*
        + 2 A_4 A_6^* + 6 A_5 A_6^*
        \right]
        \left(\cos\theta - \cos (3\theta)\right) \nonumber
        \\[6pt]
        &+ \frac{1}{4}\left(
        \left|A_1\right|^2
        - \left|A_2\right|^2
        - \left|A_3\right|^2
        - \left|A_4\right|^2
        - 2\left|A_5\right|^2
        - 6\left|A_6\right|^2
        - 8\,\re\left[A_3 A_6^*\right]
        \right)\cos (2\theta) \nonumber
        \\[6pt]
        &+ \frac{1}{16}\left(
        \left|A_2\right|^2
        - \left|A_4\right|^2
        - 4\left|A_5\right|^2
        - 4\,\re\left[A_4 A_5^*\right]
        \right)\cos (4\theta) \,.
\end{align}
The nonvanishing components of the polarization vectors are   
\begin{align}
    \tilde{B}_{N}^{y'} = &\ 
        \frac{1}{4}\,\im\left[
        3 A_2 A_3^*
        - 5 A_1 A_4^*
        - 2 A_1 A_5^*
        + 6 A_2 A_6^*
        \right]\sin\theta + \frac{1}{4}\,\im\left[
        A_2 A_4^*
        - 2 A_1 A_3^*
        + 2 A_2 A_5^*
        - 4 A_1 A_6^*
        \right]\sin(2\theta) \nonumber
        \\[6pt]
        &  - \frac{1}{4}\,\im\left[
        A_2 A_3^*
        + 2 A_2 A_6^*
        + A_1 A_4^*
        + 2 A_1 A_5^*
        \right]\sin(3\theta) - \frac{1}{8}\,\im\left[
        A_2 A_4^*
        + 2 A_2 A_5^*
        \right]\sin(4\theta)
   \,.
\end{align}
\begin{align}
    \tilde{B}_{lin}^{z'} = &
        \frac{1}{16}\left(
        -4\left|A_1\right|^2
        + 3\left|A_2\right|^2
        + 4\left|A_3\right|^2
        - 3\left|A_4\right|^2
        + 4\left|A_5\right|^2
        + 16\left|A_6\right|^2
        + 4\,\re\left[A_4 A_5^*\right]
        + 16\,\re\left[A_3 A_6^*\right]
        \right) \nonumber
        \\[6pt]
        &  + \frac{1}{4}\re\left[
        - A_1 A_2^* + A_3 A_4^* + 2 A_3 A_5^*
        + 2 A_4 A_6^* + 4 A_5 A_6^*
        \right]
        \left(\cos\theta - \cos (3\theta)\right) \nonumber
        \\ &+ \frac{1}{4}\left(
        \left|A_1\right|^2
        - \left|A_2\right|^2
        - \left|A_3\right|^2
        + \left|A_4\right|^2
        - 4\left|A_6\right|^2
        - 4\,\re\left[A_3 A_6^*\right]
        \right)\cos (2\theta) \nonumber
        \\[6pt]
        & + \frac{1}{16}\left(
        \left|A_2\right|^2
        - \left|A_4\right|^2
        - 4\left|A_5\right|^2
        - 4\,\re\left[A_4 A_5^*\right]
        \right)\cos (4\theta)
    \,. \label{bgamma}
\end{align} 
Finally, the following five   components of the spin correlation matrix are nonzero  
\begin{align}
\widetilde{C}_{lin}^{x'x'} = &\ \frac{1}{4}\Bigg[
    \im\left(
    -3 A_2 A_3^* - 3 A_3 A_4^*
    + 5 (A_1 + A_3) A_5^*
    - 3 (A_2 - A_4 + 2 A_5) A_6^*
    \right)\sin\theta \nonumber
    \\
    & \qquad 
    + \im\left[
    2 A_1 A_3^* + (A_4 - A_2) A_5^*
    + 4 (A_1 + A_3) A_6^*
    \right]\sin(2\theta) \\
    & \qquad 
    + \im\left[
    A_2 A_3^* + A_3 A_4^*
    + (A_1 + A_3) A_5^*
    + (A_2 - A_4 + 2 A_5) A_6^*
    \right]\sin(3\theta)
    + \frac{1}{2}\im\left[
    (A_2 - A_4) A_5^*
    \right]\sin(4\theta)
\Bigg].  \notag
\end{align}
\begin{align}
\tilde{C}_{lin}^{x'y'}=
& \frac{1}{4}\Bigg[
    -\re\left[
    3 A_2 A_3^* + 5 A_3 A_4^*
    + (3 A_1 - A_3) A_5^*
    + 3 (A_2 + A_4 - 2 A_5) A_6^*
    \right]\sin\theta \nonumber
    \\
  &  + \left(
    2\left(\left|A_5\right|^2 - \left|A_3\right|^2\right)
    + \re\left[
    2 A_1 A_3^* + (A_4 - A_2) A_5^*
    - 4 A_3 A_6^*
    \right]
    \right)\sin(2\theta) 
    \\
  &  + \re\left[
    A_1A^*_5+(A_2-A_4-3A_5)A_3^*+ (A_2 + A_4 - 2 A_5) A_6^*
    \right]\sin(3\theta) 
    +\left(\frac{
    \re\left[(A_2 - A_4) A_5^*\right]}{2}
    - \left|A_5\right|^2
    \right)\sin(4\theta)
\Bigg]. \notag
\end{align}
\begin{align}
\tilde{C}_{lin}^{y'z'}=& \frac{1}{4}\Bigg[
    \im\left[
    3 A_2 A_3^* + 3 A_1 A_4^* - 2 (A_1 + 4 A_3) A_5^* + 6 (A_2 + A_5) A_6^*
    \right]\sin\theta \nonumber
    \\
    & \qquad 
    + \im\left[
    -2 A_1 A_3^* + A_2 A_4^* + 2 A_2 A_5^* - 4 (A_1 + A_3) A_6^*
    \right]\sin(2\theta)
    \\
    & \qquad 
    - \im\left[
    A_2 A_3^* + A_1 A_4^* + 2 A_1 A_5^* + 2 (A_2 + A_5) A_6^*
    \right]\sin(3\theta)
    - \frac{1}{2}\im\left[
    A_2 A_4^* + 2 A_2 A_5^*
    \right]\sin(4\theta)
\Bigg]. \notag
\end{align}
\begin{align}
\tilde{C}_{lin}^{z'x'}=& \frac{1}{8}\Bigg[
    \im\left[
    4 A_1 A_3^* - (3 A_2 + 5 A_4) A_5^*
    + 4 (A_1 + A_3) A_6^*
    \right]
    + 2\im\left[
    A_2 A_3^* + A_3 A_4^*
    + (A_1 + A_3) A_5^*
    - 2 A_4 A_6^*
    \right]\cos\theta \nonumber
    \\    &\qquad 
    - 4\im\left[
    A_1 A_3^* - (A_2 + A_4) A_5^*
    + (A_1 + A_3) A_6^*
    \right]\cos(2\theta) \nn &  \qquad 
    - 2\im\left[
    A_2 A_3^* + A_3 A_4^*
    + (A_1 + A_3) A_5^*
    - 2 A_4 A_6^*
    \right]\cos(3\theta)
    - \im\left[
    (A_2 - A_4) A_5^*
    \right]\cos(4\theta)
\Bigg].
\end{align}

\begin{align}
\tilde{C}_{lin}^{z'y'} =
& \frac{1}{8}\Bigg[
    -\left(
    4\left|A_3\right|^2 + 2\left|A_5\right|^2 + 8\left|A_6\right|^2
    + \re\left[
    12 A_1 A_3^* + 3 (A_2 - A_4) A_5^* + 4 (A_1 + 3 A_3) A_6^*
    \right]
    \right) \nonumber
    \\
    & \qquad 
    + 2\re\left[
    A_2 A_3^* - A_3 A_4^* + (A_1 - 3 A_3) A_5^* - 4 A_5 A_6^*
    \right]\cos\theta 
    \\
    & \qquad 
    + 4\left(
    \left|A_3\right|^2 + 2\left|A_6\right|^2
    + \re\left[
    - A_1 A_3^* + (A_2 - A_4) A_5^* + (A_1 + 3 A_3) A_6^*
    \right]
    \right)\cos(2\theta) \nonumber
    \\
    & \qquad 
    - 2\re\left[
    A_2 A_3^* - A_3 A_4^* + (A_1 - 3 A_3) A_5^* - 4 A_5 A_6^*
    \right]\cos(3\theta)
    + \left(
    2\left|A_5\right|^2 + \re\left[(A_4 - A_2) A_5^*\right]
    \right)\cos(4\theta)
\Bigg]. \notag 
\end{align}
Using the relation (\ref{transf}) or its inverse (\ref{inverse}), one can check that $A$ and $B_N^{y'}$ agree exactly with (\ref{unpolc}) and (\ref{bn}), respectively. The other components seemingly do not agree, but this is due to the different choice of the photon polarization basis. 
 The transformation between the linear and helicity  bases can be performed by a constant matrix multiplication  
\beq
\begin{pmatrix}
\epsilon^*_{+}M\epsilon_{+}
& \epsilon^*_{+}M \epsilon_{-} \\
\epsilon^*_{-}M\epsilon_{+} & \epsilon^*_{-}M \epsilon_{-}
\end{pmatrix} 
= \frac{1}{\sqrt{2}}\begin{pmatrix} 1 & i \\ 1 & -i \end{pmatrix} \begin{pmatrix} M^{x'x'} & M^{x'y'} \\ M^{y'x'} & M^{y'y'}\end{pmatrix} \frac{1}{\sqrt{2}}\begin{pmatrix} 1 & 1 \\ -i & i \end{pmatrix} \equiv U^\dagger M U ,\label{ttm}
\eeq
for generic matrix $M$ in the photon spin space. 
Writing $M=M^0+M^a\tau^a$, we find the general relation between the two bases  
\beq
M^a =\frac{1}{2} {\rm Tr} \left[U^\dagger \tau^aU \begin{pmatrix}
\epsilon^*_{+'}M\epsilon_{+'}
& \epsilon^*_{+'}M \epsilon_{-'} \\
\epsilon^*_{-'}M\epsilon_{+'} & \epsilon^*_{-'}M \epsilon_{-'}
\end{pmatrix} \right], \qquad U^\dagger (\tau^{x'},\tau^{y'},\tau^{z'})U=(-\tau^{y'},-\tau^{z'},\tau^{x'}).  
\eeq
This immediately gives $B_{lin}^{z'}=B_{hel}^{x'}=B_\gamma^{x'}$. As for the $C$-matrix, we find the correspondence 
\begin{align}
& C_{lin}^{x'x'}=-C_{hel}^{x'y'} =C^{x'y'}, \qquad  C_{lin}^{x'y'}=-C_{hel}^{x'z'}=C^{x'z'}, \qquad C_{lin}^{y'z'}=C_{hel}^{y'x'} =C^{y'x'}, \nn 
& C_{lin}^{z'x'}=-C_{hel}^{z'y'}=C^{z'y'}, \qquad C_{lin}^{z'y'}=-C_{hel}^{z'z'}=C^{z'z'}.
\end{align}
Using again (\ref{transf}), we have checked that these relations are  indeed consistent with (\ref{bgam}) and (\ref{cmatrix}). Thus the two approaches based on $\phi$'s and $A$'s give equivalent results, although obviously the helicity amplitude formalism is much simpler.


\section{Spin density matrix with single initial  polarization}\label{app::density_singleInitialPol}

In this appendix, we list formulas for the spin density matrix  when one of the incoming particles is  polarized. Namely, in the notation of Sec.~\ref{polcom}, we compute   $B_{N,\gamma}(\vec{s}_N,0)$, $C(\vec{s}_N,0)$ (polarized nucleon) and $B_{N,\gamma}(0,\vec{s}_\gamma)$, $C(0,\vec{s}_\gamma)$ (polarized photon). The case of double initial polarization is more complicated, and we have shown only numerical results in the main text. 

We first notice that, in the denominators of (\ref{cpol})-(\ref{bpol}), 
\beq
{\rm Tr}\left[T^\dagger T\left(\frac{\mathbb{I}+\vec{s}_N\cdot \vec{\sigma} }{2}\otimes \mathbb{I}\right)\right] =\begin{cases} |\phi_1|^2+|\phi_2|^2+2|\phi_3|^2+2|\phi_4|^2+|\phi_5|^2+|\phi_6|^2\equiv D, \qquad (\vec{s}_N=\pm \hat{x},\pm \hat{z}) \\
D  \pm 2\im \left[(\phi_2-\phi_6)\phi_3^*-(\phi_1+\phi_5)\phi_4^*\right]\equiv D_N^{\pm y} .\qquad (\vec{s}_N=\pm \hat{y})
\end{cases}
\eeq
The extra term in the $\vec{s}_N=\pm \hat{y}$ case  arises only when the amplitudes are imaginary and different amplitudes interfere. In spin physics, this effect is known as    transverse  single spin asymmetry. Similarly, 
\beq
{\rm Tr}\left[T^\dagger T\left(\mathbb{I}\otimes \frac{\mathbb{I}+\vec{s}_\gamma\cdot \tau^x\vec{\tau}\tau^x}{2}\right)\right] =\begin{cases} D, \qquad (\vec{s}_\gamma = \pm \hat{y},\pm \hat{z}) \\ 
D\pm 2\re \left[(\phi_1+\phi_5)\phi_3^*+(\phi_2-\phi_6)\phi_4^*\right]\equiv D_\gamma^{\pm x}. \qquad (\vec{s}_\gamma =\pm  \hat{x}) \end{cases}
\eeq
The difference between  the $\vec{s}_\gamma=-\hat{x}$ (vertical $\perp$  polarization) and  $\vec{s}_\gamma=\hat{x}$ (horizontal $\parallel$ polarization)  cases results in the beam asymmetry \cite{Hagelstein:2015egb}\footnote{This has a sign opposite to Eq.~(34) of \cite{Huang:2001ej}  due to a different sign convention for the $\phi$'s, see Footnote 4.}
\beq
\frac{1}{2}\left(\frac{d\sigma_\perp}{d\Omega} - \frac{d\sigma_\parallel}{d\Omega}\right)  = \frac{-1}{64\pi^2s} \re \left[(\phi_1+\phi_5)\phi_3^*+(\phi_2-\phi_6)\phi_4^*\right]. 
\eeq
In the following, we use a simplified notation $\phi_{i\pm j}\equiv \phi_i\pm \phi_j$ to shorten  lengthy expressions.  

Nucleon polarized in the $\hat{x}$ direction:
\beq
&& \vec{B}_N(\pm \hat{x},0)=\frac{1}{D}\begin{pmatrix} \pm 2\re[|\phi_3|^2-|\phi_4|^2+\phi_1\phi_5^*+\phi_2\phi_6^*] \\ 
2\im[\phi_{2-6}\phi_3^*-\phi_{1+5}\phi_4^*] \\ 
\mp 2\re[\phi_{2-6}\phi_3^*+\phi_{1+5}\phi_4^*] \end{pmatrix}, \quad \vec{B}_\gamma(\pm \hat{x},0)=\frac{1}{D} \begin{pmatrix} \pm 2\im[\phi_1\phi_6^*-\phi_2\phi_5^*] \\ \pm 2\re[\phi_{1-5}\phi_4^*+\phi_{2+6}\phi_3^*] \\ 2\re[\phi_{1+5}\phi_3^*+\phi_{2-6}\phi_4^*] \end{pmatrix},
\nn
&& C(\pm \hat{x},0) =\frac{1}{D}    \begin{pmatrix} 2\im[\phi_1\phi_2^*+\phi_5\phi_6^*] & 2\re[\phi_{2+6}\phi_3^*-\phi_{1-5}\phi_4^*]  & \pm 2\re[\phi_{1+5}\phi_3^*-\phi_{2-6}\phi_4^*] \\ \pm 2\re[\phi_{1-5}\phi^*_3-\phi_{2+6}\phi_4^*] & \pm 2\im[\phi_1\phi_5^*+\phi_2\phi_6^*] & 2\im[\phi_5\phi_6^*-\phi_1\phi_2^*-2\phi_3\phi_4^*] \\ 2\im[\phi_{1-5}\phi_3^* + \phi_{2+6}\phi_4^*] & |\phi_5|^2+|\phi_6|^2-|\phi_1|^2-|\phi_2|^2 & \pm 2\re[\phi_1\phi_6^* -\phi_2\phi_5^* -2\phi_3\phi_4^*]
\end{pmatrix}. 
\eeq

Nucleon polarized in the $\hat{y}$ direction:
\beq
&&\hspace{-7mm} \vec{B}_N(\pm \hat{y},0)=\frac{2}{D_N^{\pm y}} \begin{pmatrix} 0 \\ \begin{array}{c} \pm \re[|\phi_3|^2+ |\phi_4|^2+\phi_1\phi_5^*-\phi_2\phi_6^*] \\ +\im[\phi_{2-6}\phi_3^* -\phi_{1+5}\phi_4^*]\end{array} \\ 0\end{pmatrix}, \quad \vec{B}_\gamma(\pm \hat{y},0)=\frac{2}{D_N^{\pm y}} \begin{pmatrix} 0\\ 0\\  \begin{array}{c} \re[\phi_{1+5}\phi_3^*+\phi_{2-6}\phi_4^*] \\ \pm \im[\phi_1\phi_6^*+\phi_2\phi_5^*-2\phi_3\phi_4^*]\end{array} \end{pmatrix} , \nn
&& \hspace{-7mm} C(\pm \hat{y},0) =\frac{1}{D_N^{\pm y}}    \begin{pmatrix} \begin{array}{c} \mp 2\re[\phi_{1-5}\phi_3^*+\phi_{2+6}\phi_4^*] \\ +2\im[\phi_1\phi_2^*+\phi_5\phi_6^*] \end{array} & \begin{array}{c} 2\re[\phi_{2+6}\phi_3^* -\phi_{1-5}\phi_4^*]  \\ \pm 2\im[\phi_2\phi_6^* -\phi_1\phi_5^*] \end{array} & 0 \\ 0 & 0 & \begin{array}{c} \pm 2\re[\phi_{1+5}\phi_3^*+\phi_{2-6}\phi_4^*]\\ +2\im[\phi_5\phi_6^*-\phi_1\phi_2^*-2\phi_3\phi_4^*] \end{array} \\ \begin{array}{c}  2\im[\phi_{1-5}\phi_3^*+\phi_{2+6}\phi_4^*] \\ \mp 2\re[\phi_2\phi_5^*+\phi_1\phi_6^*]  \end{array}  & \begin{array}{c} |\phi_5|^2+|\phi_6|^2-|\phi_1|^2-|\phi_2|^2 \\ \pm 2\im[(\phi_{1-5}\phi_4^*-\phi_{2+6}\phi_3^*] \end{array} & 0
\end{pmatrix}.  
\eeq

Nucleon polarized in the $\hat{z}$ direction:
\beq
&& \hspace{-4mm} \vec{B}_N(\pm \hat{z},0) = \frac{1}{D} \begin{pmatrix} \pm 2\re[\phi_{2-6}\phi_3^*+\phi_{1+5}\phi_4^*] \\ 2\im[\phi_{2-6}\phi_3^* -\phi_{1+5}\phi_4^*] \\ \begin{array}{c} \pm(|\phi_1|^2-|\phi_2|^2+2|\phi_3|^2-2|\phi_4|^2\\ +|\phi_5|^2 -|\phi_6|^2) \end{array} \end{pmatrix} ,\quad \vec{B}_\gamma(\pm \hat{z},0)=\frac{1}{D} \begin{pmatrix} \pm 2\im[\phi_{1-5}\phi_3^*-\phi_{2+6}\phi_4^*]  \\ \pm( |\phi_2|^2+|\phi_5|^2 -|\phi_1|^2- |\phi_6|^2)  \\ 2\re[\phi_{1+5}\phi_3^*+\phi_{2-6}\phi_4^*]
\end{pmatrix}  , \nn
&& C(\pm \hat{z},0) =\frac{1}{D}    \begin{pmatrix} 2\im[\phi_1\phi_2^*+\phi_5\phi_6^*] & 2\re[\phi_{2+6}\phi_3^*-\phi_{1-5}\phi_4^*]  & \pm 2\re[\phi_1\phi_2^*+2\phi_3\phi_4^*-\phi_5\phi_6^*] \\ \pm 2\re[\phi_1\phi_2^*+\phi_5\phi_6^*] & \pm 2\im[\phi_{1-5}\phi_4^*+\phi_{2+6}\phi_3^*] & 2\im[\phi_5\phi_6^*-\phi_1\phi_2^* - 2\phi_3\phi_4^*] \\ 2\im[\phi_{1-5}\phi_3^* +\phi_{2+6}\phi_4^*] & |\phi_5|^2+|\phi_6|^2-|\phi_1|^2-|\phi_2|^2 & \pm 2\re[\phi_{1+5}\phi_3^*-\phi_{2-6}\phi_4^*]
\end{pmatrix}.  
\eeq

Photon linearly polarized  $\hat{s}_\gamma =\pm\hat{x}$:
\beq
&& \hspace{-5mm} \vec{B}_N(0, \pm \hat{x})= \frac{2}{D_\gamma^{\pm x}} \begin{pmatrix} 0 \\   \begin{array}{c}\im[\phi_{2-6}\phi_3^* -\phi_{1+5}\phi_4^* \\ 
\pm (\phi_1\phi_6^*+\phi_2\phi_5^*-2\phi_3\phi_4^*)] \end{array} \\  0\end{pmatrix}, \quad  \vec{B}_\gamma(0,\pm  \hat{x})= \frac{2}{D_\gamma^{\pm x}} \begin{pmatrix} 0 \\ 0 \\  \begin{array}{c} \re[\phi_{1+5}\phi_3^* +\phi_{2-6}\phi_4^* \\ 
\pm (\phi_1\phi_5^*-\phi_2\phi_6^*+|\phi_3|^2+|\phi_4|^2)] \end{array} \end{pmatrix},
\nn
&&C(0,\pm \hat{x})= \frac{1}{D_\gamma^{\pm x}} \begin{pmatrix} \begin{array}{c} 2\im[ \phi_1\phi_2^*+\phi_5\phi_6^*  \\ \pm(\phi_{1-5}\phi_4^*-\phi_{2+6}\phi_3^*)] \end{array} &  \begin{array}{c} 2\re[\phi_{2+6}\phi_3^*-\phi_{1-5}\phi_4^* \\ 
\pm (\phi_1\phi_6^*+\phi_2\phi_5^*)] \end{array} & 0 \\ 0& 0 & \begin{array}{c} 2\im[\phi_5\phi_6^*-\phi_1\phi_2^*-2\phi_3\phi_4^* \\ 
\pm(\phi_{2-6}\phi_3^*-\phi_{1-5}\phi_4^*)] \end{array} \\ \begin{array}{c} 2\im[\phi_{1+5}\phi_3^*-\phi_{2+6}\phi_4^* \\ \pm (\phi_1\phi_5^* -\phi_2\phi_6^*)] \end{array} & \begin{array}{c} |\phi_5|^2+|\phi_6|^2-|\phi_1|^2-|\phi_2|^2 \\ \mp 2\re [ \phi_{1-5}\phi_3^* +\phi_{2+6}\phi_4^*] \end{array}  & 0 \end{pmatrix}.
\eeq

Photon linearly polarized $\vec{s}_\gamma =\pm\hat{y}$:
\beq
&& \vec{B}_N(0, \pm\hat{y})= \frac{2}{D} \begin{pmatrix}  \pm \im[\phi_1\phi_6^*-\phi_2\phi_5^*]  \\ \im[\phi_{2-6}\phi_3^*-\phi_{1+5}\phi_4^*] \\ \pm \im[\phi_{2+6}\phi_4^*-\phi_{1-5}\phi_3^* ] \end{pmatrix}, \quad 
\vec{B}_\gamma(0,\pm\hat{y})= \frac{2}{D} \begin{pmatrix} \pm \re[\phi_1\phi_5^*+\phi_2\phi_6^*-|\phi_3|^2+|\phi_4|^2] \\ \pm \im[\phi_{1+5}\phi_3^*-\phi_{2-6}\phi_4^* ] \\ \re[\phi_{1+5}\phi_3^* +\phi_{2-6}\phi_4^*]
\end{pmatrix}, \nn
&&  C(0,\pm\hat{y})=\frac{1}{D}   \begin{pmatrix}  2\im[\phi_1\phi_2^*+\phi_5\phi_6^*] & 2\re[\phi_{2+6}\phi_3^* -\phi_{1-5}\phi_4^*] &  \mp 2\im[\phi_{2+6}\phi_3^* +\phi_{1-5}\phi_4^*] \\  \mp 2\im[\phi_{2-6}\phi_3^* +\phi_{1+5}\phi_4^*] &  \mp 2\re[ 2\phi_3\phi_4^*-\phi_2\phi_5^*+\phi_1\phi_6^*] &  2\im[\phi_5\phi_6^* -\phi_1\phi_2^* -2\phi_3\phi_4^*] \\ 2\im[\phi_{1-5}\phi_3^*+\phi_{2+6}\phi_4^*] & |\phi_5|^2+|\phi_6|^2-|\phi_1|^2-|\phi_2|^2 &  \mp 2\im[\phi_1\phi_5^*+\phi_2\phi_6^*] .
\end{pmatrix} 
\eeq

Photon circularly polarized  $\vec{s}_\gamma =\pm\hat{z}$:
\beq
&& \hspace{-5mm} \vec{B}_N(0, \pm\hat{z} ) 
= \frac{1}{D} \begin{pmatrix}  \mp 2\re[\phi_{1-5}\phi_4^*+\phi_{2+6}\phi_3^*]  \\ 2\im[\phi_{2-6}\phi_3^*-\phi_{1+5}\phi_4^*] \\ \pm (|\phi_2|^2+|\phi_5|^2-|\phi_1|^2-|\phi_6|^2) \end{pmatrix},\quad 
\vec{B}_\gamma(0, \pm \hat{ z})
=\frac{1}{D} \begin{pmatrix} \pm 2\im [\phi_{2-6}\phi_4^*-\phi_{1+5}\phi_3^*] \\
\begin{array}{c}\pm(|\phi_1|^2-|\phi_2|^2-2|\phi_3|^2+2|\phi_4|^2\\
+|\phi_5|^2-|\phi_6|^2)
\end{array}
\\ 2\re[\phi_{1+5}\phi_3^* + \phi_{2-6}\phi_4^* ] 
\end{pmatrix}, \nn
&& C(0, \pm\hat{z}) =\frac{1}{D}   \begin{pmatrix} 2\im[\phi_1\phi_2^*+ \phi_5\phi_6^*] & 2\re[\phi_{2+6}\phi_3^* -\phi_{1-5}\phi_4^*] &  \mp 2\re[\phi_1\phi_2^*+\phi_5\phi_6^*] \\ \pm 2\re[2\phi_3\phi_4^*+\phi_5\phi_6^*-\phi_1\phi_2^*] & \mp 2\im[\phi_{1+5}\phi_4^*+\phi_{2-6}\phi_3^*] &   2\im[\phi_5\phi_6^* -\phi_1\phi_2^* -2\phi_3\phi_4^*] \\  2\im[\phi_{1-5}\phi_3^*+\phi_{2+6}\phi_4^*] & |\phi_5|^2+|\phi_6|^2-|\phi_1|^2-|\phi_2|^2 &  \pm 2\re[\phi_{2+6}\phi_4^*-\phi_{1-5}\phi_3^* ]
\end{pmatrix}.
\eeq


\bibliography{bibliography}
	
\end{document}